\documentclass{aa}
\usepackage[varg]{txfonts}
\usepackage{graphics}
\usepackage{graphicx}
\usepackage{amssymb}
\usepackage{amsmath}
\usepackage{slashed}
\usepackage{float}
\usepackage{subcaption}
\newcommand{\be}{\begin{eqnarray}}
\newcommand{\ee}{\end{eqnarray}}
\newcommand{\beq}{\begin{equation}}
\newcommand{\eeq}{\end{equation}}

\newcommand{\bemul}{\begin{multline}}
\newcommand{\eemul}{\end{multline}}
\begin{document}

\titlerunning{Flavor composition of neutrinos from choked gamma-ray bursts}
\authorrunning{M.M. Reynoso and F.A. Deus}
\title{Flavor composition of neutrinos from choked gamma-ray bursts}

\author{Matías M. Reynoso\inst{1}
  \and Florencia A. Deus\inst{1} } 

\offprints{M.M. Reynoso, \email{mreynoso@mdp.edu.ar}}
\institute{Instituto de Investigaciones F\'{\i}sicas de Mar del Plata (IFIMAR – CONICET), and Departamento de F\'{\i}sica, Facultad de Ciencias Exactas y Naturales, Universidad Nacional de Mar del Plata, Funes 3350, (7600) Mar del Plata, Argentina} 

\date{Received 20 January 2023 }

\abstract {The nature of the astrophysical sources responsible for producing the observed high energy neutrinos have yet to be unveiled. Choked gamma-ray bursts (CGRBs) are sources that have been proposed as being capable of generating the flux detected by IceCube, since no accompanying gamma-ray signal is expected from them, as required by observations. {We focus on obtaining the neutrino flux and flavor composition corresponding to CGRBs under different assumptions for the target photon density and the magnetic field of the emission region.} We consider the injection of both electrons and protons into the internal shocks of CGRBs, and using a steady-state transport equation, we account for all the relevant cooling processes. In particular, we include the usually adopted background of soft photons, which is a fraction of the thermalized emission originated at the shocked jet head. Additionally, we consider the synchrotron photons emitted by the electrons co-accelerated with the protons at the internal shocks in the jet. We also obtain the distribution of charged pions, kaons, and muons using the transport equation to account for the cooling effects due not only to synchrotron emission but also interactions with the soft photons in the ambient. {We integrate the total diffuse flux of neutrinos of different flavors and compute the flavor ratios to be observed on Earth. As a consequence of the losses suffered mainly by pions and muons, we find these ratios to be dependent on the energy: for energies above $\sim ( 10^{5}-10^6)\, {\rm GeV}$ (depending on the magnetic field, proton-to-electron ratio, and jet power), we find that the electron flavor ratio decreases and the muon flavor ratio increases, while the tau flavor ratio increases only moderately.} {Our results are sensitive to the mentioned key physical parameters of the emitting region of CGRBs. Hence, the obtained flavor ratios are to be contrasted with cumulative data from ongoing and future neutrino instruments in order to assess the contribution of these sources to the diffuse flux of astrophysical neutrinos.} }

\maketitle



\section{Introduction}

Since 2014, neutrinos of astrophysical origin have been detected by IceCube in the energy range $ 10^4{\rm GeV}\lesssim E_\nu\lesssim 10^6{\rm GeV}$ \citep{aartsen2014}. However, the sources corresponding most of these neutrinos could not be directly associated with any known sources, such as the typically proposed active galactic nuclei \citep[AGN; e.g.,][]{stecker1991}, gamma-ray bursts \citep[GRBs; e.g.,][]{waxman1997}, tidal disruption events \citep[TDEs; e.g.,][]{dai2017}, and starburst galaxies \citep[e.g.,][]{loeb2006}. The few exceptions are as follows: First, the neutrino event IC-170922A was associated with the blazar TXS 0506+056, which was almost simultaneously detected in gamma rays by Fermi, VERITAS, and MAGIC \citep{aartsen2018}. Second, another AGN, the Seyfert galaxy NGC 1068, was recently detected by IceCube without a gamma-ray counterpart \citep{abbasi2022}. And third, the TDEs AT2019dsg \citep{stein2021} and AT2019fdr \citep{reusch2022} were detected in the optical/UV bands by the Zwicky Transient Facility, and they are very plausibly the IceCube counterparts of the muon track events IC191001A and IC200530A, respectively. 

Still, the origin of a great majority of the observed astrophysical neutrinos cannot be identified yet, and it can only be recognized that the sources should be extragalactic, given that the incoming directions in the sky are consistent with an isotropic emission. Among the candidate sources, choked gamma-ray bursts (CGRBs) arise as an interesting possibility. A CGRB is generated when a jet launched inside a collapsing massive star fails to emerge from the stellar mantle, yielding no gamma-ray counterpart. This phenomenon is expected to take place in some core-collapse supernovae (SNe), such as those of the types II \citep{macfadyen2001} and Ib/c. Although the fraction of such sources presenting jets is still uncertain \citep{piran2019}, the link between SNe and CGRBs is conjectured in a similar manner to the case of the confirmed connection between type Ib/c SNe and long GRBs with flat spectrum after the observation of several SNe at the same location previously detected as long GRB \citep{hjorth2012}. 
 Similar to the proposed scenarios of neutrino production in GRB jets \citep[e.g.,][]{waxman1997,murase2006,petropoulou2014}, the study of CGRBs as  hidden neutrino sources has been realized under different considerations \citep{meszaros2001,razzaque2004,murase2013,senno2016,he2018,fasano2021}. Recently,
\cite{chang2022} performed a statistical analysis using data from observed SNe Ib/c, and although they found no significant correlation with IceCube data regarding muon track events, they concluded that choked jets in SNe can still be the main contributors to the diffuse neutrino flux.
 
 We therefore have the relevant physical scenario well-established and we explore it to study the flavor composition of the emitted neutrinos from CGRBs in more detail and with the aim of obtaining specific signatures consistent with their assumed description. In general, it has been recognized that analyzing the flavor composition of the neutrino fluxes can be useful \citep{beacom2003,bustamante2015}, and particularly in highly magnetized sources, the synchrotron cooling of muons can gradually modify the emitted flavor ratio  $(f_{e}:f_{\mu}:f_{\tau})_{s}$ from $(1:2:0)$ to $(0:1:0)$ at high energies \citep{lipari2007,baerwald2012}. Along this line, a detailed statistical analysis was performed  in \cite{bustamente2020} using the available neutrino data along with a generic description of the neutrino sources, characterized by their comoving magnetic field $B$ and Lorentz factor $\Gamma$. In \cite{fiorillo2021}, they also studied the effect of different distributions of target photons and the impact on the flavor ratios for different benchmark parameters describing typical sources.
  
In this work, we focus on the particular case of CGRBs and consider different aspects that we believe can play a role and should be taken into account in the neutrino production model. First, we solve a stationary transport equation for primary electrons as well as for protons, which are both supposed to be injected in the region with internal shocks of the outflow. The synchrotron emission of these electrons is an additional contribution to the soft photon background that serves as a target for the high energy particles. The other contribution is the commonly assumed one, i.e., the thermalized emission of another electron population, which is accelerated by the reverse shocks at the jet head. A fraction of these photons is considered to escape to the internal shock region and also serve as targets for the high energy particles there. Specifically, the possible interactions with the background photons include: $p\gamma$ interactions (which produce pions and kaons) and inverse Compton (IC) interactions with electrons as well as with muons from the decay of pions. Taking these cooling processes into account, plus the synchrotron one, we also solve a transport equation to obtain the distributions of pions, muons, and kaons. This allows us to finally compute the neutrino emission consistently with the cooling processes affecting all the particle species involved in the production process. Hence, the generated neutrino flux is sensitive to important physical parameters, such as the jet power, magnetic field, and proton-to-electron ratio. {Although the values of the latter, as well as that of the energy budget \citep{eichler2010}, are uncertain, we aim to explore plausible combinations of them in order to determine their effect on the neutrino emission and flavor ratios.}

In particular, for energies above $\sim 10^5-10^6\,{\rm GeV,}$ the electron flavor ratio to be observed on Earth gradually decreases, the muon flavor ratio increases, and the tau one increases only slightly. Experimental data from present and future detectors will therefore help in determining whether CGRBs, as they are normally conceived, can significantly contribute to the diffuse neutrino flux.

The rest of the manuscript is organized as follows. In Sect. \ref{sec:model}, we present the basic scenario of the model adopted, and in Sect. \ref{sec:heparticles} we describe the method of calculation used to characterize the particle distributions and neutrino emissivity. In Sect. \ref{sec:fluxandflavor}, we present our results for the neutrino flux and flavor ratios as a function of the energy and also integrated over the energy. Finally, in Sect. \ref{sec:discussion}, we conclude with a discussion.


\section{Basic scenario}\label{sec:model}

\begin{figure}[tbp]
\includegraphics[width=.67\textwidth,trim=100 200 0 35,clip]{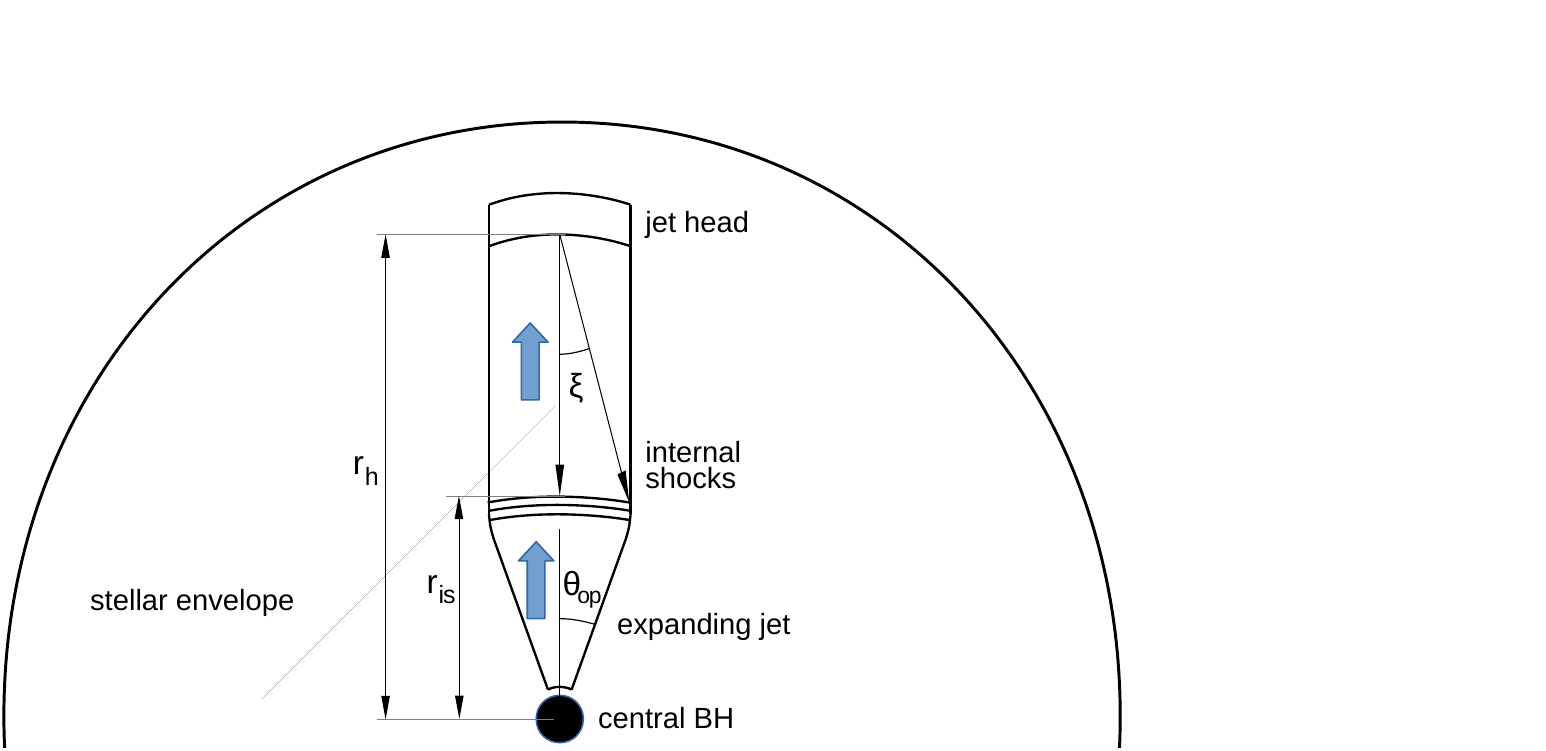}
\caption{Schematic view (not to scale) of the jet propagating inside a star. \label{fig1:sketch}}
\end{figure}
\begin{figure*}[h!]                            
\centering
\  \centering
    \begin{subfigure}[t]{0.4\textwidth}
        \centering                          
        \includegraphics[width=0.5\linewidth,trim= 130 30 210 0]{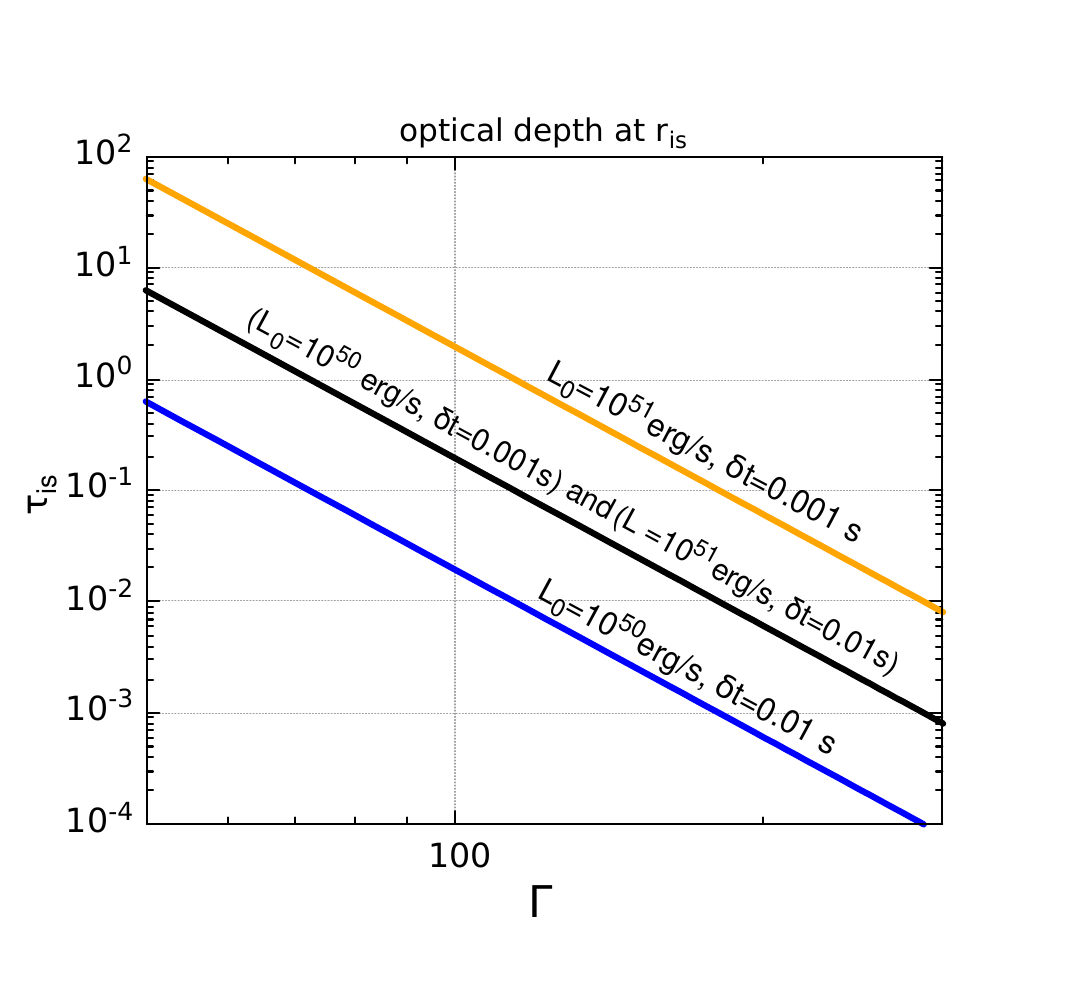} 
    \end{subfigure}
    \hfill
    \begin{subfigure}[t]{0.4\textwidth}
        \centering
        \includegraphics[width=0.5\linewidth,trim= 210 30 130 0]{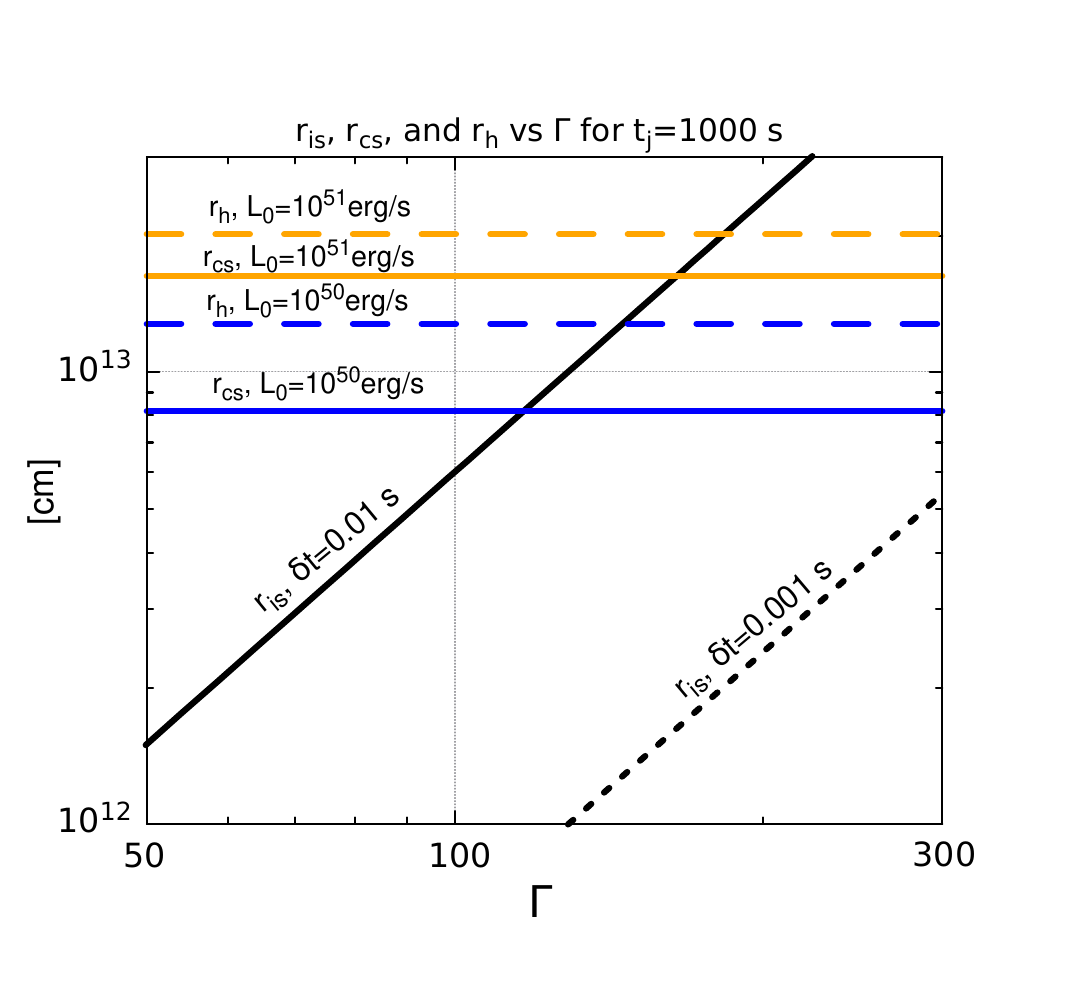} 
    \end{subfigure}
    \caption{Left panel: Optical depth at the internal shock region as a function of the Lorentz factor of the jet for $(L_0,\delta t)=(10^{50}{\rm erg/s},0.001 \,{\rm s})$ and $(L_0,\delta t)=(10^{51}{\rm erg/s},0.01 \,{\rm s})$, both marked by the black curve, while the blue line corresponds to  $(L_0,\delta t)=(10^{50}{\rm erg/s},0.01 \,{\rm s})$  and the orange one to $(L_0,\delta t )=(10^{51}{\rm erg/s},0.001 {\rm s})$. Right panel: Positions of the jet head ($r_{\rm h}$) as a function of the Lorentz factor of the jet for $L_0=10^{50}{\rm erg/s}$ and $L_0=10^{51}{\rm erg/s}$ marked by long-dashed blue and orange lines, respectively. Solid blue and orange lines correspond to the positions of collimation shock ($r_{\rm cs}$) for $L_0=10^{50}{\rm erg/s}$ and $L_0=10^{51}{\rm erg/s}$, respectively. The positions of the internal shocks ($r_{\rm is}$) for $\delta t=0.001\,{\rm s}$ and $\delta t=0.01\,{\rm s}$ are indicated with black short-dashed and solid lines, respectively. 
    }\label{fig2:tauis_rcs}
\end{figure*}

In this section, we briefly describe the model applied, which is based in the same set of assumptions adopted for CGRBs in several works \citep[e.g.,][]{meszaros2001,murase2013,senno2016,fasano2021}. {In particular,  our basic scenario is the same as in \cite{fasano2021}, since we assume similar values for the jet power, Lorentz factor, and the density of the stellar envelope. The main differences arise in the treatment applied to describe the high-energy particles involved in the neutrino production process, which in our case implies the use of a transport equation to account for the cooling processes, including $\pi\gamma$ and muon IC interactions. Additionally, as mentioned above, we also include high energy electrons as a population of particles that is co-accelerated with protons by the internal shock.}
  
Typically, the jet of a CGRB has a Lorentz factor $\Gamma\sim 100-300$ and a half-opening angle $\theta_{\rm op}\approx 0.2 {\rm rad}$, and it carries a power $L_{\rm j}$, which corresponds to an isotropic equivalent power 
\be 
L_{0}= 2L_{\rm j}/(1-\cos\theta)= 10^{50-51}{\rm erg \, s^{-1}}. 
\ee
Fixing a variability timescale $\delta t$, the distance from the central black hole (BH) to the position where internal shocks develop is
\be
r_{\rm is}= 2\Gamma^2c\delta t\simeq 6\times 10^{12}{\rm cm}\ {\delta t_{-2}} \Gamma_{2},
\ee
where $\Gamma_2=\Gamma/100$ and $\delta t_{-2}=\delta t/(0.01 {\rm s})$.
There, the comoving number density of cold protons is
\be
n'_{\rm j}= \frac{L_{0}}{4\pi\Gamma^2 r^2_{\rm is} m_p c^3} \simeq 4.9\times 10^{12}{\rm cm^{-3}}L_{0, 51}\Gamma_2^{-6}\delta t_{-2}^{-2}. 
\ee The jet is magnetized, and the magnetic energy density is usually taken to be a fraction $\epsilon_{B}$ of the kinetic energy density $m_p c^2 n'_{\rm j}$, that is:
\be
B'=\sqrt{\frac{2 \,\epsilon_B L_{0}}{\Gamma^2 r^2_{\rm is}  c}}= 1.36\times 10^{5}{\rm G} \ \epsilon_{B,-1}^{{1}/{2}}L_{0,51}^{{1}/{2}} \Gamma_{2}^{-2} \delta t_{-2}^{-1}. 
\ee
We also consider the case of equipartition, $\epsilon_B=1$, to explore a situation with a higher magnetic field and the consequences for neutrino production. 

As the jet propagates through the massive star, inside the extended hydrogen envelope up to a radius $r_{\rm ext}\approx 10^{13.5}{\rm cm}$, a forward shock and a reverse shock develop, and the jet can be stalled. {This has been studied through simulations and analytically by \cite{bromberg2011} and \cite{mizuta2013}.} Specifically, the part of the jet that is affected by the reverse shock is usually called the "jet head", and in the present context, it has been found that the Lorentz factor of this region is $\Gamma_{\rm h}\simeq 1$ {\citep[e.g.,][]{he2018},} and its position is given by \citep[e.g.,][]{mizuta2013,murase2013}
\be 
  r_{\rm h}\simeq 1.3\times 10^{13}{\rm cm} \, t_{\rm j,3}^{3/5}\left(\frac{L_0}{10^{50}{\rm erg/s}}\right)^{1/5}\left(\frac{\theta_{\rm op}}{0.2}\right)^{-4/5}\rho_{{\rm ext},-7}^{-1/5}.
\ee 
Here, the lifetime of the jet is $t_{\rm j}$, with $t_{\rm j,3}=t_{\rm j}/(1000\,{\rm s})$, and the density of the envelope $\rho_{\rm ext}$ at $r_{\rm ext}$ appears scaled with respect to the typical value, that is, $\rho_{\rm ext,-7}=\rho_{\rm ext}/(10^{-7}{\rm g \ cm^{-3}})$.
{Before reaching the jet head, the jet is expected to stop expanding and become collimated, keeping a cylindrical shape up to the position of the jet head $r_{\rm h}$. In these conditions, a collimation shock is generated at a distance from the BH \citep{bromberg2011,mizuta2013}}
\be 
r_{\rm cs}\simeq 8.2\times 10^{12}{\rm cm} \,t_{\rm j,3}^{2/5}\left(\frac{L_0}{10^{50}{\rm erg/s}}\right)^{3/10} \rho_{\rm ext,-7}^{-3/10}.
\ee

\begin{table*}[h!]
\caption{Parameters of the CGRB model }              
\label{table:params}      
\centering                                      
\begin{tabular}{c c c c}          
\hline\hline                        
input parameter &  description & values \\    
\hline                                   
    $L_0$[erg/s]  &  isotropic power  & $(10^{50};10^{51})$\\      
    $t_{\rm j}$[s] &  duration of typical CGRB &  $1000$ \\
    $\epsilon_{\rm rel}$ & fraction of power in relativistic particles & 0.1 \\
    $ a_{ep}$   & proton-to-electron ratio  & $(1; 100)$ \\
    $\Gamma$ & Lorentz factor of internal shock region & $(100;300)$ \\
    $\delta t {\rm[s]}$ & variability timescale  & $(0.001; 0.01)$ \\
    $\epsilon_B$  & magnetic-to-kinetic energy ratio & $(0.01; 0.1; 1)$  \\
\hline
\end{tabular}
\end{table*}

It is then assumed that a population of electrons in the shocked jet head emits synchrotron and IC radiation that thermalizes due to a high optical depth \citep[e.g.,][]{he2018}, 
\be
\tau_{\rm h}&\approx& (4\Gamma_{\rm rel}+3)n_{\rm j}(r_{\rm h}) \, \sigma_T  \nonumber \\ 
&\simeq&  1170 \,  \Gamma_2^{-2}L_{0,51}^{3/5}\rho_{\rm ext,-7}^{2/5}r_{\rm ext,13.5}^{5/2}t_{\rm j,3}^{3/5},
\ee
where $\Gamma_{\rm rel}\approx \Gamma/(2\Gamma_{\rm h})$ is the Lorentz factor of the jet head with respect to the jet.
 Specifically, a fraction $\epsilon_e\approx 0.1$ of the kinetic energy of the flow in the jet head is supposed to be carried by electrons so that the total radiation density of the thermalized emission is \citep[e.g.,][]{meszaros2001,he2018}: 
\be
a {T_{\rm h}}^4 = \epsilon_e 4(\Gamma_{\rm rel}+ 3)(\Gamma_{\rm rel}-1)n_j m_p c^2,  
\ee
which implies that
\begin{equation}
T_{\rm h}= \left[\epsilon_e(\Gamma_{\rm rel}+ 3)(\Gamma_{\rm rel}-1) \left(\frac{L_{\rm iso}}{a\,\pi \Gamma^2 r_{\rm h}^2  c}\right)\right]^\frac{1}{4}. 
\end{equation}
Hence, the distribution of these photons in the jet head frame is
\be
N_{\rm ph,h}(E_{\rm ph,h})= \frac{2 E_{\rm ph,h}^2}{(hc)^3\left[\exp\left(\frac{E_{\rm ph,h}}{kT_\gamma}\right)-1\right]} \ [\rm erg^{-1}cm^{-3}sr^{-1}], 
\ee
and it is normally assumed that a fraction $f_{\rm esc}=1/\tau_{\rm h}$ of them escape to the internal shock region. However, instead of making the approximation that all the escaping photons carry the same boosted energy $\approx\Gamma_{\rm rel}(2.8 k_B T)$ in the internal shock frame, we transform the photon distribution to the internal shock frame and integrate on a solid angle element, as we describe in the next paragraph.

In the internal shock frame, the photon energy $E_{\rm ph}$ is such that $E_{\rm ph,h}=\Gamma_{\rm rel}E_{\rm ph}(1- \beta_{\rm rel}x)$, where $x$ is the cosine of the angle between the relative velocity of the flow, $\vec{\beta}_{\rm rel}c$, and the photon momentum in the internal shock frame. Considering the Lorentz invariant \citep{dermer2002},
\be
\frac{1}{E\sqrt{E^2-m^2c}}\frac{d{\mathcal{N}}}{dV\,dE\, d\Omega},
\ee
we can obtain the distribution of photons in the internal shock frame as
\be
N_{\rm ph}(E_{\rm ph},x)= \frac{N_{\rm ph,h}(\Gamma_{\rm rel}E_{\rm ph}(1-\beta_{\rm rel}x))}{\Gamma_{\rm rel}^2(1-\beta_{\rm rel}x)^2} \ [\rm erg^{-1}cm^{-3}sr^{-1}].
\ee
We then obtain the differential photon density as 
\be
n_{\rm ph}(E_{\rm ph})=2\pi \int_{-1}^{x_{\rm is}}dx N_{\rm ph}(E_{\rm ph},x)   \ [\rm erg^{-1}cm^{-3}], \label{nphIS}
\ee
where we perform the integration on the solid angle subtended by the internal shock region with respect to the jet head position (where the thermal photons are produced), $\Delta\Omega_{\rm is}= 2\pi(x_{\rm is}+1)$. Here, $x_{\rm is}=\cos(\pi-\xi)$, with the angle $\xi$ being such that (see Fig. \ref{fig1:sketch})
$$\tan{\xi}=\frac{r_{\rm is}\tan{\theta_{\rm op}}}{(r_{\rm h}-r_{\rm is})/{\Gamma}}.$$ 
Therefore, the differential photon density of Eq.(\ref{nphIS}) characterizes the external photon background that could be relevant, in particular, for the neutrino producing $p\gamma$ interactions of the protons accelerated by the internal shocks.

\begin{figure*}[]                            
\centering
\  \centering
    \begin{subfigure}[t]{0.49\textwidth}
        \centering                          
        \includegraphics[width=0.5\linewidth,trim= 170 30 180 35]{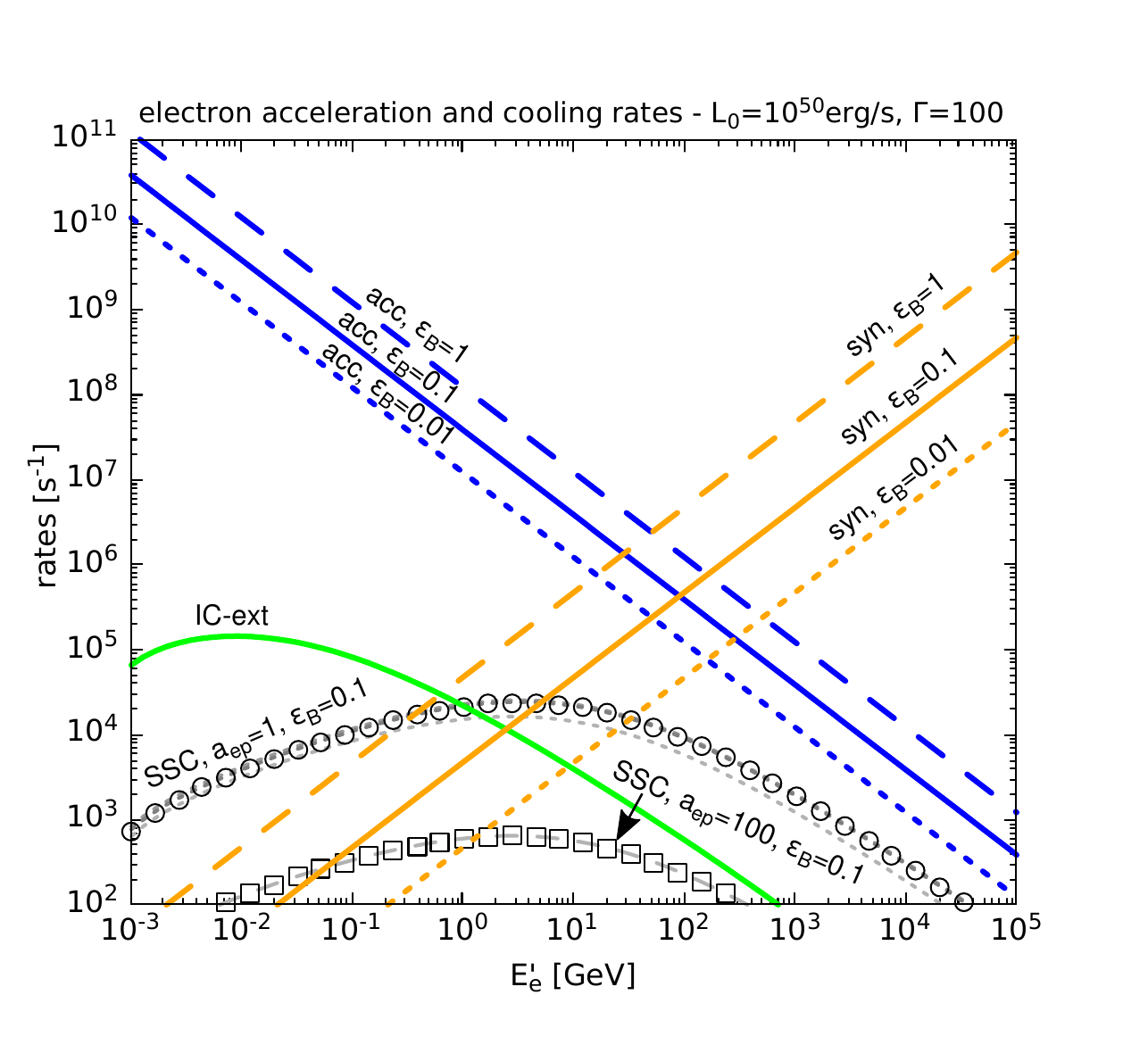} 
    \end{subfigure}
    \hfill
    \begin{subfigure}[t]{0.49\textwidth}
        \centering
        \includegraphics[width=0.5\linewidth,trim= 170 30 180 35]{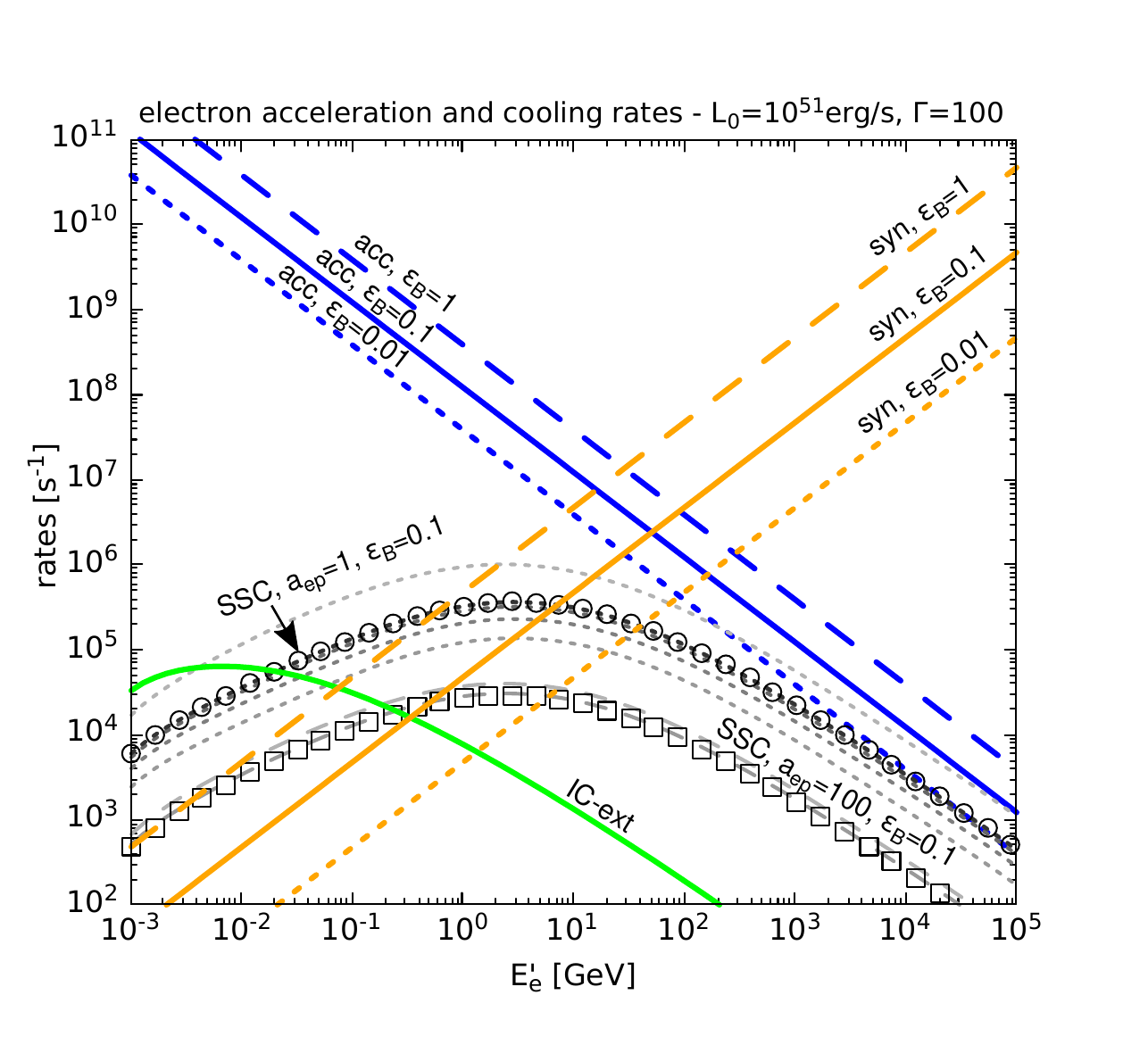} 
    \end{subfigure}

    \begin{subfigure}[t]{0.49\textwidth}
        \centering                          
        \includegraphics[width=0.5\linewidth,trim= 170 30 180 35]{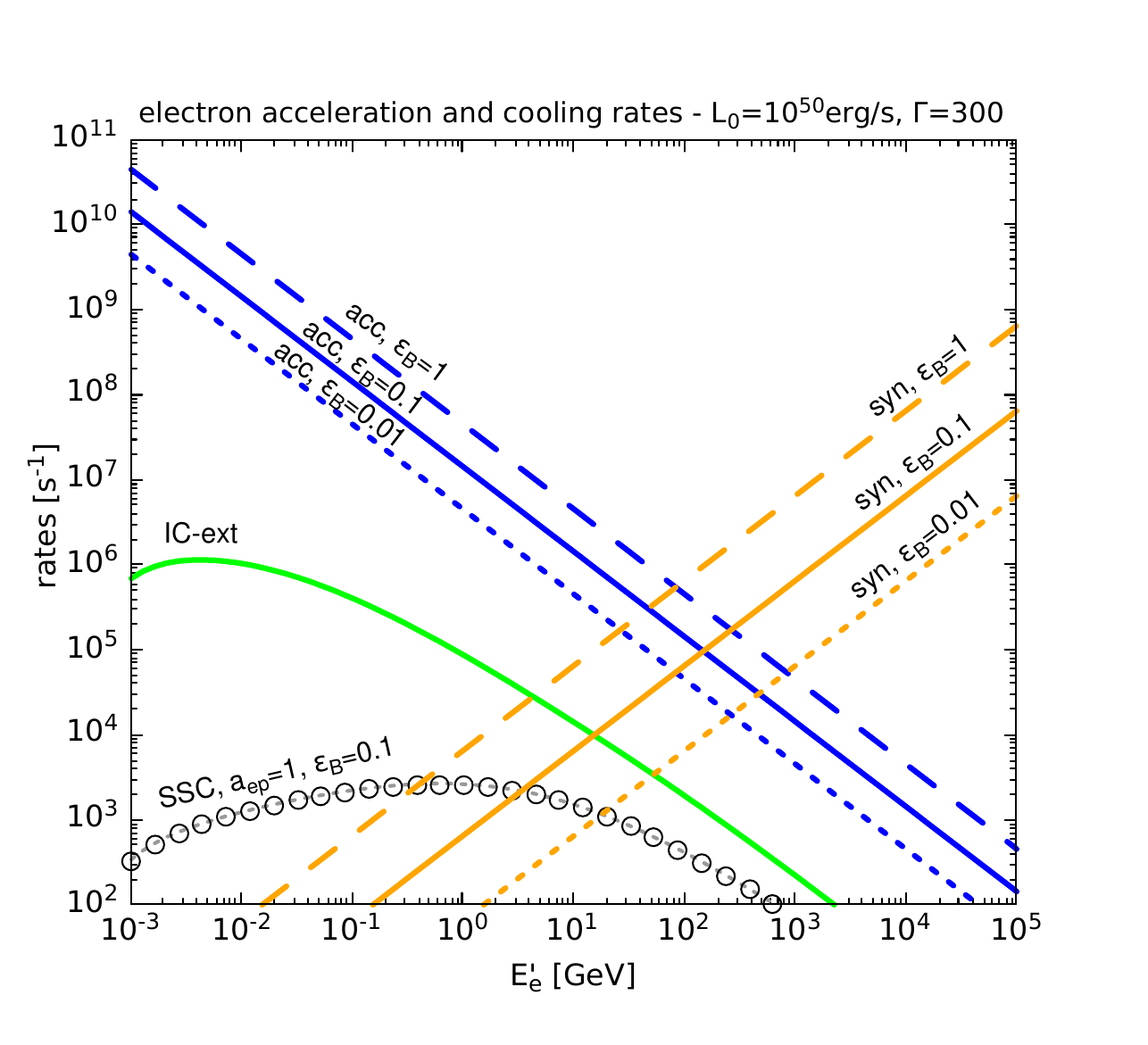} 
    \end{subfigure}
    \hfill
    \begin{subfigure}[t]{0.49\textwidth}
        \centering
        \includegraphics[width=0.5\linewidth,trim= 170 30 180 30]{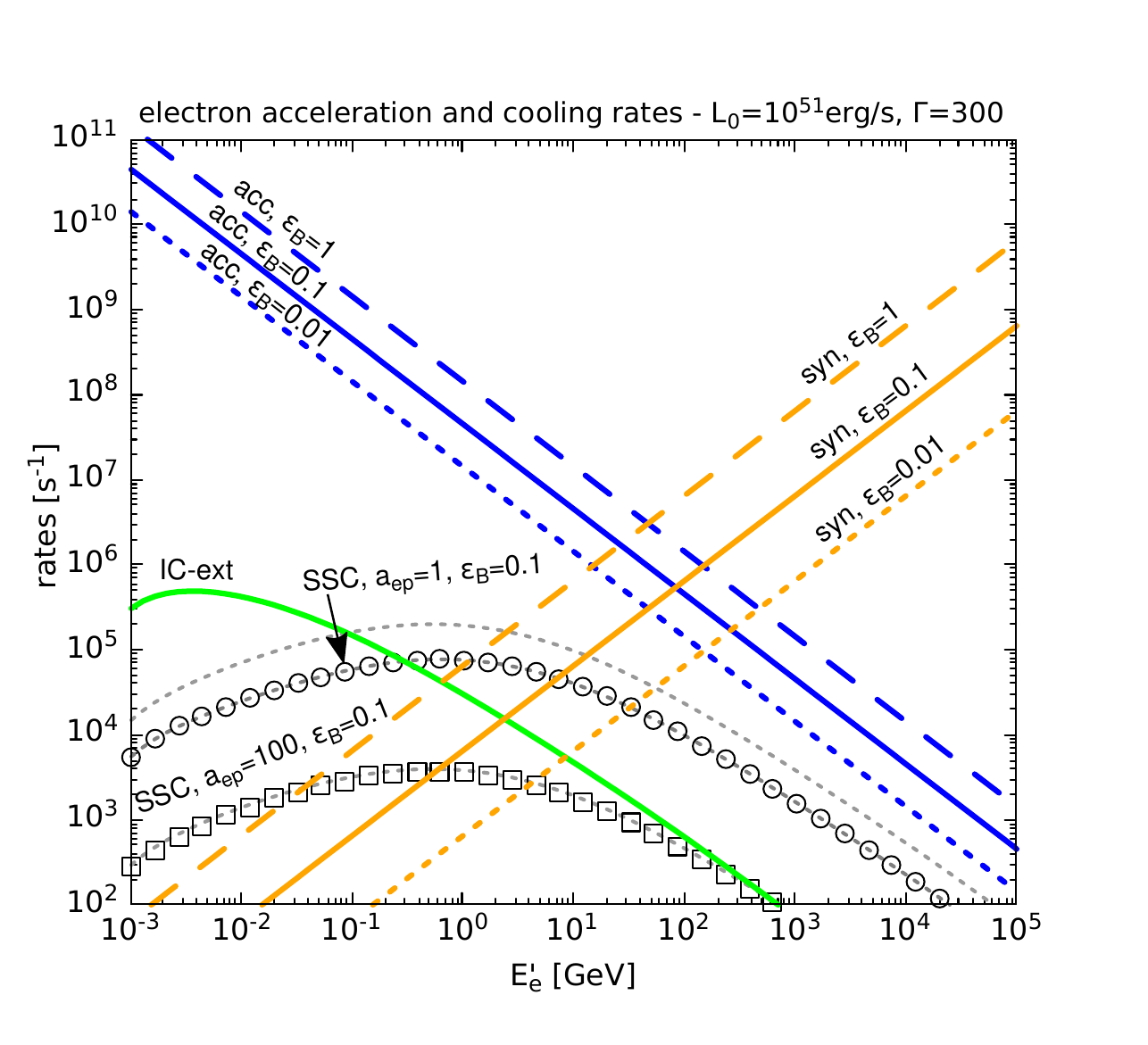} 
   \end{subfigure}     
    \caption{Cooling and acceleration rates for electrons at the internal shocks of CGRBs for $L_0=10^{50}{\rm erg/s}$ and $L_0=10^{51}{\rm erg/s}$ in the left and right panels, respectively. {Top panels correspond to $\Gamma=100$ and the bottom ones to $\Gamma=300$}. {The different processes are indicated with the following curves: Dashed lines are for $\epsilon_B=1$, solid lines for  $\epsilon_B=0.1$, and short-dashed lines for $\epsilon_B=0.01$. The blue curves refer to acceleration, orange ones to synchrotron, and green ones to external IC. The circles refer to SSC with $(a_{ep}=1,\epsilon_B=0.1)$ and squares to SSC with $(a_{ep}=100, \epsilon_B=0.1)$. With dashed gray lines, we also show the auxiliary results used to obtain the SSC rate with the iterative method mentioned in the text.}}\label{fig3:e-rates}
\end{figure*}

\begin{figure*}                            
\centering
\  \centering
    \begin{subfigure}[t]{0.49\textwidth}
        \centering                          
        \includegraphics[width=0.5\linewidth,trim= 185 30 185 0]{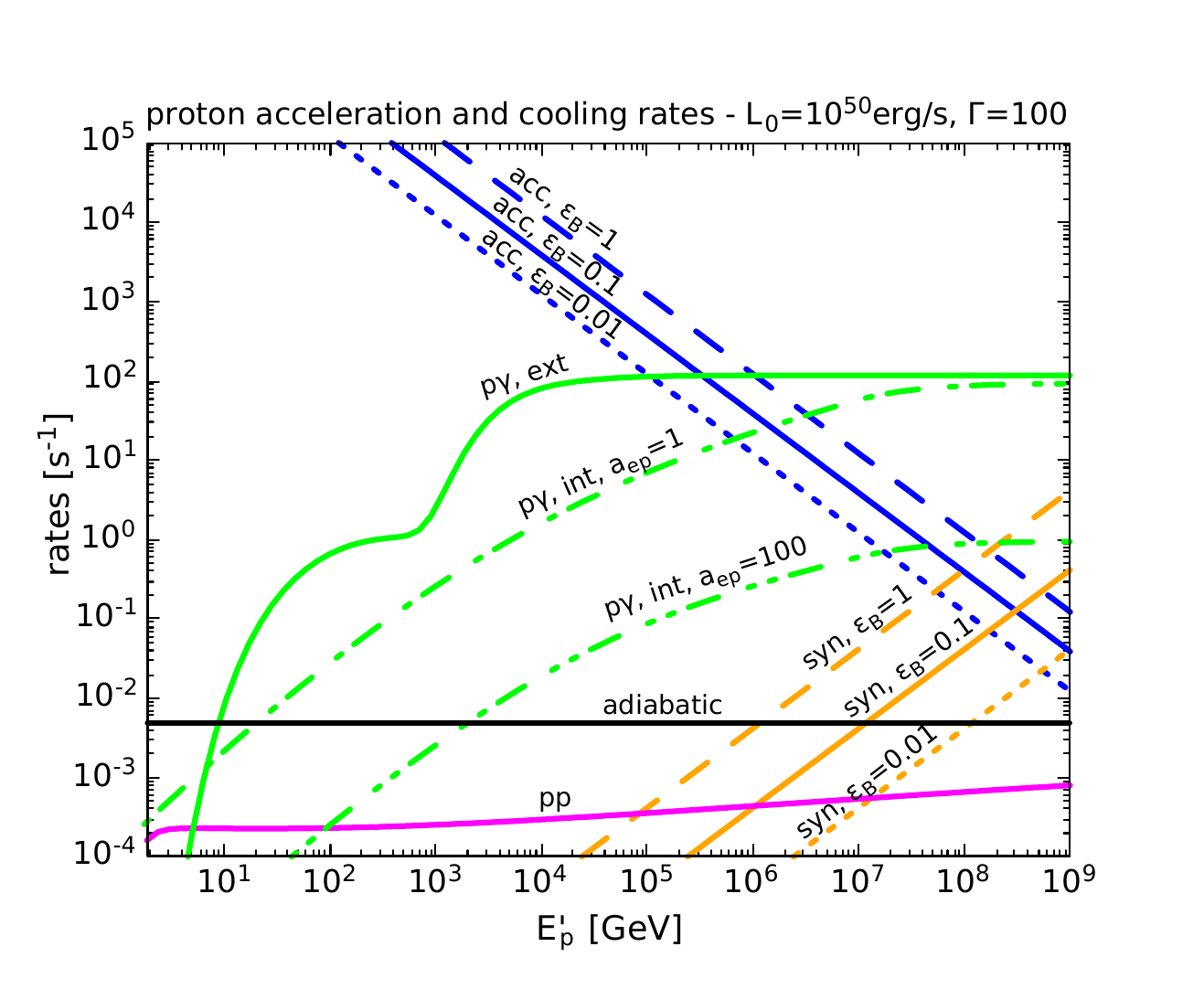} 
    \end{subfigure}
    \hfill
    \begin{subfigure}[t]{0.49\textwidth}
        \centering
        \includegraphics[width=0.5\linewidth,trim= 185 30 185 0]{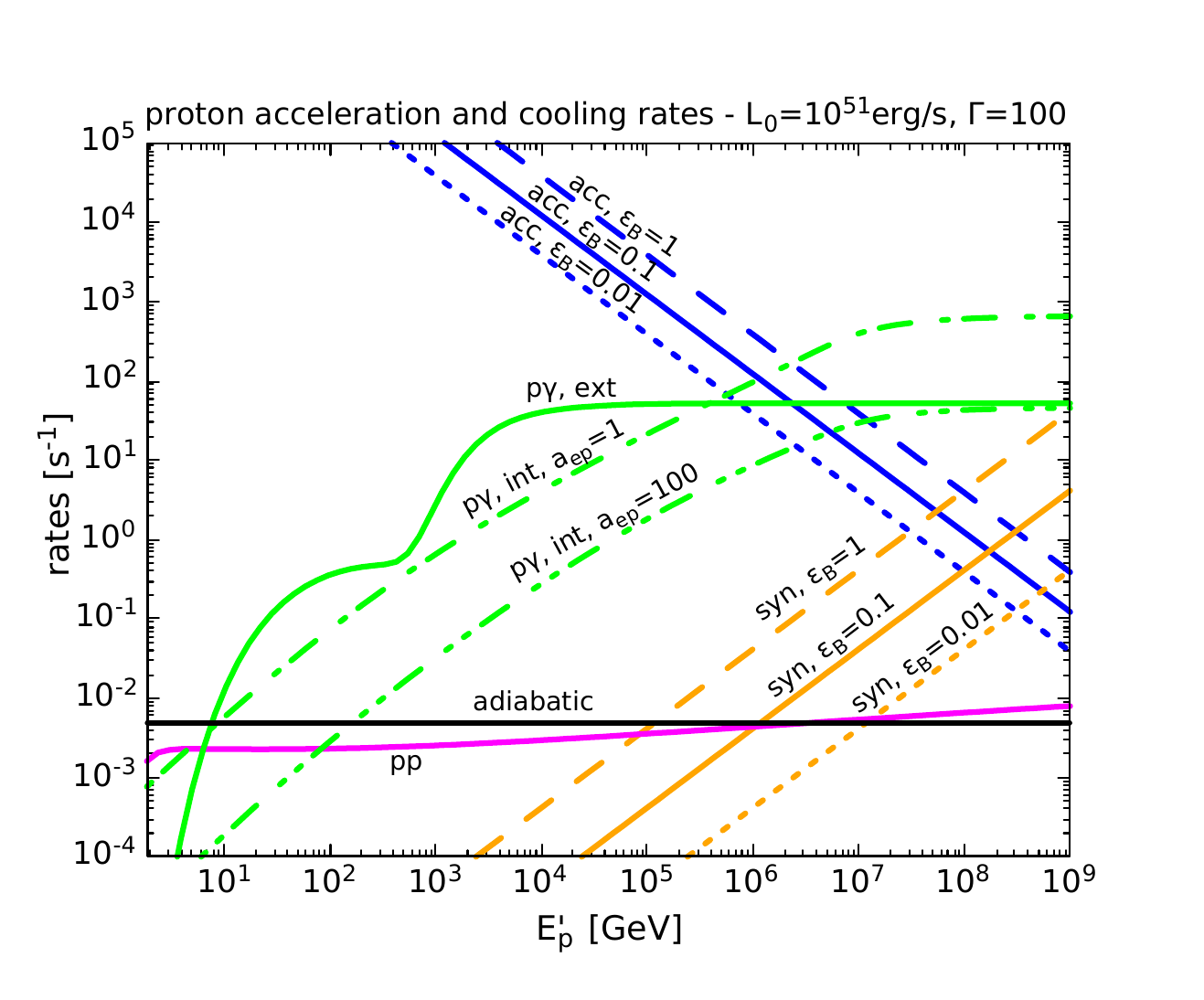} 
    \end{subfigure}
    \begin{subfigure}[t]{0.49\textwidth}
        \centering                          
        \includegraphics[width=0.5\linewidth,trim= 185 30 185 0]{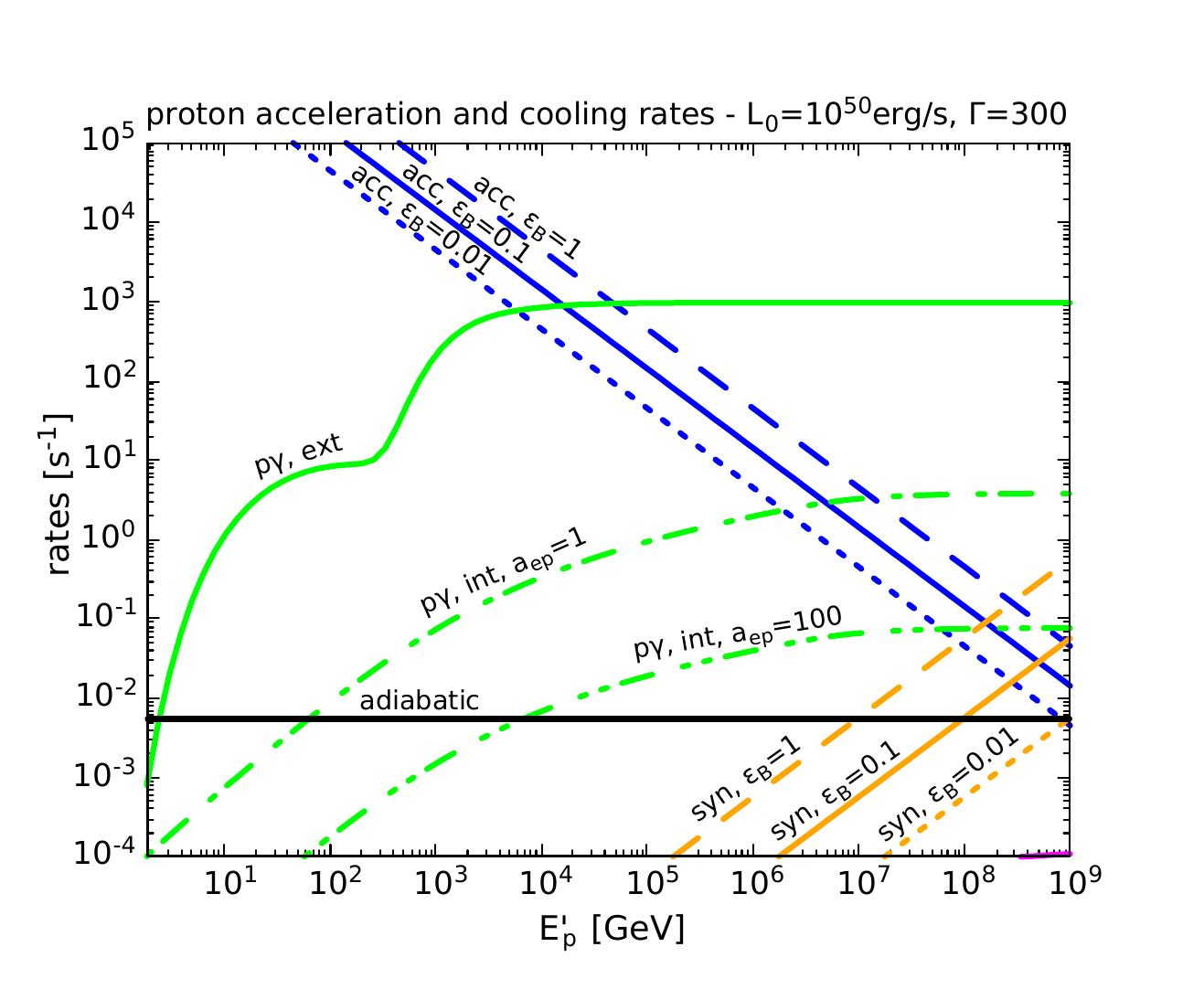} 
    \end{subfigure}
    \hfill
    \begin{subfigure}[t]{0.49\textwidth}
        \centering
        \includegraphics[width=0.5\linewidth,trim= 185 30 185 0]{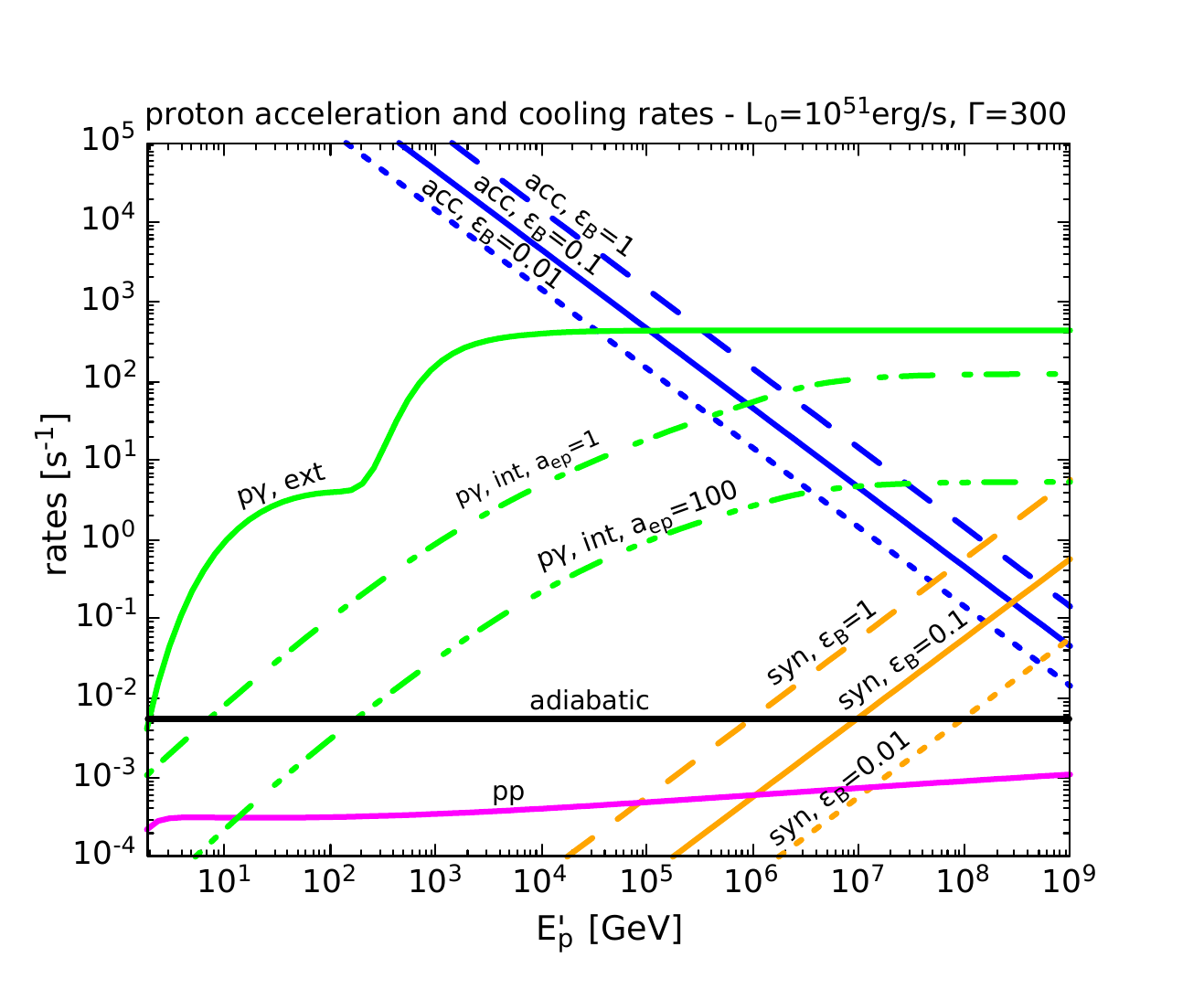} 
    \end{subfigure}
    \caption{Cooling and acceleration rates for protons at the internal shocks of CGRBs for $L_0=10^{50}{\rm erg/s}$ and $L_0=10^{51}{\rm erg/s}$ in the left and right panels, respectively. {Top panels correspond to $\Gamma=100$ and the bottom ones to $\Gamma=300$. The different processes are indicated with the following curves: Dashed lines are for $\epsilon_B=1$, solid lines for  $\epsilon_B=0.1$, and short-dashed lines for $\epsilon_B=0.01$. The blue curves refer to acceleration, orange ones to synchrotron, magenta ones to $pp$ interactions, and black ones to adiabatic cooling. With green, we mark the $p\gamma$ cooling rates corresponding to a target of external photons (solid lines)  and to internal $p\gamma$ interactions with electron synchrotron photons as targets for $a_{ep}=1$ (long-short dashed lines) and $a_{ep}=100$ (long-short-short dashed lines)} }
\label{fig4:p-rates}       
\end{figure*}
 
\section{High energy particles and their interactions} \label{sec:heparticles}

Particle acceleration is normally expected to take place at the internal shocks mentioned above, provided they do not become radiation dominated. As discussed by \cite{murase2013}, this can be satisfied if the comoving size of the shocked region $r_{\rm is}/\Gamma$ is smaller than the photon interaction length $1/(n'_{\rm j}\sigma_{\rm T})$, that is, if
\be
 \tau_{\rm is}= \frac{r_{\rm is}}{\Gamma} n'_{\rm j} \sigma_{\rm T} \simeq 0.196 \, L_{0,51}\Gamma_{2}^{-5}\delta t_{-2}^{-1}< 1,
\ee 
and this condition is fulfilled in the cases considered in the present work, {namely, for $\Gamma=100$ and $\Gamma=300$, as can be seen in the left panel of Fig. \ref{fig2:tauis_rcs}. In the figure, we plot the optical depth as a function of $\Gamma$ in the cases of $L_0=10^{50}{\rm erg/s}$ and $L_0=10^{51}{\rm erg/s}$ for both $\delta t=0.001 \,{\rm s}$ and $\delta t= 0.01 \,{\rm s}$. In particular, for the typically adopted value $\Gamma=100$, the optical depth is less than one for both $L_0=10^{50}{\rm erg/s}$ and $L_0=10^{51}{\rm erg/s}$, and the variability timescale is $\delta t= 0.01$ s. If this timescale is shorter, such as $\delta t=0.001\,{\rm s}$ \citep[e.g.,][]{senno2016}, and setting $L_0=10^{51}{\rm erg/s}$, we obtain low optical depths for higher values of $\Gamma$.}

 {As an additional condition to avoid radiation dominated shocks, we required that the position where the particles are accelerated by internal shocks be placed inside the collimation shock, that is, $r_{\rm is}<r_{\rm cs}$, as the post-shock region would otherwise be dominated by radiation \citep{murase2013}. As can be seen in the right panel of Fig. \ref{fig2:tauis_rcs}, this condition can be satisfied for a range of values of $\Gamma$ that depend on jet power and other parameters, such as envelope density and jet duration. Adopting typical values for these parameters \citep[e.g.,][]{fasano2021}, $t_{\rm j}=t_{\rm j,3}$ , $\rho_{\rm ext}= \rho_{\rm ext,-7}$, we observe that the internal shock region is inside the collimation shock, in particular for $\Gamma=100$ and $\delta t=0.01\,{\rm s}$ as well as for $\Gamma=300$ along with $\delta t=0.001\,{\rm s}$. Hence, we retain these two combinations of basic parameters, which are consistent with an efficient particle acceleration, and we explore the impact of different values of other key physical parameters, such as  jet power,  proton-to-electron ratio, and jet magnetization, on neutrino production.}

We therefore assume the particle acceleration rate to be
\be
t_{\rm acc}^{-1}(E'_i)=\eta\,\frac{  e\, B \,c}{E'_i},
\ee 
where $i=\{e,p\}$ stands for the primary electrons and protons, respectively, and we consider the acceleration efficiency coefficient to take a high value, namely, $\eta= 0.1$. In addition, we assume that the accelerated particles carried a power $L_p+L_e=\epsilon_{\rm rel} L_k$, that is, a fraction $\epsilon_{\rm rel}\approx 0.1$ of the total kinetic power carried by the jet. As mentioned above, we explore two different possibilities for the proton-to-ratio $a_{ep}=L_p/L_e$: an equal power injected for both protons and electrons ($a_{ep}=1$) and a situation similar to what is inferred for the cosmic-ray flux arriving on Earth ($a_{ep}=100$). 

The key parameters of the model and the values assumed in the present work are shown in Table \ref{table:params}. We remark that given the adopted typical duration of a CGRB event, {$t_{\rm j}=1000\,{\rm s}$, a higher energy budget is clearly required in the case of $L_{0}=10^{51}{\rm erg/s}$. However, such conditions can still be attained, for instance, in blue supergiant (BSG) progenitors, where the corresponding jet breakout time can be estimated as \citep{murase2013}}
\be
t_{\rm bo}\simeq 9800\,{\rm s} \, L_{0,51}^{-1/3}\left(\frac{\theta_{\rm op}}{0.2}\right)^{2/3}\left( \frac{R_*}{10^{13.5}{\rm cm} }\right). 
\ee
{Since this value is well above the assumed jet duration time, the jet would effectively be choked.}

 {We proceed, then, by computing} the injection of primary particles in the comoving frame as
\be 
Q'_i(E'_i)= \frac{d\mathcal{N}_i}{dE'_i d\Omega'\,dV'dt'}=K_iE'_i \exp\left(-\frac{E'_i}{E'_{i,\rm max}}\right),
\ee
where $E'_{i,\rm max}$ is a maximum energy obtained by the balance of the acceleration rate with the energy loss rate, which we describe below. The normalizing constant $K_i$ is therefore obtained from the following relation: 
\be
\Delta V \int_{4\pi}d\Omega \int_{E_{i,\rm min}}^{\infty}dE_i E_iQ_i(E_i)= L_i, 
\ee
where $\Delta V= 4\pi r_{\rm is}^2 \Delta d$ is the volume of the region with internal shocks, and 
$\Delta d\simeq c \delta t_v$ is the corresponding thickness in the central BH rest frame. The injection in this frame is obtained by taking into account that 
$$
 \frac{E_i}{p^2}\frac{d\mathcal{N}_i}{dV\,dp\,d\Omega\,dt}\propto 
  \frac{Q_i}{\sqrt{E_i^2-m_i^2c^4}}
$$
is a Lorentz invariant \citep{dermer2002} and that the energy in the comoving frame is related to the one in the BH frame by the usual Lorentz transformation $E'_i=\Gamma(E_i-\beta\mu \sqrt{E_i^2-m_i^2})$, where $\mu= \cos{\theta_i}$ is the cosine of the angle between the particle momentum and the jet velocity in the BH frame. As for the minimum energy of the accelerated particles, our results are not very sensitive to its exact value, but we fixed it at $E'_{i,\rm min}=2m_i c^2$ in the comoving frame, as this accounts for the fact that diffusive shock acceleration is effective only for suprathermal particles, that is, particles with energies well above those corresponding to the mean energy of the thermal distribution \citep{boschramon2006}.

We obtain the particle distribution by solving the steady-state transport equation
\be
\frac{d\left[b'_{i,\rm loss}(E'_i) N'_i(E'_i)\right]}{dE'_i}+  \frac{N'_i(E'_i)}{T_{i,\rm esc}}= Q'_i(E'_i),\label{Eq_transport_i}  
\ee
where the $b'_{i,\rm loss}$ is the total energy loss for particles of the type $i$, which is related to the sum of the cooling rates $t'^{-1}_{i,j}$ corresponding to all the processes $j$ as
\be 
b'_{i,\rm loss}= \left.\frac{dE'_i}{dt'}\right|_{\rm loss}\equiv - E'_i \sum_j t'^{-1}_{i,j}(E'_{i}).
\ee
The escape timescale is conservatively considered to be the Bohm diffusion time corresponding to the size $\Delta d'$, such that 
\be
T^{-1}_{\rm esc}&=&  \frac{2 E'_i}{3(\Delta d')^2(e\,B'\,c)} \\ 
  &\simeq & 1.7\times 10^{-5}{\rm s}^{-1} \left(\frac{E'_i}{10^4{\rm GeV}}\right)\left(\frac{\epsilon_B}{0.01}\right)^{-\frac{1}{2}} \left(\frac{L_0}{10^{51}{\rm erg \, s^{-1}}}\right)^{-\frac{1}{2}}\delta t_2^{-1}.\nonumber
\ee
This escape rate is generally well below the cooling rates involved, which are presented in the next subsection. 

The solution of Eq.(\ref{Eq_transport_i}) is formally given by
\begin{multline}
N'_i(E'_i)= \frac{1}{b_{i,\rm loss}(E'_i)}\int_{E'_i}^\infty dE'Q'_i(E') \\ \times \exp\left[-\int_{E'_i}^{E'}\frac{dE''}{T_{\rm esc}(E'')b_{i,\rm loss}(E'')}\right]\label{solution_Eq_transport}
\end{multline}
and is applied to the different particle species $(e, \, p, \, \pi^{\pm}, \, \mu^{\pm})$ while taking into account the corresponding cooling processes and injections in each case.

\begin{figure*}
    \centering
    \begin{subfigure}[t]{0.49\textwidth}
        \centering                          
        \includegraphics[width=0.5\linewidth,trim= 145 30 183 20]{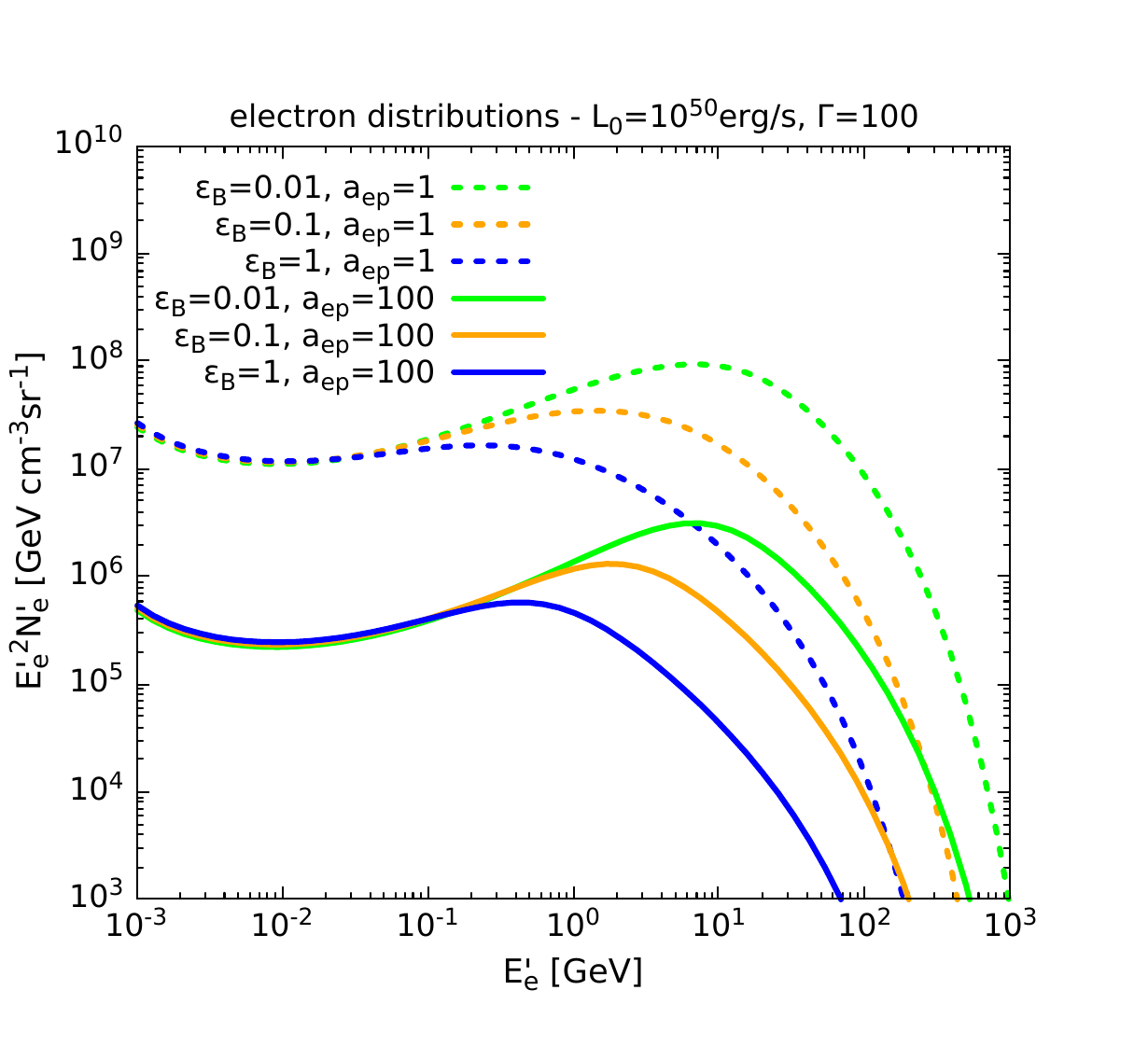} 
    \end{subfigure}
    \hfill
    \begin{subfigure}[t]{0.49\textwidth}
        \centering
        \includegraphics[width=0.5\linewidth,trim= 145 30 183 20]{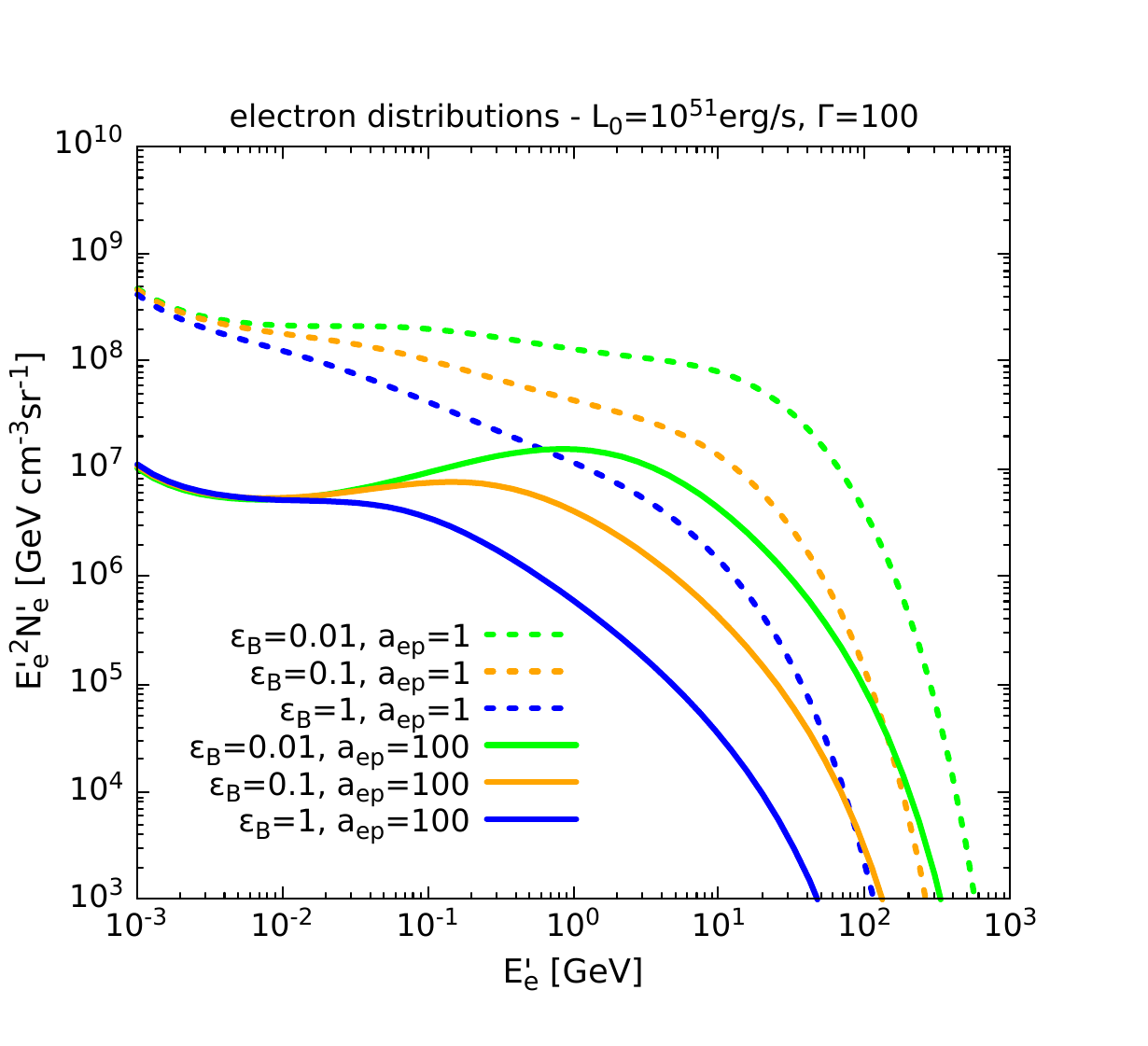} 
    \end{subfigure}
    \centering
    \begin{subfigure}[t]{0.49\textwidth}
        \centering                          
        \includegraphics[width=0.5\linewidth,trim= 145 30 183 20]{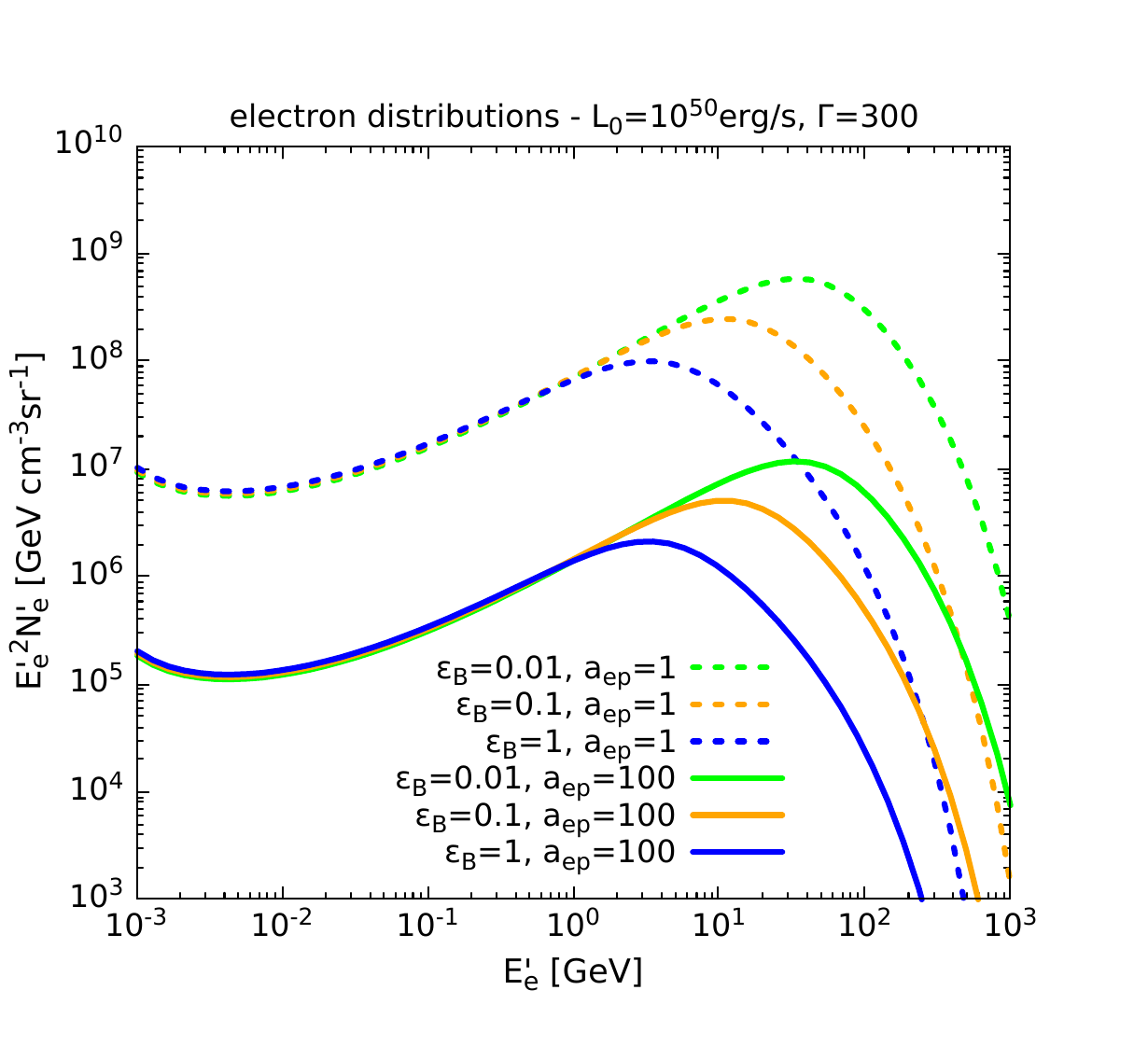} 
    \end{subfigure}
    \hfill
    \begin{subfigure}[t]{0.49\textwidth}
        \centering
        \includegraphics[width=0.5\linewidth,trim= 145 30 183 20]{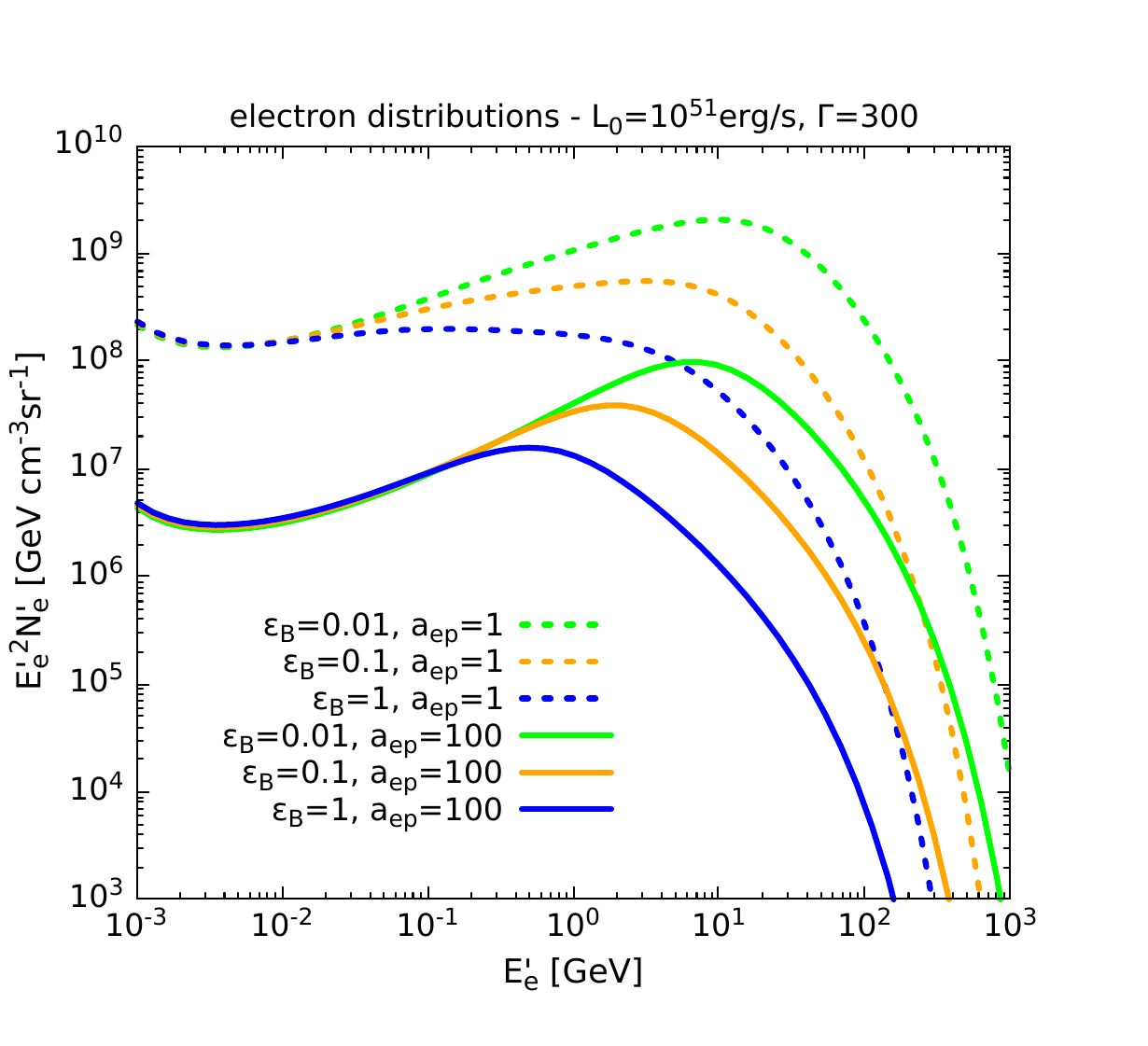} 
    \end{subfigure}
    \caption{Electron distributions for $L_0=10^{50}{\rm erg/s}$ (left panel) and $L_0=10^{51}{\rm erg/s}$ (right panel). {The top panels correspond to $\Gamma=100$ and the bottom ones to $\Gamma=300$. The green, orange, and blue curves refer to the cases of $\epsilon_B=0.01$, $\epsilon_B=0.1$, and $\epsilon_B=1$, respectively. The dashed curves correspond to $a_{ep}=1$ and the solid ones to $a_{ep}=100$. }}\label{fig5:Ne}
\end{figure*}
%
\begin{figure*}
    \centering
    \begin{subfigure}[t]{0.49\textwidth}
        \centering                          
        \includegraphics[width=0.5\linewidth,trim= 125 30 173 30]{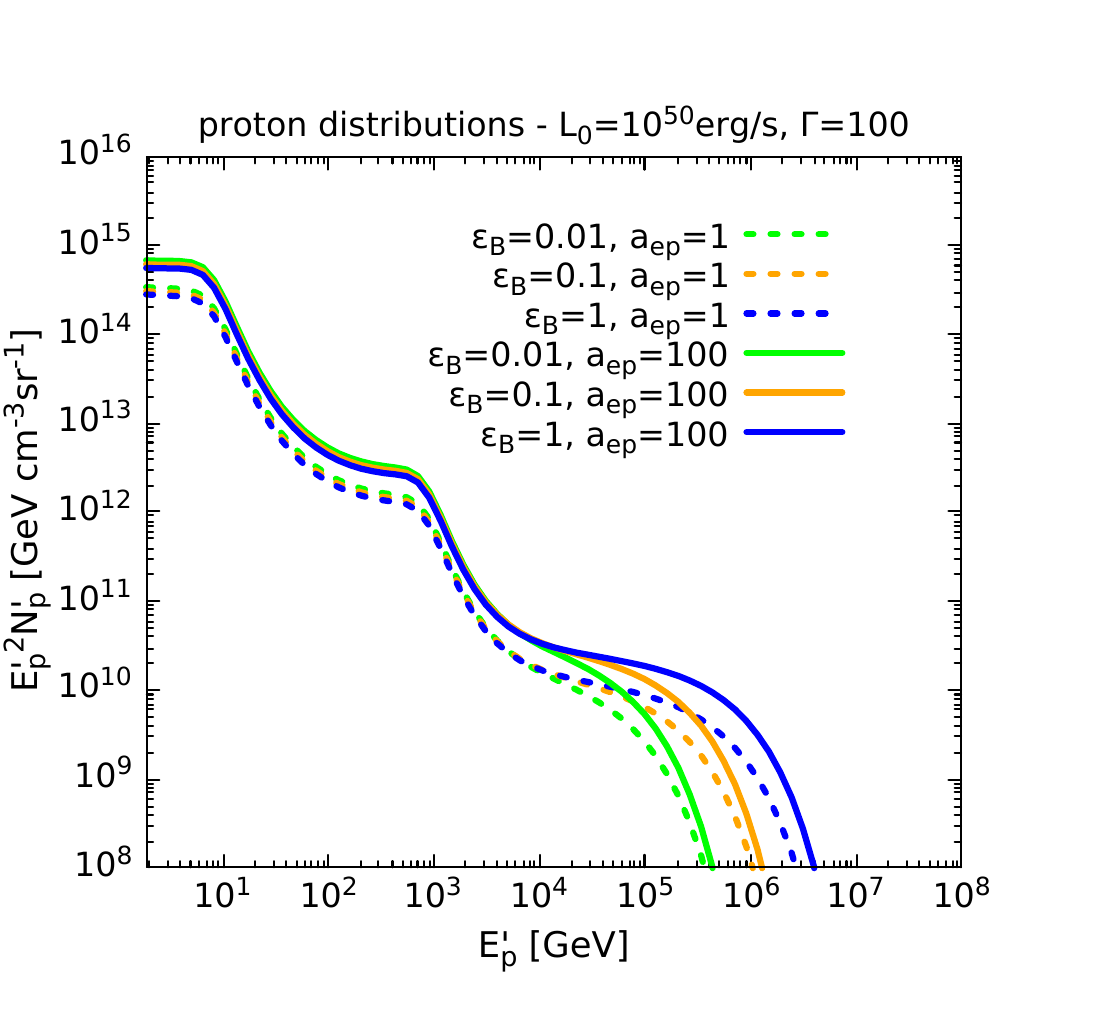} 
    \end{subfigure}
    \hfill
    \begin{subfigure}[t]{0.49\textwidth}
        \centering
        \includegraphics[width=0.5\linewidth,trim= 125 30 173 30]{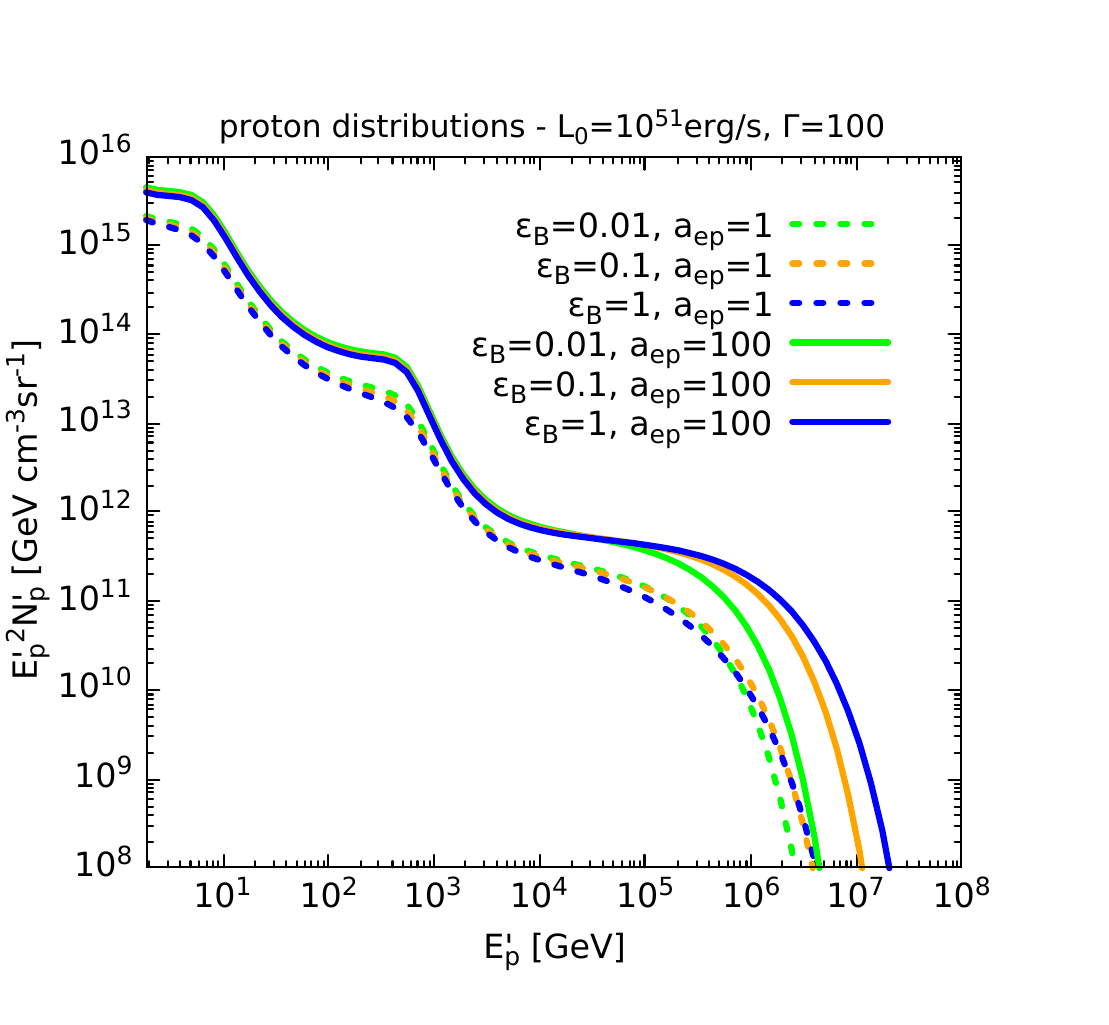} 
    \end{subfigure}
    \centering
    \begin{subfigure}[t]{0.49\textwidth}
        \centering                          
        \includegraphics[width=0.5\linewidth,trim= 125 30 173 30]{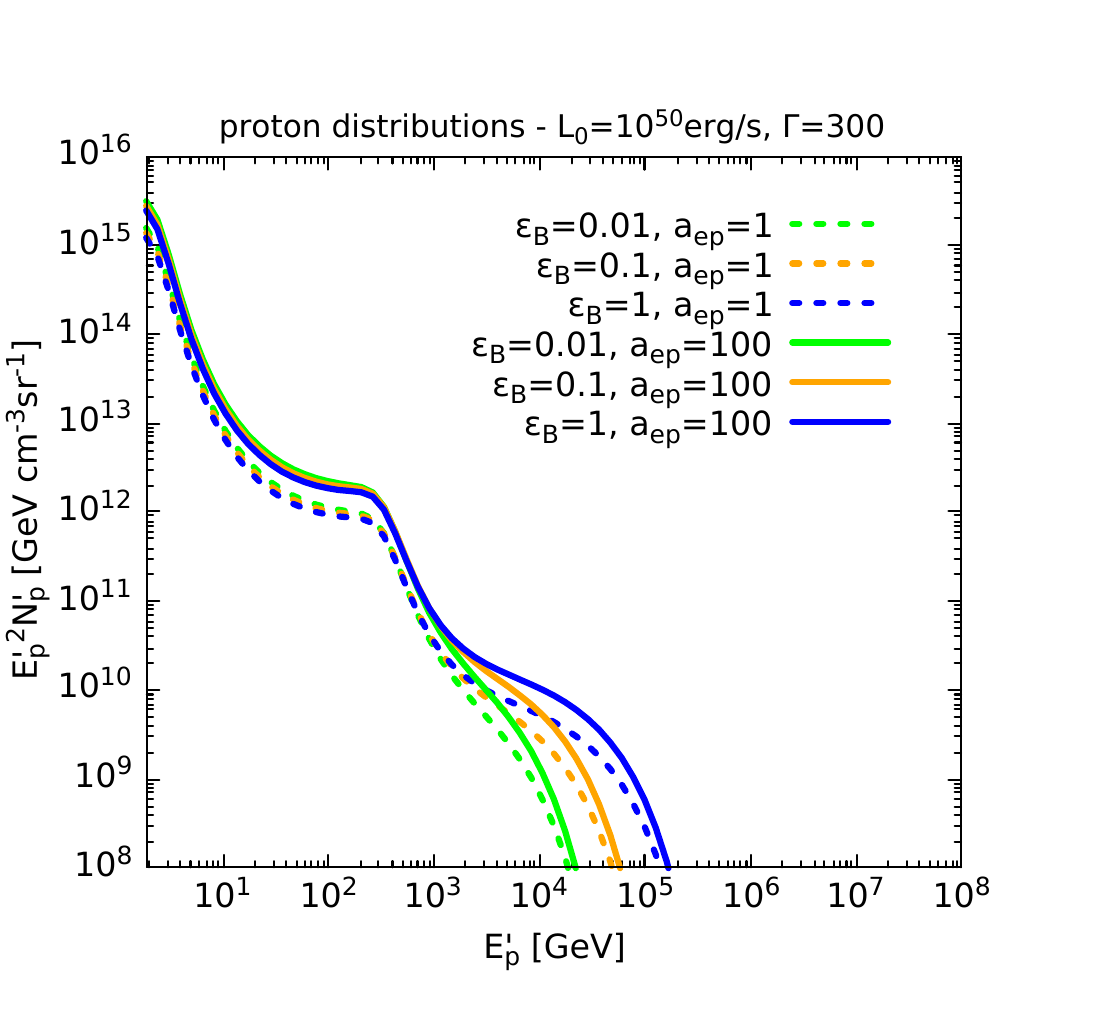} 
    \end{subfigure}
    \hfill
    \begin{subfigure}[t]{0.49\textwidth}
        \centering
        \includegraphics[width=0.5\linewidth,trim= 125 30 173 30]{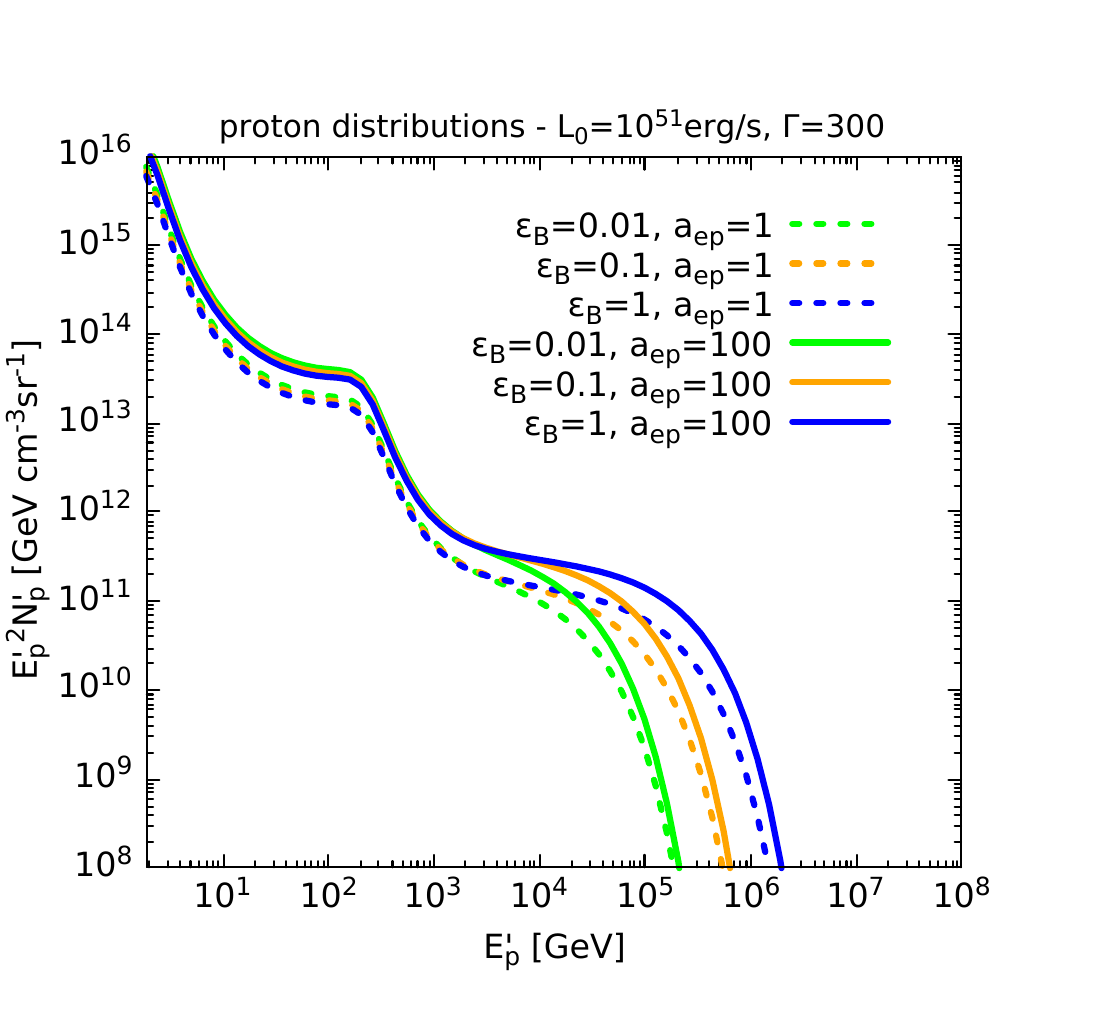} 
    \end{subfigure}
    \caption{Proton distributions for $L_0=10^{50}{\rm erg/s}$ (left panel) and $L_0=10^{51}{\rm erg/s}$ (right panel). {The top panels correspond to $\Gamma=100$ and the bottom ones to $\Gamma=300$. The green, orange, and blue curves refer to the cases of $\epsilon_B=0.01$, $\epsilon_B=0.1$, and $\epsilon_B=1$, respectively. The dashed curves correspond to $a_{ep}=1$ and the solid ones to $a_{ep}=100$. }}\label{fig6:Np}
\end{figure*}


\begin{figure}
    \centering
        \includegraphics[width=0.5\linewidth,trim= 162 30 197 0]{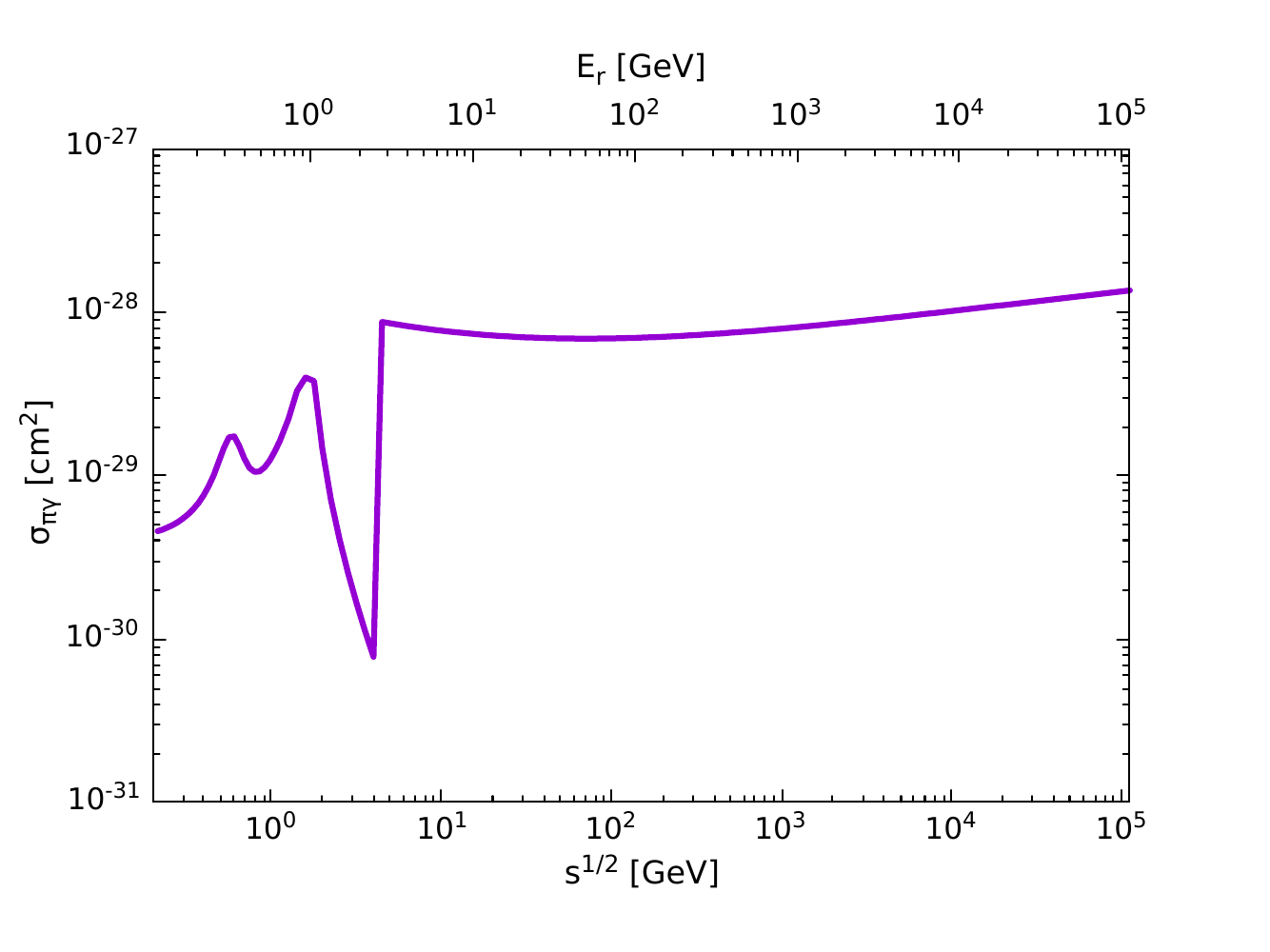} 
      \caption{Pion-photon cross section adopted to obtain $t_{\pi\gamma}^{-1}$. The bottom x-axis refers to the the square root of the center-of-mass energy $s^{1/2}$, and the top one marks the corresponding photon energy in the pion rest frame, $E_r$.}\label{fig7:sigmapig}
\end{figure} 
 
\begin{figure*}[]
    \centering
    \begin{subfigure}[t]{0.49\textwidth}
        \centering                          
        \includegraphics[width=0.5\linewidth,trim= 135 30 160 30]{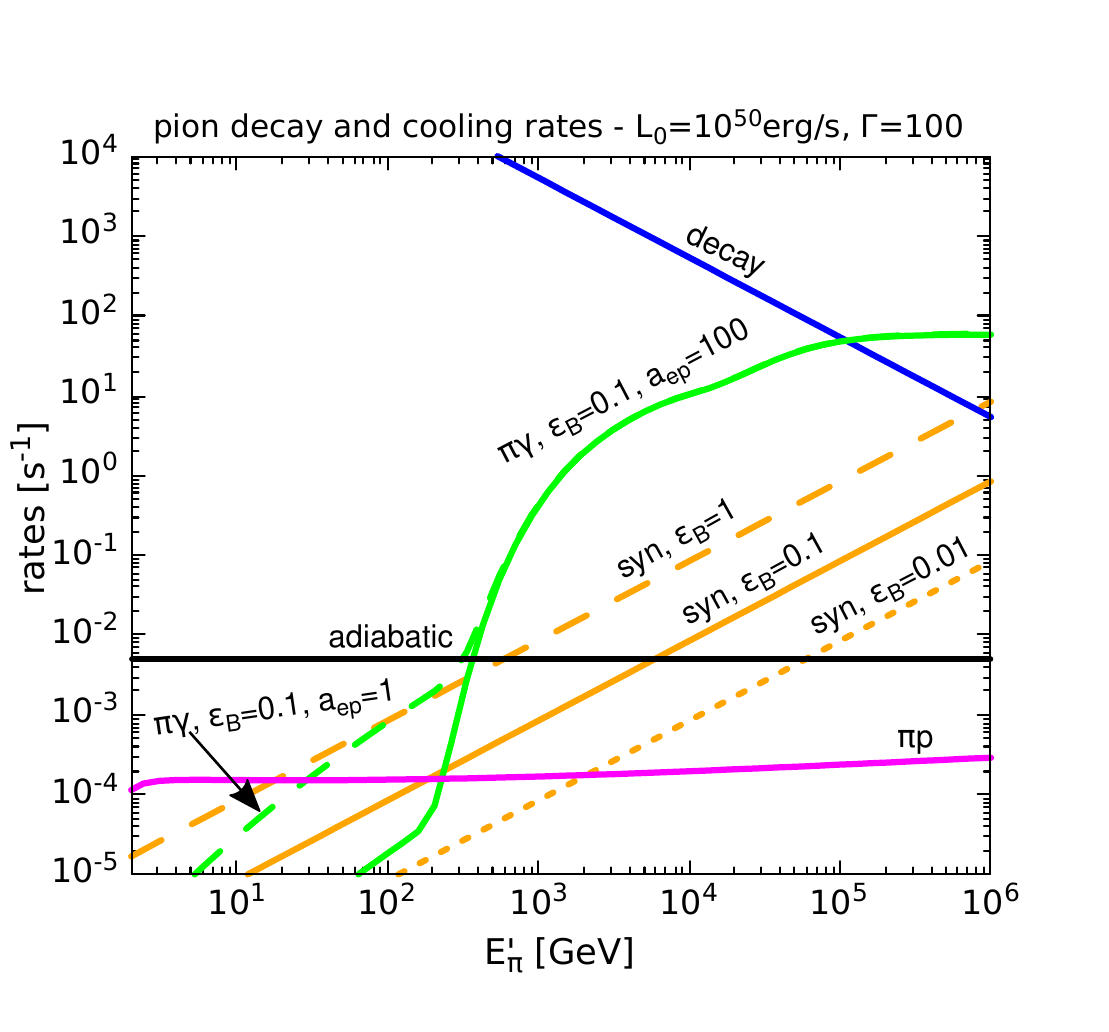} 
    \end{subfigure}
    \hfill
    \begin{subfigure}[t]{0.49\textwidth}
        \centering
        \includegraphics[width=0.5\linewidth,trim= 135 30 160 30]{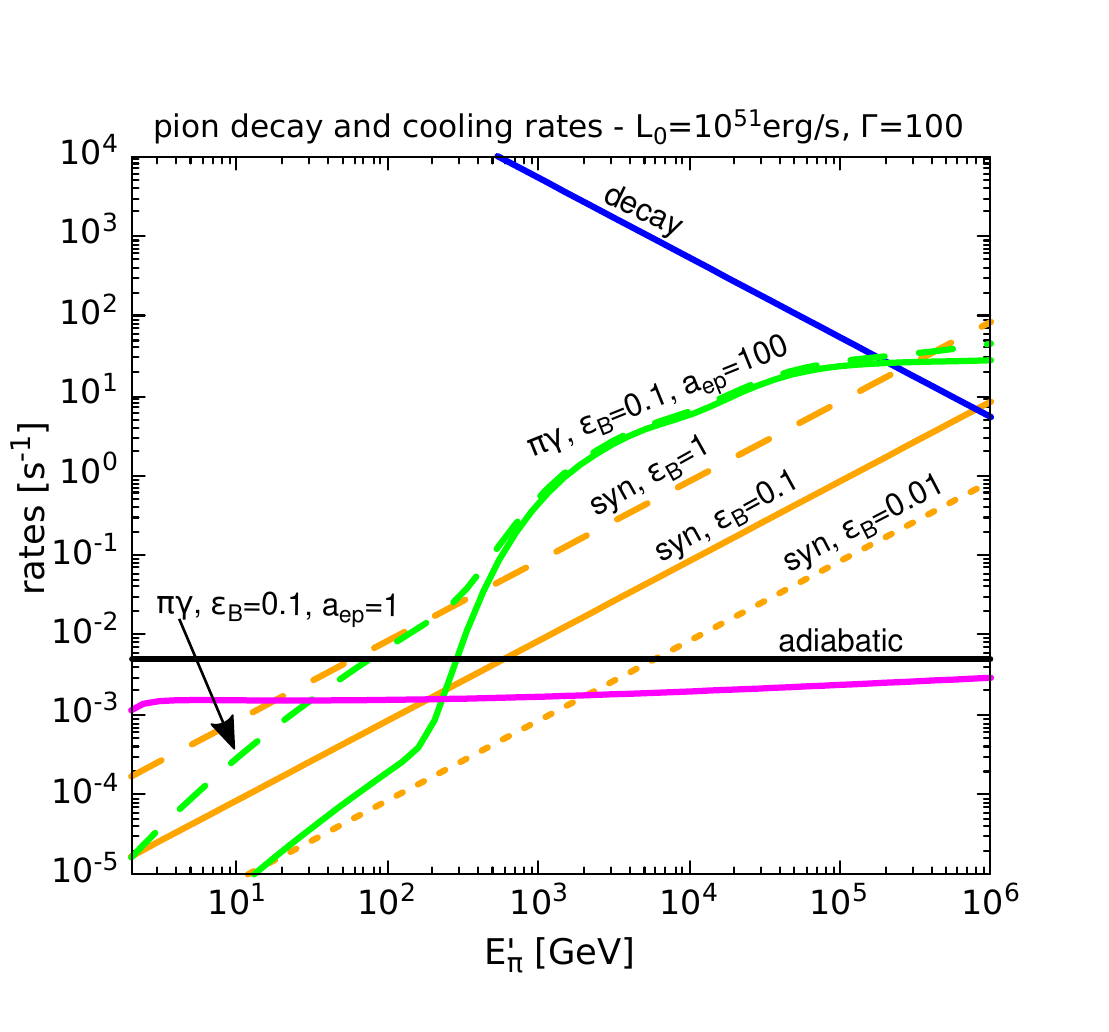} 
    \end{subfigure}
    \centering
    \begin{subfigure}[t]{0.49\textwidth}
        \centering                          
        \includegraphics[width=0.5\linewidth,trim= 135 30 160 30]{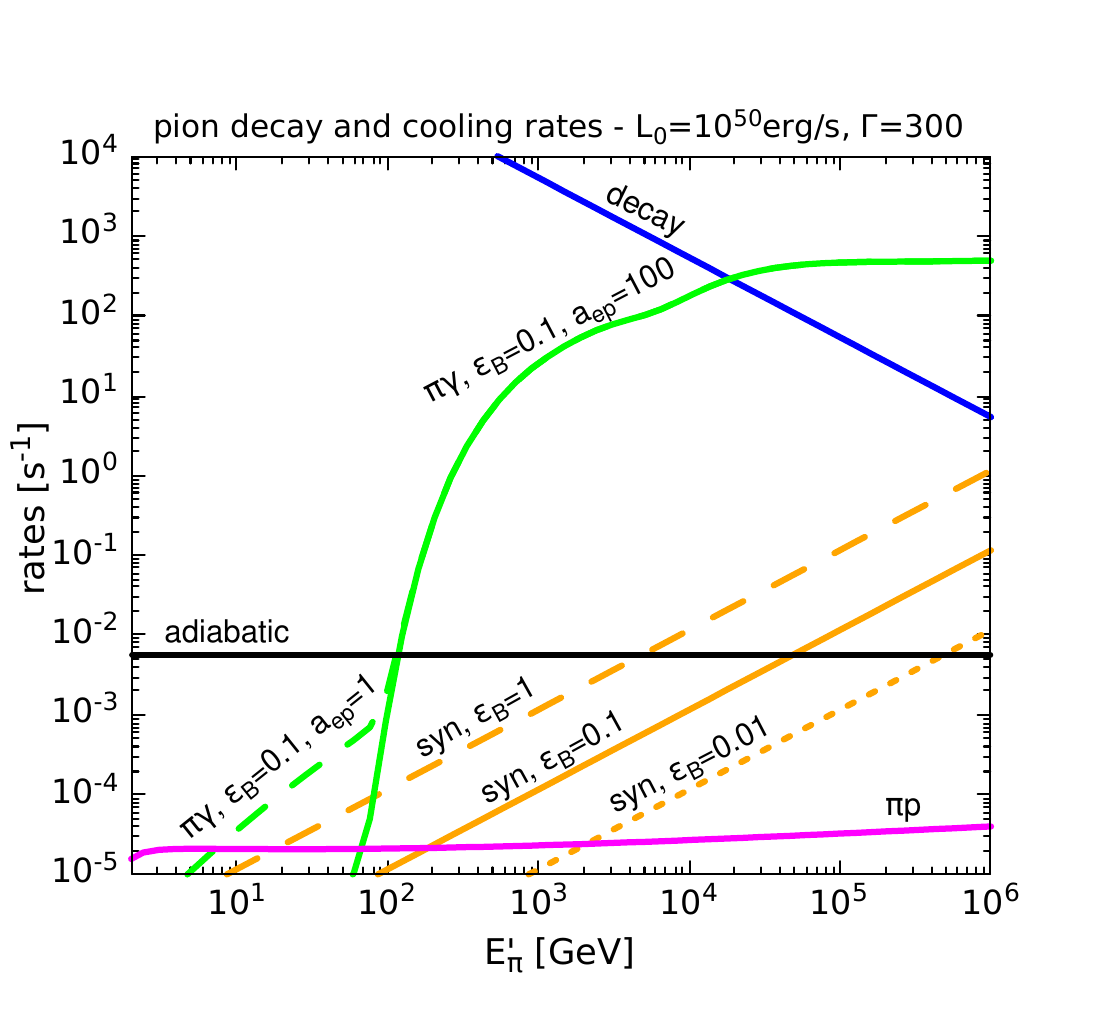} 
    \end{subfigure}
    \hfill
    \begin{subfigure}[t]{0.49\textwidth}
        \centering
        \includegraphics[width=0.5\linewidth,trim= 135 30 160 30]{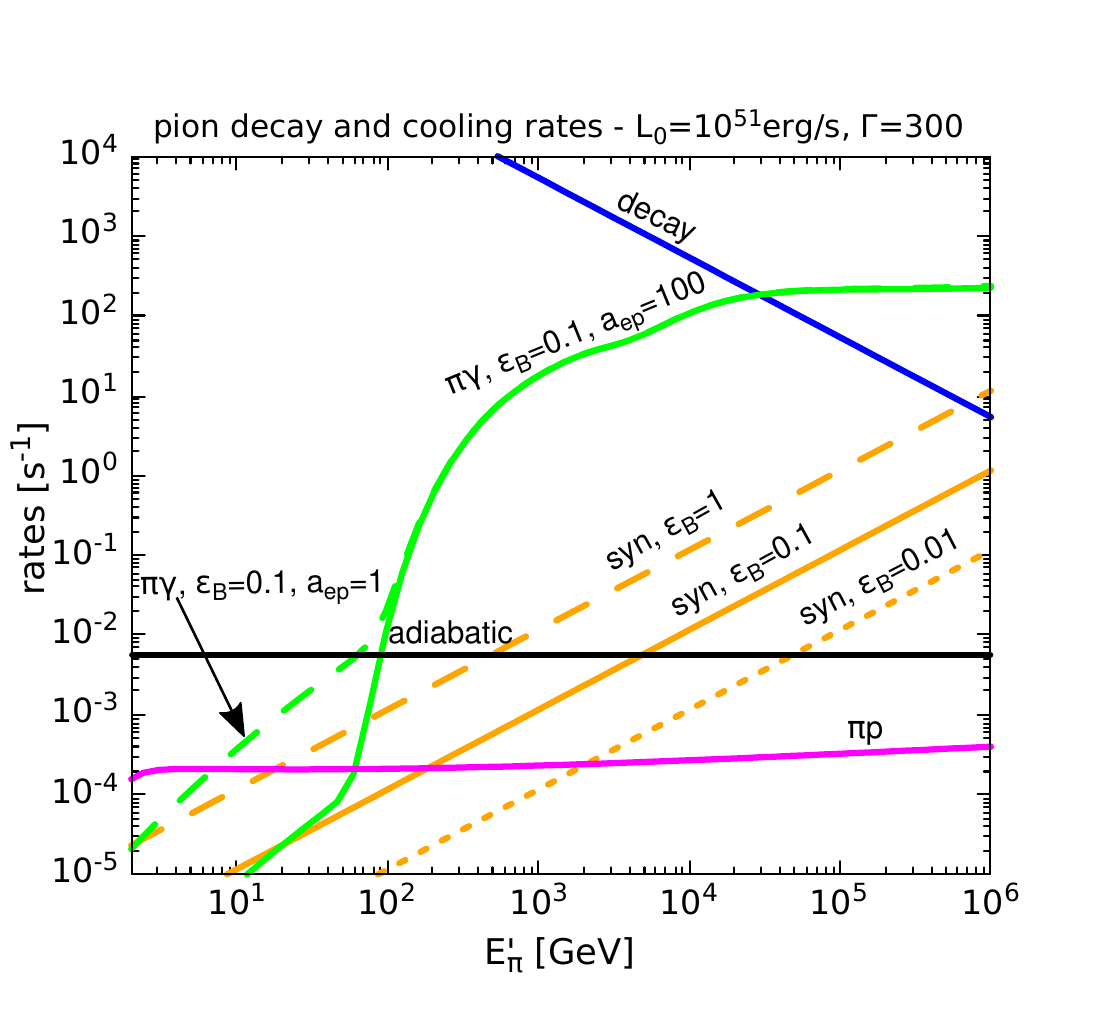} 
    \end{subfigure}
    \caption{Pion decay and cooling rates for $L_0=10^{50}{\rm erg/s}$ (left panel) and $L_0=10^{51}{\rm erg/s}$ (right panel). {The top panels correspond to $\Gamma=100$ and the bottom ones to $\Gamma=300$. The blue curves correspond to the pion decay rate, and the green curves represent the $\pi\gamma$ interaction with external and internal photons with $a_{ep}=1$ (dashed lines) and with $a_{ep}=100$ (solid lines). The pion synchrotron cooling is marked with orange curves for $\epsilon_B=0.01$ (short-dashed lines), $\epsilon_B=0.1$ (solid lines), and $\epsilon_B=1$ (long-dashed lines). The solid black curves mark the adiabatic cooling rate.}}\label{fig8:pi-rates}
\end{figure*} 
\begin{figure*}[h!]
    \centering
    \begin{subfigure}[t]{0.49\textwidth}
        \centering                          
        \includegraphics[width=0.5\linewidth,trim= 130 30 170 0]{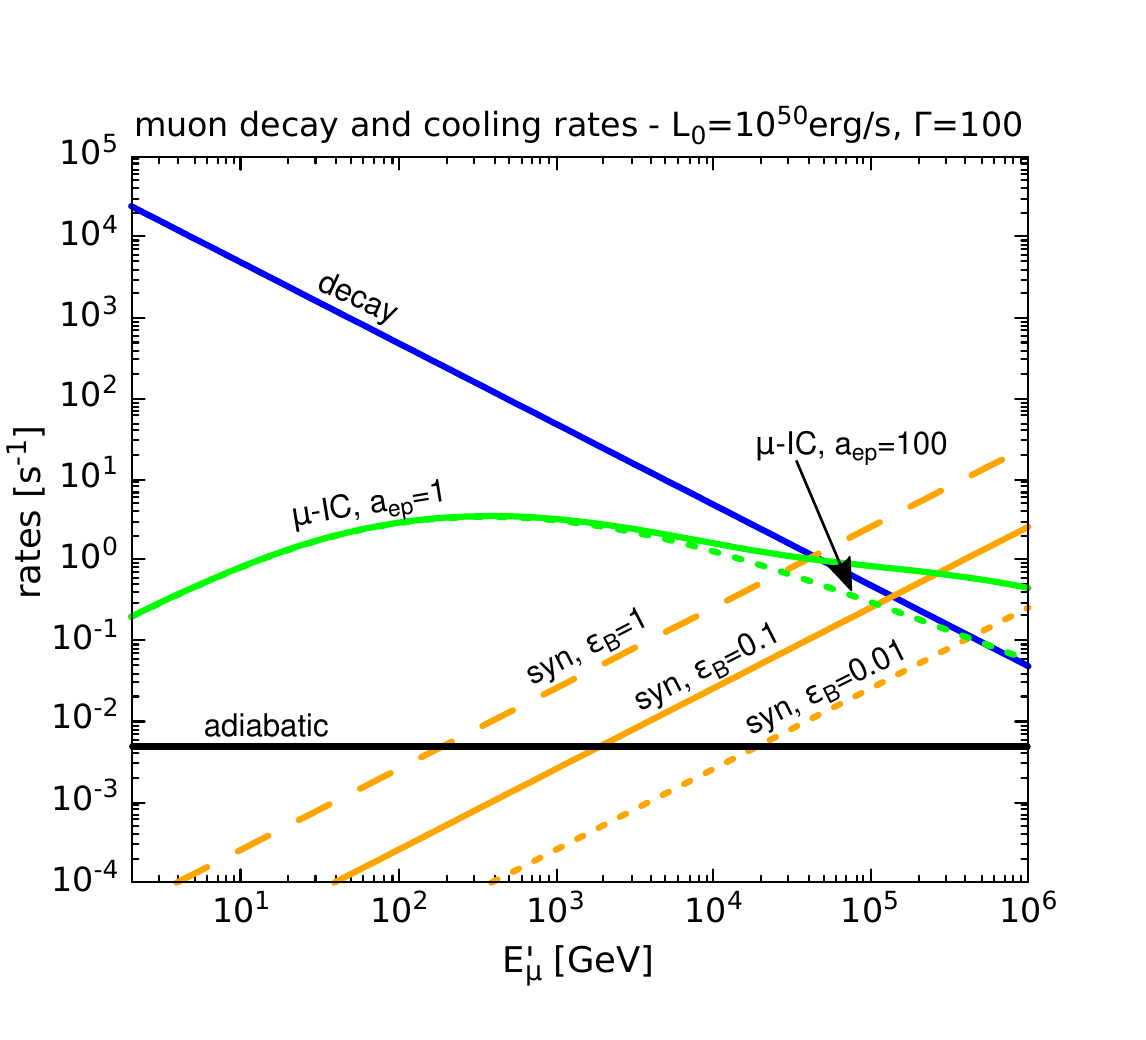} 
    \end{subfigure}
    \hfill
    \begin{subfigure}[t]{0.49\textwidth}
        \centering
        \includegraphics[width=0.5\linewidth,trim= 130 30 170 0]{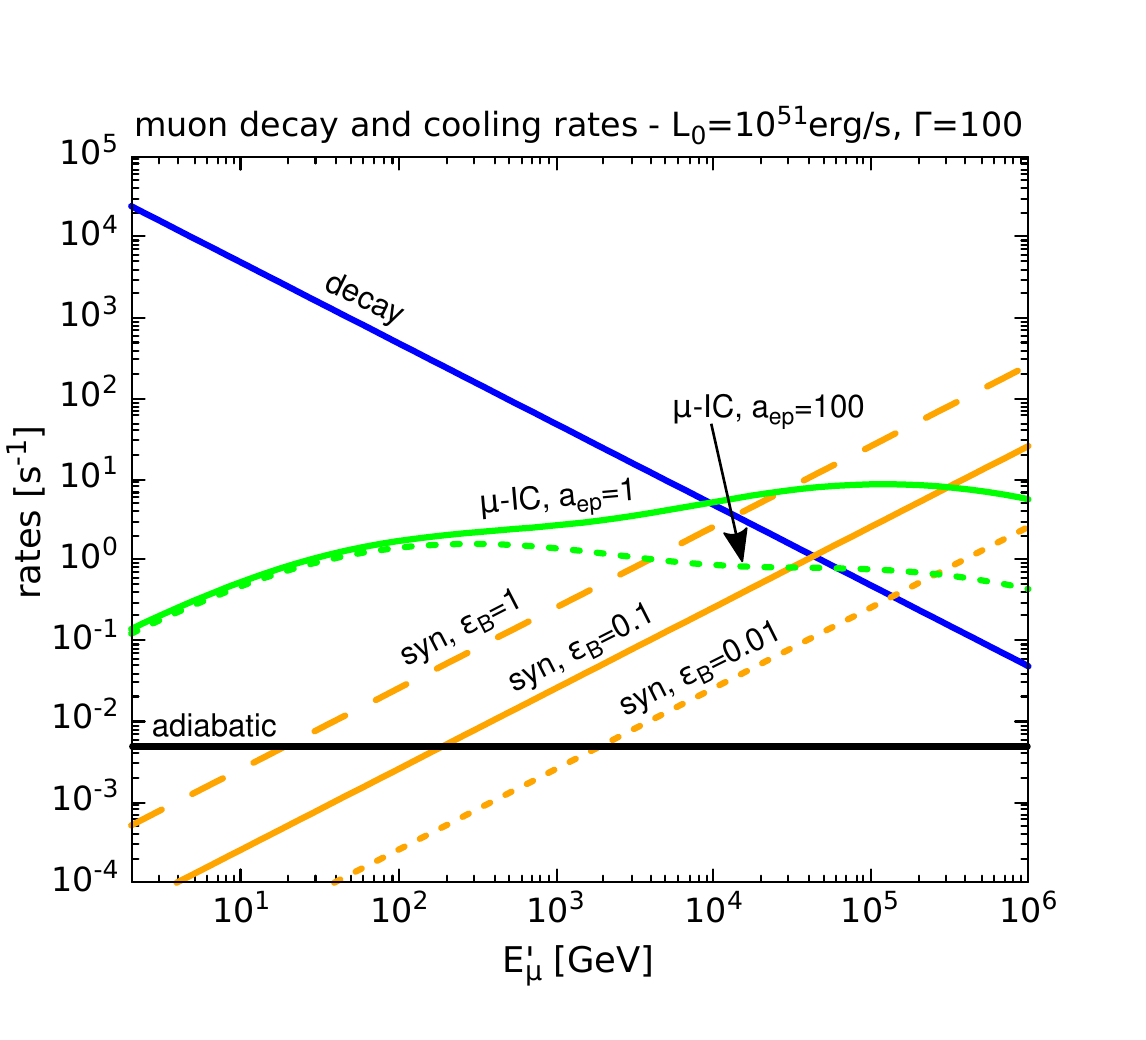} 
    \end{subfigure}
    \centering
    \begin{subfigure}[t]{0.49\textwidth}
        \centering                          
        \includegraphics[width=0.5\linewidth,trim= 130 30 170 0]{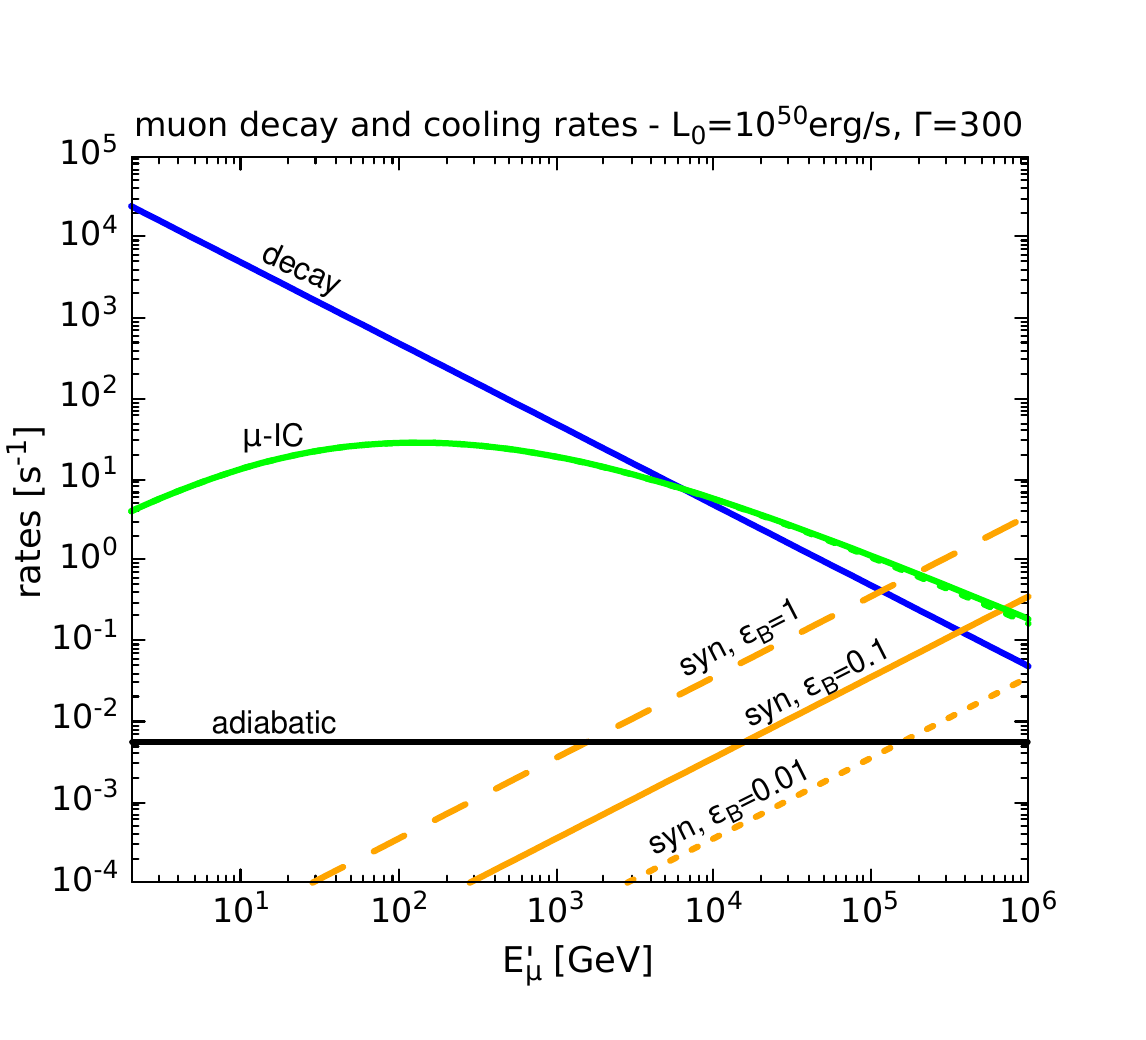} 
    \end{subfigure}
    \hfill
    \begin{subfigure}[t]{0.49\textwidth}
        \centering
        \includegraphics[width=0.5\linewidth,trim= 130 30 170 0]{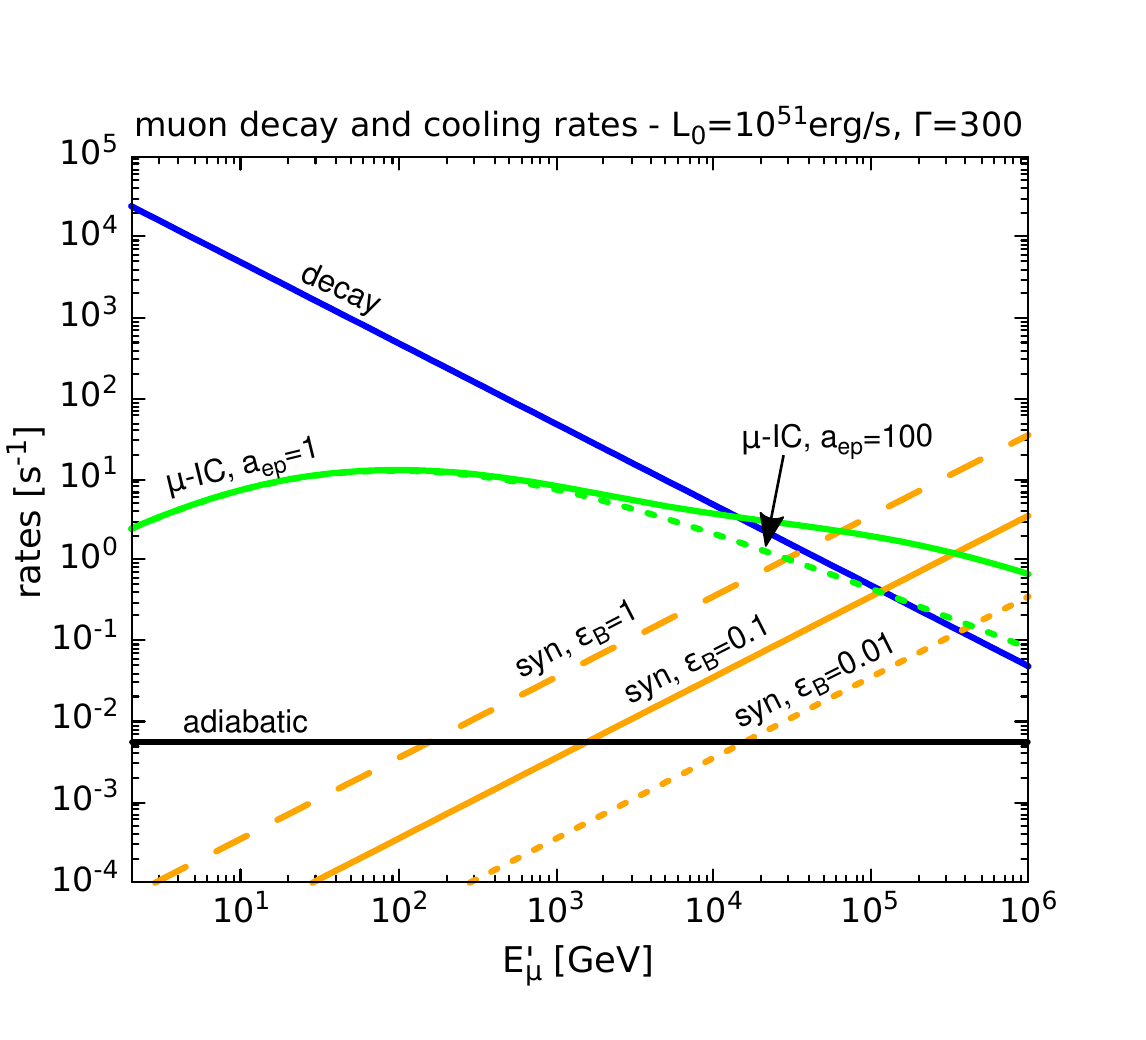} 
    \end{subfigure}
    \caption{Muon decay and cooling rates for $L_0=10^{50}{\rm erg/s}$ (left panel) and $L_0=10^{51}{\rm erg/s}$ (right panel). {The top panels correspond to $\Gamma=100$ and the bottom ones to $\Gamma=300$. The blue curves correspond to the muon decay rate, and the green curves represent the IC interaction with external and internal photons with $a_{ep}=1$ (dashed lines) and with $a_{ep}=100$ (solid lines). The muon synchrotron cooling is marked with orange curves for $\epsilon_B=0.01$ (short-dashed lines), $\epsilon_B=0.1$ (solid lines), and $\epsilon_B=1$ (long-dashed lines). The solid black curves mark the adiabatic cooling rate.}}\label{fig9:mu-rates}
\end{figure*} 

%

\subsection{Cooling processes}

In this section, we discuss the interactions and corresponding cooling rates for the high energy charged particles at the internal shock region. 
As discussed above, there is background of external soft photons (i.e., produced outside the internal shock region). These photons, characterized by the differential density of Eq.\ref{nphIS}, enter the internal shock zone and can serve as targets for IC and $p\gamma$ interactions. In this work, we also account for the possible interaction of high energy particles with synchrotron photons emitted by electrons in the internal shock zone itself. 

In general, for any charged particle of mass $m_i$, the synchrotron cooling rate is given by
\be
t_{i,{\rm syn}}^{-1}(E'_i)= \frac{4}{3} \left(\frac{m_e}{m_i}\right)^3 \frac{\sigma_T B^2}{m_e c \ 8\pi} \frac{E'_i}{m_i c^2}. 
\ee
We apply this equation to electrons, protons, pions, and muons in order to obtain their corresponding particle distribution in the internal shock zone.

The differential density of synchrotron photons is computed as
\begin{equation}
n_{\rm ph,syn}= \frac{\epsilon_{\rm syn}(E_{\rm ph})}{E_{\rm ph}} \frac{\Delta d'}{c},
\end{equation}
where $\Delta d'=\Delta d/\Gamma$ is the comoving thickness of the internal shock region, and $\epsilon_{\rm syn}$ is the power per unit volume per unit energy of the photons,
\begin{equation}
\epsilon_{\rm syn}(E_{\rm ph}) = \left( \frac{1-e^{-\tau_{\rm SSA}(E_{\rm ph}))}}{\tau_{\rm SSA}(E_{\rm ph})} \right) \int_{m_e c^2}^{\infty} dE 4\pi P_{\rm syn} N'_e(E).
\end{equation}
Here,
$P_{\rm syn}(E_{\rm ph})$ is defined as in \citep{blumenthal1970}, and $\tau_{\rm SSA}$ stands for the optical depth given by \citep{rybicki2004}.

The synchrotron emission of electrons in the internal shock region can provide an additional target for IC interactions of the electrons themselves, that is, the so-called synchrotron self-Compton (SSC) interactions.  We also note that IC interactions of muons can be relevant, since they provide an additional cooling process (which may become even more significant than the synchrotron one) that decreases the neutrino output at high energies, as we show below in the next subsection. The IC cooling rate for a charged lepton of mass $m_l$ is computed following the standard expressions of \cite{blumenthal1970}, {which can account for both the Thomson and the Klein-Nishina regimes}: 
\begin{multline}
t_{\rm IC}^{-1}(E'_l)= \frac{3m_l^2c^4\sigma_{\rm T}}{4E^3} \int_{E_{\rm ph}^{\rm (min)}}^{E} \frac{dE_{\rm ph}}{E_{\rm ph}} n_{\rm ph}(E_{\rm ph}) \\ 
     \times \int_{E_{\rm ph}}^{\frac{\Gamma_l}{\Gamma_l + 1}E}dE_\gamma F_{\rm IC}(q) \left[E_\gamma - E_{\rm ph}\right],\label{tIC}
\end{multline}
where $E_{\rm ph}^{\rm (min)}$ is the lowest energy of the available background of target photons, $q=E_\gamma(\Gamma_l(E_l-E_\gamma))$, with $\Gamma_l=4\,E_{\rm ph}E_l/(m_l^2c^4)$, and the function $F_{\rm IC}(q)$ is given by
\begin{multline}
F_{\rm IC}(q)=2q\ln\,q + (1+2q)(1-q)
+\frac{1}{2}(1-q)\frac{(q\Gamma_l)^2}{1+\Gamma_l q}.
\end{multline}

In Fig. \ref{fig3:e-rates}, we present the obtained acceleration and cooling rates for the injected primary electrons {in the cases of $L_0=10^{50}{\rm erg \, s^{-1}}$ in the left panels and of $L_0=10^{51}{\rm erg \, s^{-1}}$ in the right panels. The top panels correspond to $\Gamma=100$ and the bottom ones to $\Gamma=300$.} Except for the IC cooling rate due to the external photons, the rest of the rates are sensitive to the magnetic field, for which we consider three different values determined by fixing the parameter $\epsilon_B$ at $0.01,$ $0.1$, and $1$. 

As for the SSC process, since the density of the target synchrotron photons depends on the electron distribution, the SSC cooling rate does as well. It is therefore a non-trivial issue to address this process self-consistently through a single expression for the cooling rate \citep[e.g.,][in the Thomson regime]{schlickeiser2009}. 
To handle this situation, we obtain a first approximation of the electron distribution $N'_{e,1}$ while assuming a nil value for the SSC cooling rate. With $N'_{e,1}$ we compute a first approximation of the SSC cooling rate $t^{-1}_{\rm SSC,1}$ using Eq. (\ref{tIC}), and we use it to obtain a second approximation for the electron distribution, $N'_{e,2}$, and then $t^{-1}_{\rm SSC,2}$. We repeatd this iterative method several times until the ratio between $t^{-1}_{\rm SSC,j}$ and $t^{-1}_{\rm SSC,j-1}$ was reduced to less than a few percent at low energies. For illustration of this, we include in Fig. \ref{fig3:e-rates} our successive approximations of the SSC cooling rate obtained in the case of $\epsilon_B=0.1$, using short-dashed gray lines for $a_{ep}=1$ and long-dashed gray lines for $a_{ep}=100$. As can be seen in the figure, for a higher injected power in electrons and $\Gamma=100$ (i.e., in the top-right panel), the SSC becomes dominant as compared to the external IC rate, and more iterations are necessary, particularly for $a_{ep}=1$. {In the cases of $\Gamma=300$, the external IC rate is higher than for $\Gamma=100$. This is due to an increased boosting effect of the corresponding target density of photons.} 

For protons, we compute the
 $p\gamma$ cooling rate as \citep{atoyan2003}:
\begin{multline}
t_{p\gamma}^{-1} (E'_p) = \int_{E_{\rm th}/2\gamma_p}^{\infty} dE_{\rm ph} \frac{c\, n_{\rm ph}(E_{\rm ph})}{2 \gamma_p^2 E_{\rm ph}^2} \\ \times\int_{E_{\rm th}}^{2 E_{\rm ph} \gamma_p} dE_{\rm r} \sigma_{p\gamma}(E_{\rm r}) K_{p\gamma}(E_{\rm r}) E_{\rm r}, \label{tpg}
\end{multline}
where $\gamma_p=E'_p/(m_pc^2)$, $E_{\rm th}=2m_ec^2$ {is the threshold energy corresponding to $e^{+}e$-pairs (Bethe-Heitler process),} or $E_{\rm th}\simeq 150{\,\rm MeV}$ in the case of pion production. The corresponding cross sections $\sigma_{p\gamma}$ and inelasticity coefficients $K_{p\gamma}$ are taken as in  \cite{begelman1990}.

For completeness, we also consider the cooling due to $pp$ interactions with a rate
\be
t_{pp}^{-1}(E'_p)\simeq \frac{1}{2} n'_{\rm j} c\,\sigma_{pp}(E'_p),
\ee
where the cross section $\sigma_{pp}$ is taken from \cite{kelner2006}. {In turn, the adiabatic cooling rate, which can be estimated as $$t_{\rm ad}^{-1}\approx \frac{c}{r_{\rm is}},$$ gives a generally negligible effect for high energy protons in the present context.} 
 
In Fig. \ref{fig4:p-rates}, we show the obtained acceleration and cooling rates for protons in the cases of $L_0=10^{50}{\rm erg \, s^{-1}}$  and $L_0=10^{51}{\rm erg \, s^{-1}}$ in the left and right panels, respectively. Again, the top panels correspond to $\Gamma=100$ and the bottom ones to $\Gamma=300$. As can be seen in the figure, the $p\gamma$ interactions are always the dominant process. In the plots, we also include the contributions to the cooling rate due to the external photons produced in the jet head and the contribution due to the synchrotron emission of electrons in the internal shock region for $\epsilon_B=0.1$. {It can also be seen that the $p\gamma$ interactions with external photons are dominant for $\Gamma=300$, while for $\Gamma=100,$ the interactions with electron synchrotron photons can become relevant only for very high energy protons if $a_{ep}=1$.}
 
We show in Figs. \ref{fig5:Ne} and \ref{fig6:Np} the obtained electron and proton distributions for $L_0=10^{50}{\rm erg \, s^{-1}}$  and $L_0=10^{51}{\rm erg \, s^{-1}}$. {The top and bottom panels correspond to $\Gamma=100$ and $\Gamma=300$, respectively.} In the figures, it can be seen that the electron distributions decrease with $\epsilon_B$. This is because the synchrotron cooling becomes more relevant as the magnetic field increases. The maximum electron energy also decreases with the magnetic field, since the acceleration rate grows with $B$, but the synchrotron cooling rate grows with $B^2$. {For $\Gamma=300$ and $L_0=10^{50}{\rm erg/s}$, the cases with different $a_{ep}$ only differ by a constant factor, reflecting only the fact that a different power is injected in the electrons since the SSC process does not play a significant role. In the other cases, SSC can be relevant for $a_{ep}=1$ and high electron energies.}

For protons, the synchrotron cooling is not so relevant, and the effect of increasing the magnetic field leads to a faster acceleration rate. 
{Only in the cases with $L_0=10^{51}{\rm erg/s}$ and $\Gamma=100$ can this be compensated by a faster cooling via $p\gamma$ interactions with an increased target of photons due significant electron synchrotron emission, particularly for $a_{ep}=1$. In the rest of the cases, such additional interactions do not play a relevant role.}

\begin{figure*}[]
    \centering
    \begin{subfigure}[t]{0.49\textwidth}
        \centering                          
        \includegraphics[width=0.5\linewidth,trim= 125 30 170 50]{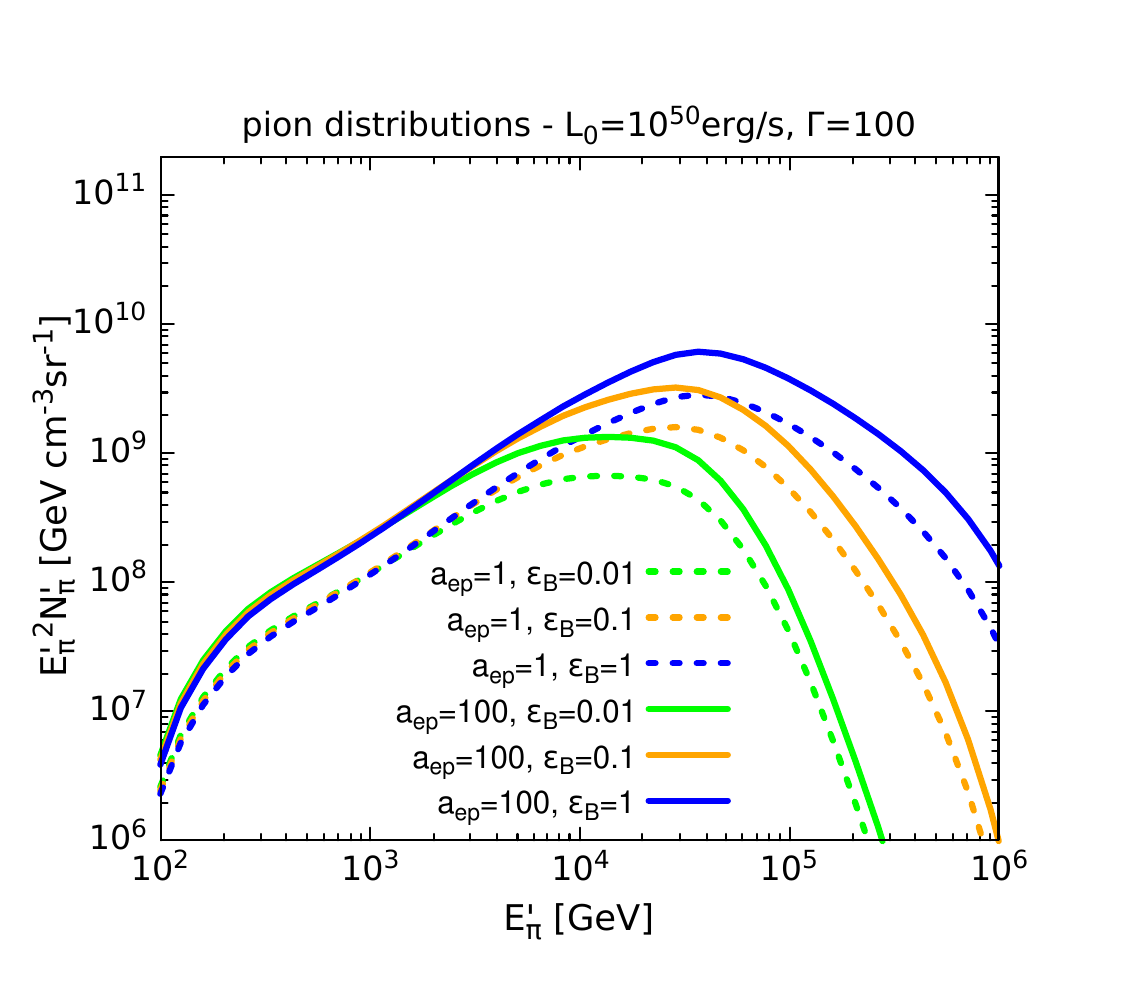} 
    \end{subfigure}
    \hfill
    \begin{subfigure}[t]{0.49\textwidth}
        \centering
        \includegraphics[width=0.5\linewidth,trim= 125 30 170 50]{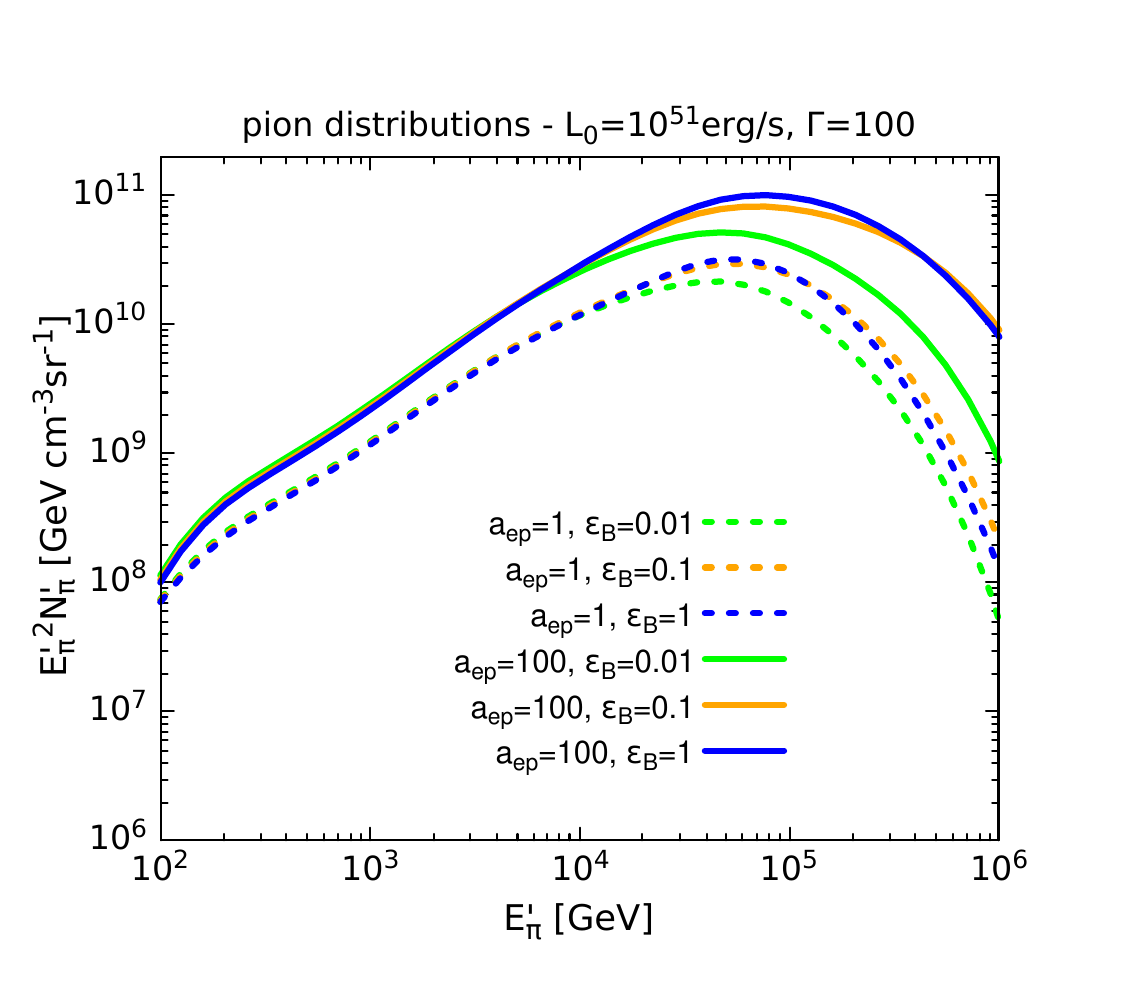} 
    \end{subfigure}
    \begin{subfigure}[t]{0.49\textwidth}
        \centering                          
        \includegraphics[width=0.5\linewidth,trim= 127 30 167 35]{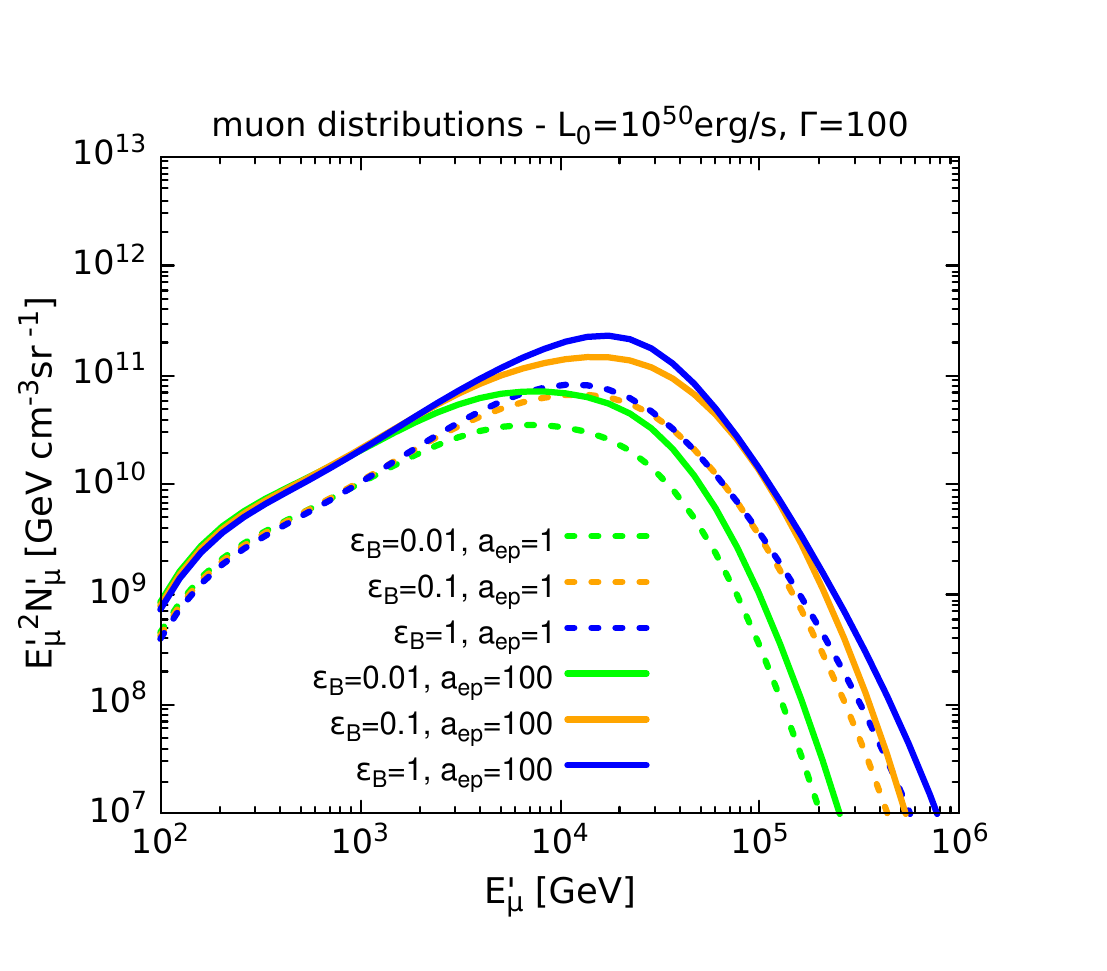} 
    \end{subfigure}
    \hfill
    \begin{subfigure}[t]{0.49\textwidth}
        \centering
        \includegraphics[width=0.5\linewidth,trim= 127 30 167 35]{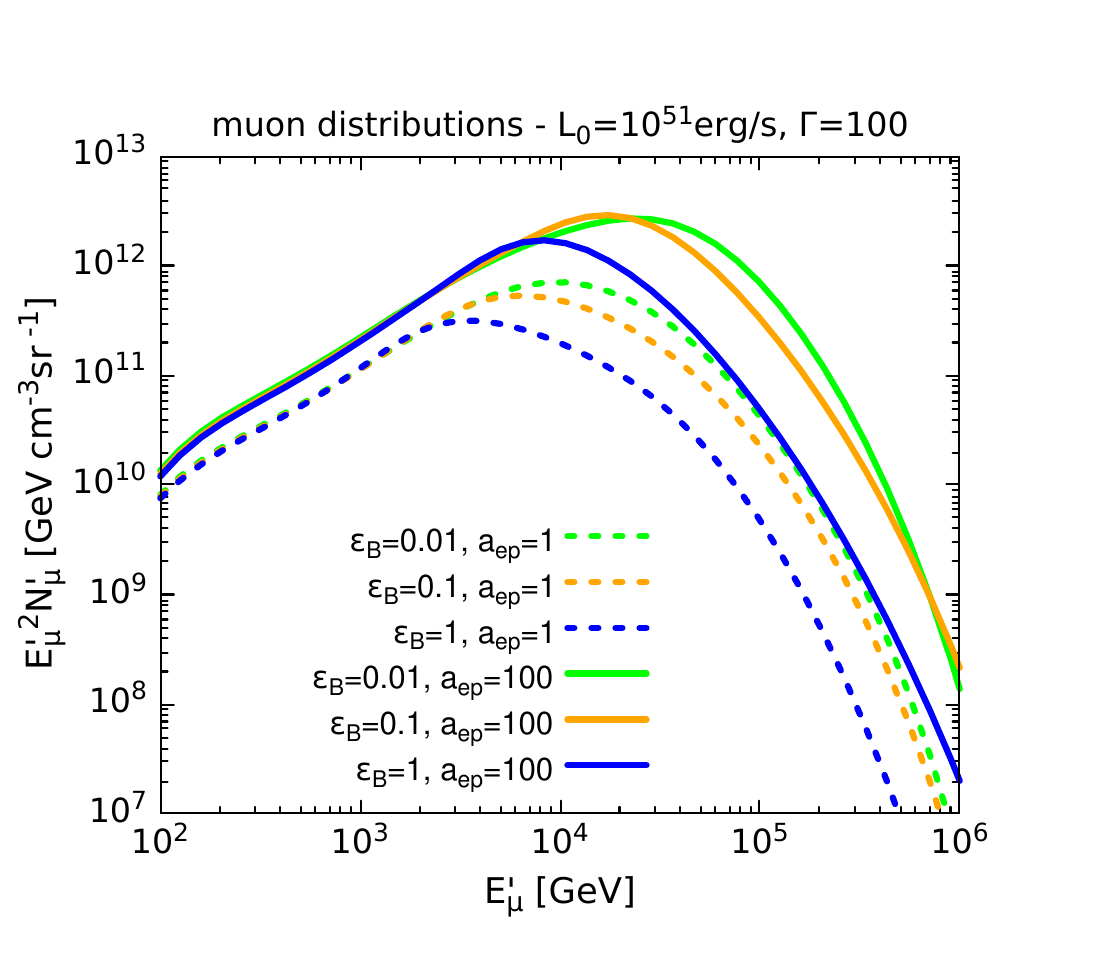} 
    \end{subfigure}     
    \begin{subfigure}[t]{0.49\textwidth}
        \centering                          
        \includegraphics[width=0.5\linewidth,trim= 125 30 170 35]{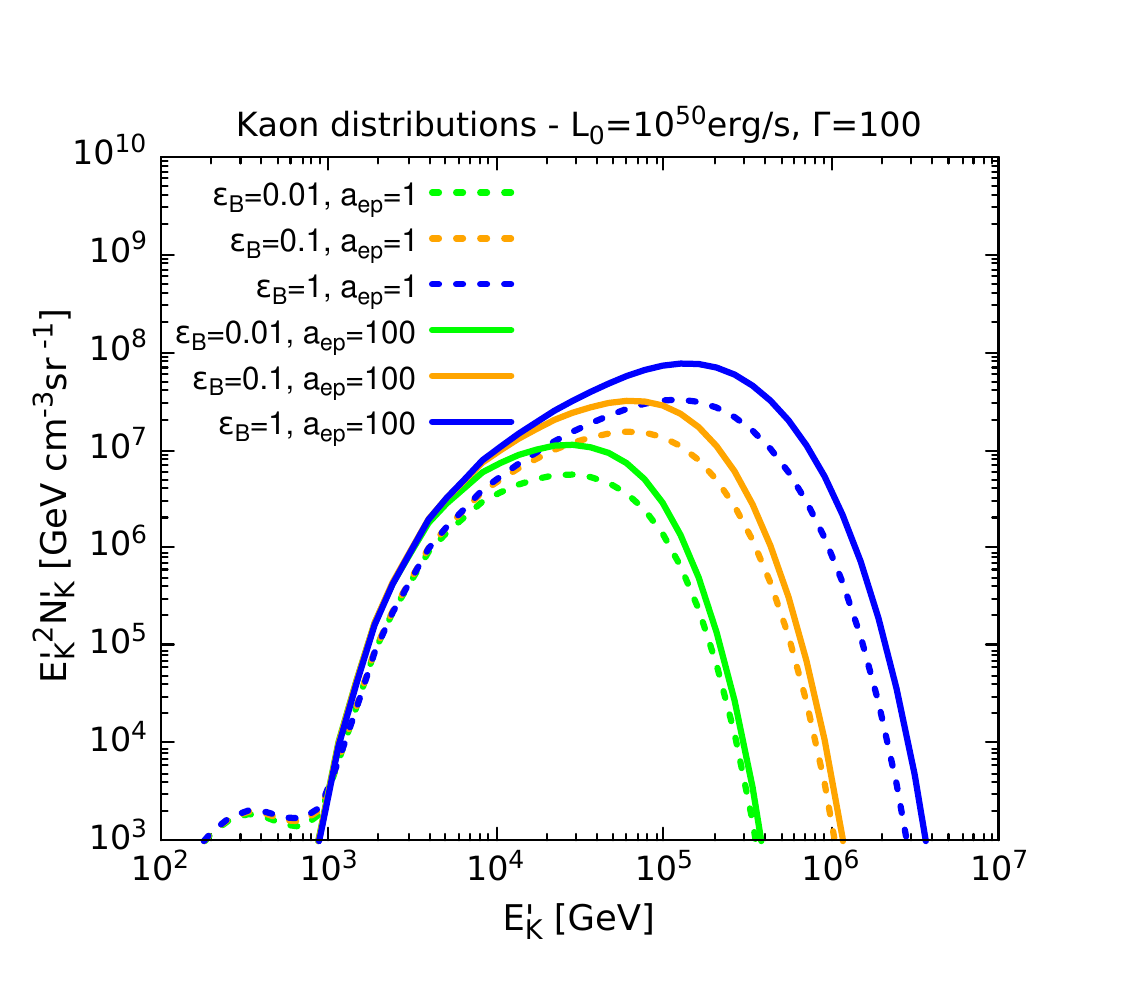} 
    \end{subfigure}
    \hfill
    \begin{subfigure}[t]{0.49\textwidth}
        \centering
        \includegraphics[width=0.5\linewidth,trim= 125 30 170 35]{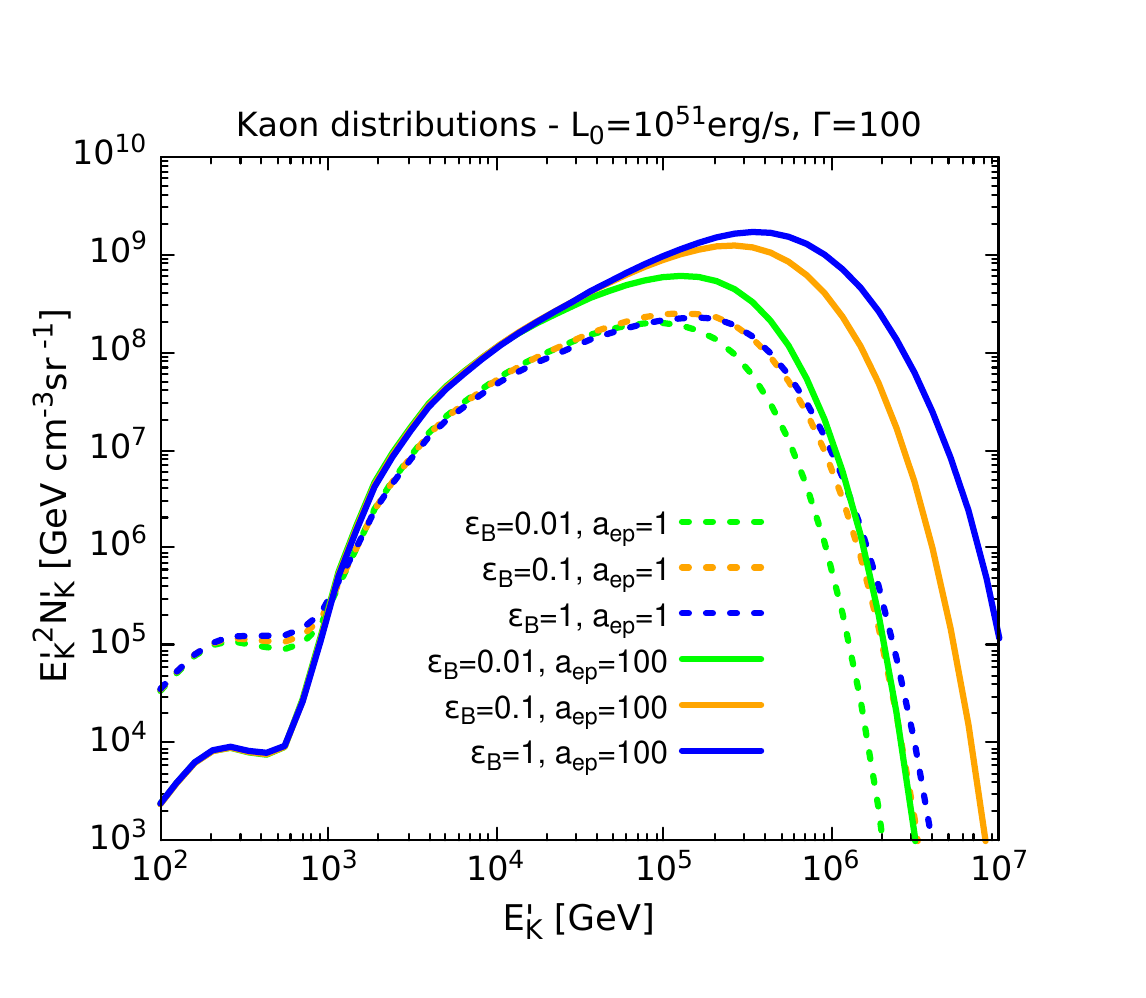} 
    \end{subfigure}     
    \caption{Distributions of pions (top), muons (middle), and charged kaons (bottom) {for $\Gamma=100$}. The left plots correspond to $L_0=10^{50}{\rm erg/s}$ and the right plots to $L_0=10^{51}{\rm erg/s}$. {The green, orange, and blue curves refer to the cases of $\epsilon_B=0.01$, $\epsilon_B=0.1$, and $\epsilon_B=1$, respectively. The dashed curves correspond to $a_{ep}=1$ and the solid curves to $a_{ep}=100$.}}\label{fig10:NpiNmuNK}
\end{figure*} 
\begin{figure*}[]
    \centering
    \begin{subfigure}[t]{0.49\textwidth}
        \centering                          
        \includegraphics[width=0.5\linewidth,trim= 130 30 170 50]{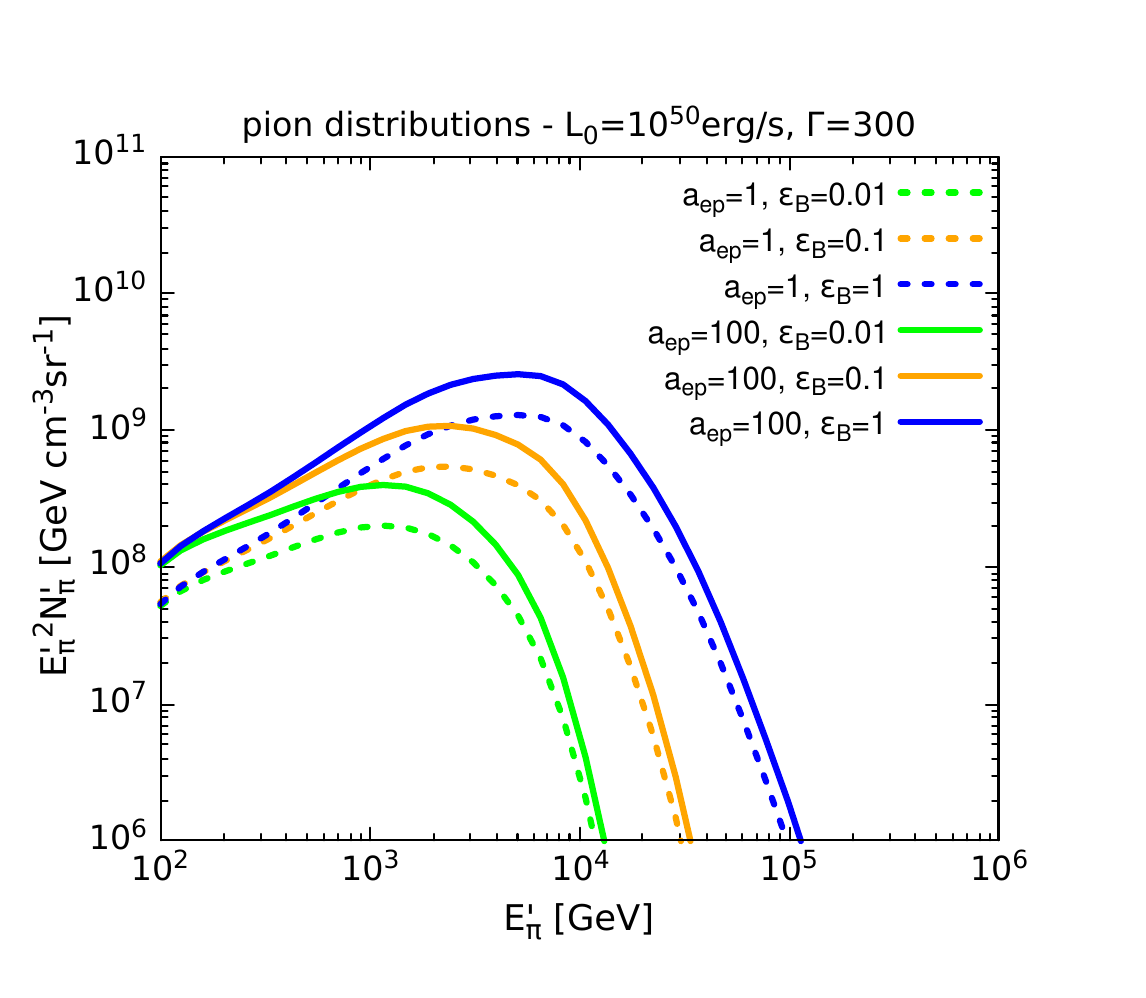} 
    \end{subfigure}
    \hfill
    \begin{subfigure}[t]{0.49\textwidth}
        \centering
        \includegraphics[width=0.5\linewidth,trim= 130 30 170 50]{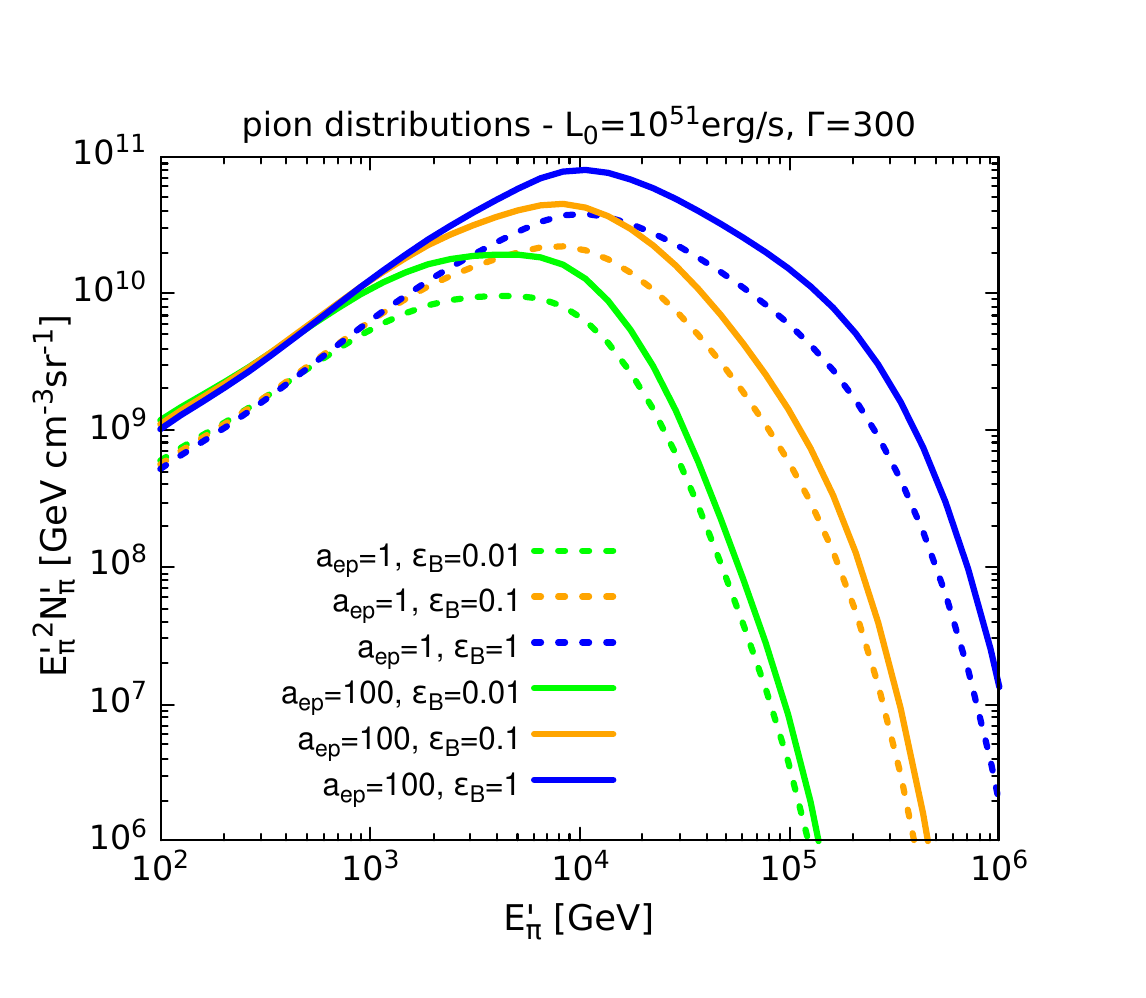} 
    \end{subfigure}
    \begin{subfigure}[t]{0.49\textwidth}
        \centering                          
        \includegraphics[width=0.5\linewidth,trim= 127 30 167 35]{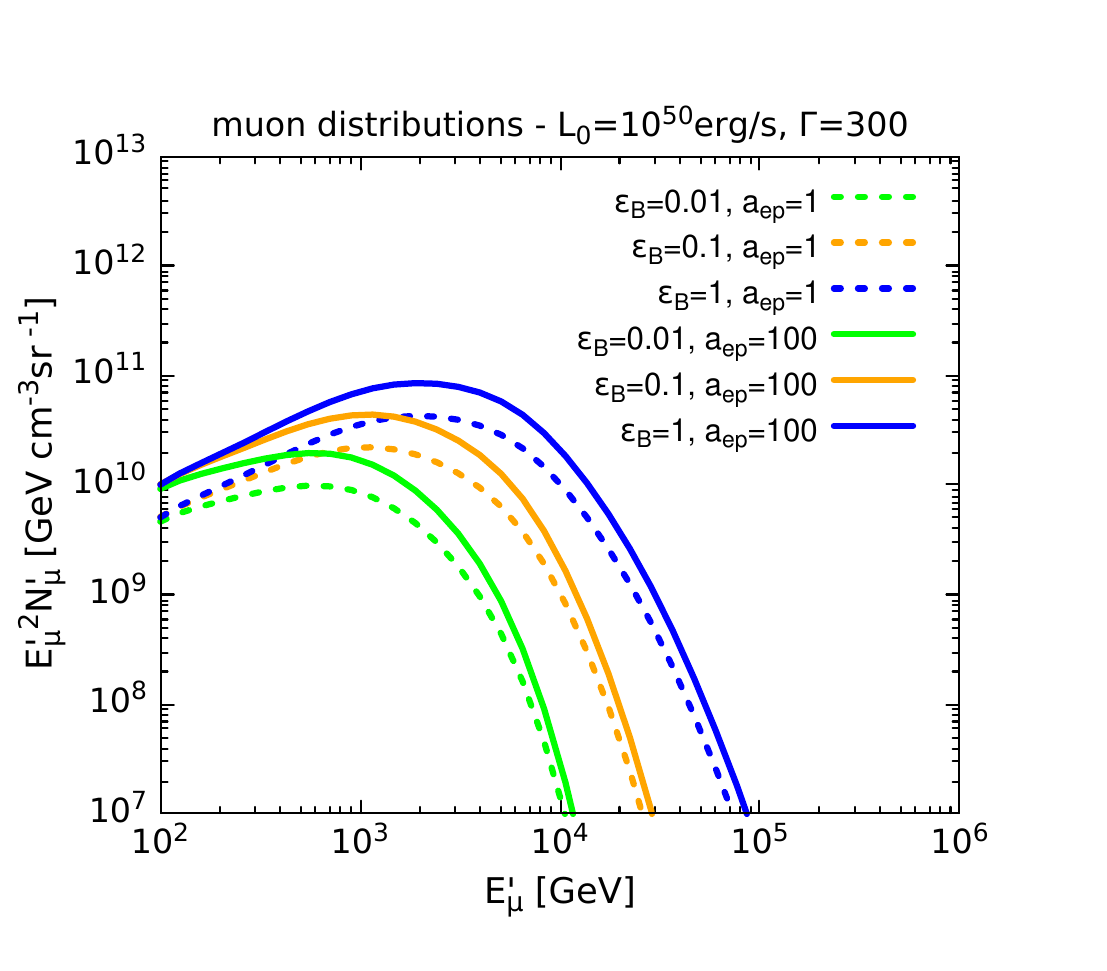} 
    \end{subfigure}
    \hfill
    \begin{subfigure}[t]{0.49\textwidth}
        \centering
        \includegraphics[width=0.5\linewidth,trim= 127 30 167 35]{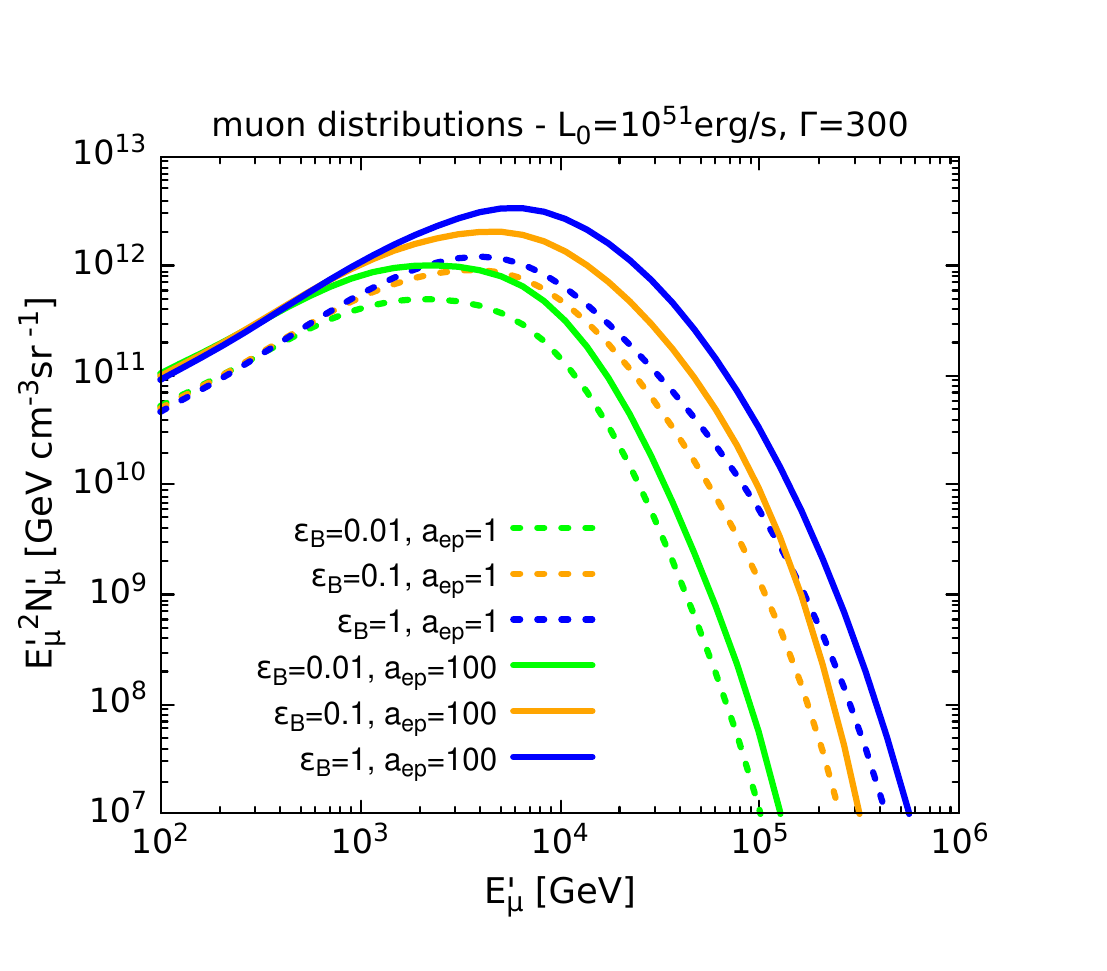} 
    \end{subfigure}     
    \begin{subfigure}[t]{0.49\textwidth}
        \centering                          
        \includegraphics[width=0.5\linewidth,trim= 130 30 170 35]{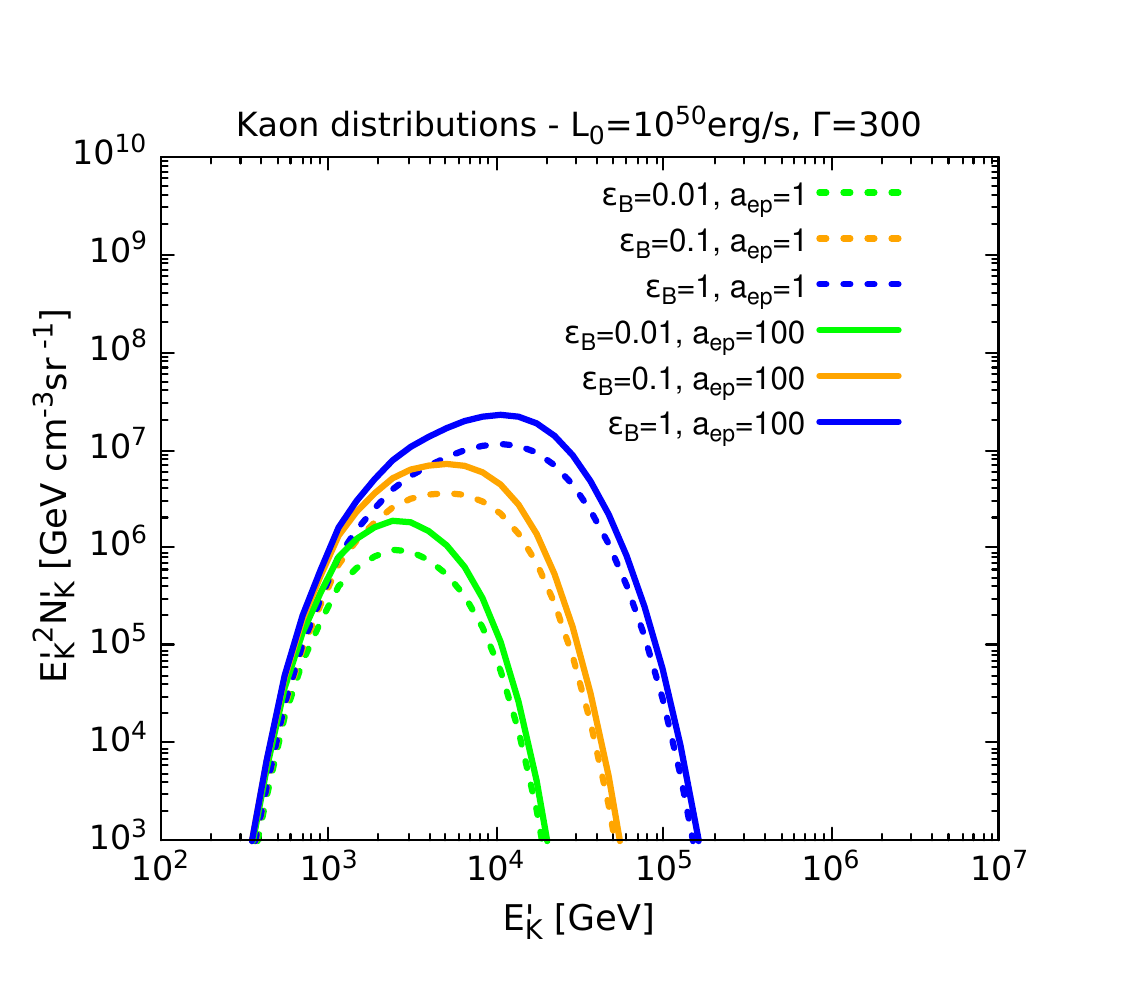} 
    \end{subfigure}
    \hfill
    \begin{subfigure}[t]{0.49\textwidth}
        \centering
        \includegraphics[width=0.5\linewidth,trim= 130 30 170 35]{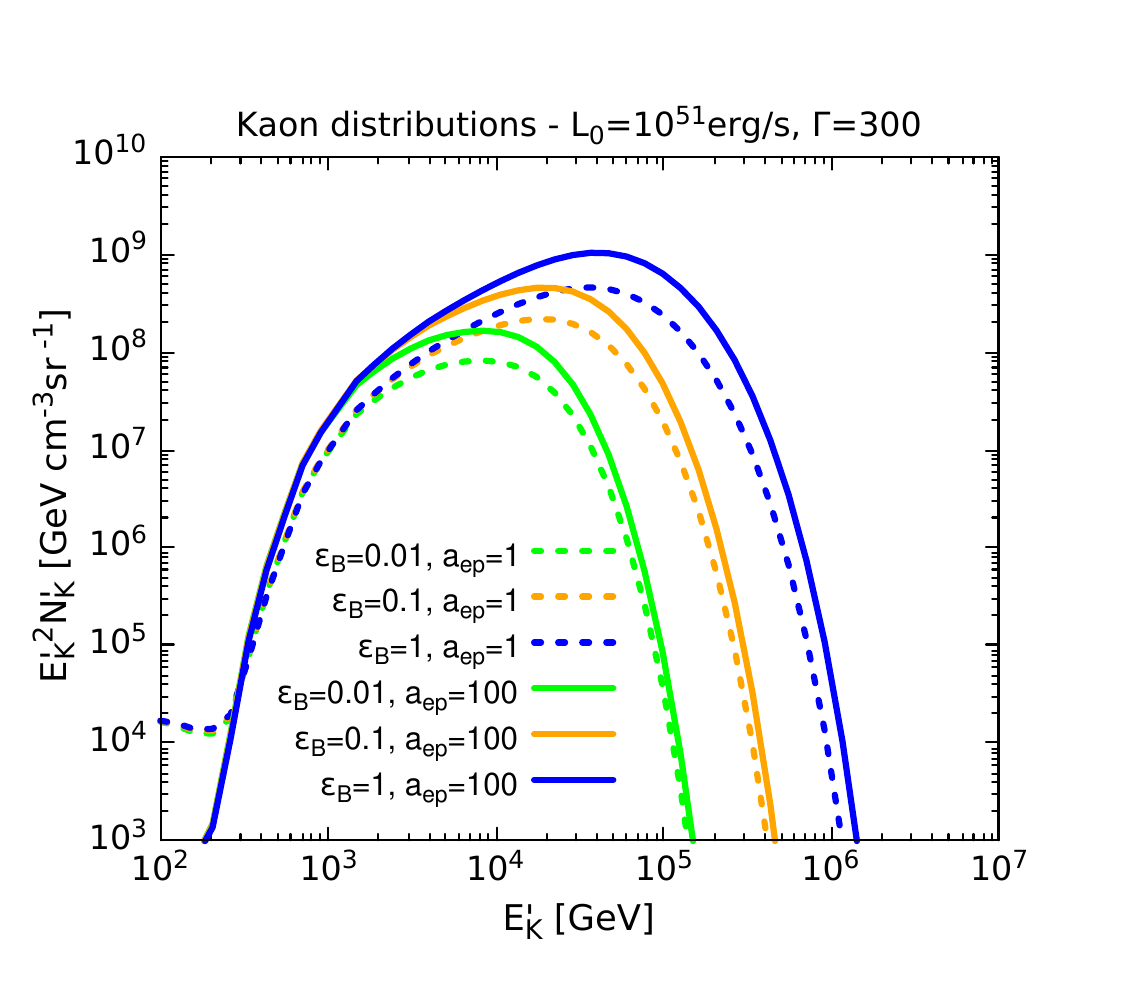} 
    \end{subfigure}     
    \caption{Distributions of pions (top), muons (middle), and charged kaons (bottom) {for $\Gamma=300$}. The left plots correspond to $L_0=10^{50}{\rm erg/s}$ and the right plots to $L_0=10^{51}{\rm erg/s}$. {The green, orange, and blue curves refer to the cases of $\epsilon_B=0.01$, $\epsilon_B=0.1$, and $\epsilon_B=1$, respectively. The dashed curves correspond to $a_{ep}=1$ and the solid ones to $a_{ep}=100$.}}\label{fig11:NpiNmuNK}
\end{figure*}
 \subsection{Pions, muons, and kaons}
 Pions are produced mainly through $p\gamma$ interactions, and we compute the corresponding injection following \cite{hummer2010}:
\begin{multline}
Q'_{p\gamma\rightarrow\pi^\pm}(E'_\pi)= \sum_{{\rm it}=\{\rm LR,NR\}}N'_{p}\left(\frac{E'_\pi}{\chi_{\rm it}}\right) \int_{E_{\rm th,\pi}/2}^\infty n_{\rm ph}\left(\frac{y\,\chi_{\rm it}m_pc^2}{E'_\pi}\right)\frac{M_\pi^{\rm it}}{2y^2} dy \\ \times \int_{E_{{\rm th},\pi}}^{2y}dE_rE_r \sigma_{p\gamma}(E_r),    
\end{multline}
where the interaction channels "${\rm it}$" refers to are the low and high energy resonances (LR and HR). The corresponding parameters $\chi_{LR,HR}$ and $M_{\pi}^{LR,HR}$ are listed in \cite{hummer2010}. Additionally, charged kaons are produced in $p\gamma$ collisions, although with a subdominant contribution at high energies. The injection of $K^+$ can be computed as
\be
Q'_{K^+}(E'_K)=  N'_{p}\left(\frac{E'_K}{\chi_{K}}\right) \int_{E_{\rm th,K}/2}^\infty n_{\rm ph}\left(\frac{y\,\chi_{K}m_pc^2}{E'_K}\right)\frac{1}{2y^2} dy \\ \times \int_{E_{{\rm th},K}}^{2y}dE_rE_r \sigma_{p\gamma}(E_r),
\ee
where $\chi_K=0.35$ and $E_{{\rm th},K}=1\, {\rm GeV}$. For the negative pions $K^-$, we make the approximation $Q_{K^-}\approx 0.5 Q_{K^+}$, which yields results consistent with the ones obtained by \cite{lipari2007}. The neutral kaons $K_L$ are also produced at an intermediate level between the $K^-$ and the $K^+$ injections. However, we neglected the $K_L$ contribution, since they only decay to neutrinos via three body channels, which implies that the neutrino energy is low and therefore the corresponding emissivity is expected to be negligible compared to those of the rest of the channels \citep[see the discussion in][]{hummer2010,petropoulou2014}.

Charged pions can also interact with matter and photons. In the former case, we obtain $t_{\pi p}^{-1}$ making the simple approximation that $\sigma_{\pi p}\approx \frac{2}{3}\sigma_{pp}$, which is inspired by the fact that while protons are made up of three valence quarks, pions are formed by only two of them \citep{gaisser1990}. As for pion-photon interactions, we adopt an approach similar to \cite{lipari2007} and consider the production of the resonances $\rho[770]$, $a_1[1260]$, $a_2[1320]$, and $b_1[1235]$, since they all have a possible decay channel to $\pi \gamma$ according to the Particle Data Group (PDG) review \citep{workman2022}. Therefore, for low values of center-of-mass energies, $\sqrt{s}<2{\, \rm GeV}$, we describe the cross section as the sum of the corresponding contributions using the Breit-Wigner expression for the mentioned resonances \citep{wigner1946}. For higher values of $s$, we adopted the general parametrization derived for hadronic interactions \citep{block2004} along with the fitting constants derived in the PDG review \citep{patrignani2016} to account for the total cross sections for the processes $pp$, $p\gamma$, and $\pi p$. These can be related to the $\pi\gamma$ cross section if we apply the factorization relation that follows from the Regge theory \citep[e.g.,][]{gribov2022}
\be 
  \sigma_{\pi\gamma}=\sqrt{ \sigma_{\pi\pi}\sigma_{\gamma\gamma} }  \simeq \sigma_{\pi p}\sqrt{\frac{\sigma_{\gamma\gamma}}{\sigma_{pp}}}, \label{sigmapig}
\ee   
where we also use $\sigma_{\pi\pi}= \sigma_{\pi p}^2/\sigma_{pp}$. The resulting cross section is shown in Fig. \ref{fig7:sigmapig}, and it was adopted to estimate the $\pi\gamma$ cooling rate with an expression analogous to that of Eq. (\ref{tpg}) for $p\gamma$ interactions, taking $K_{\pi\gamma}\approx 0.5$ for the inelasticity.

 We show all the obtained cooling rates for pions in Fig. \ref{fig8:pi-rates}, with $L_0=10^{50}{\rm erg/s}$ and $L_0=10^{51}{\rm erg/s}$ in the left and right panels, respectively, {and $\Gamma=100$ in the top panels and $\Gamma=300$ in the bottom ones}. In these plots, the decay is shown  to be faster than the cooling for energies $E'_\pi\lesssim 10^5{\rm GeV}$ for $\Gamma=100$, and for $E'_\pi\lesssim 2\times 10^4{\rm GeV}$ for $\Gamma=300$. As for the pion cooling, we unified the $\pi\gamma$ cooling rate in the figure and present the sum of the contribution due to collisions with internal photons from the synchrotron emission of electrons in the internal shock zone as well as the one corresponding to external photons produced in the jet head. {These $\pi\gamma$ interactions provide the dominant cooling mechanism for high energy pions regardless of the value of $a_{ep}$, except for the cases with $\Gamma=100$ and $L_0=10^{51}{\rm erg/s}$ at energies greater than $10^{5}{\rm GeV}$, which is where pion synchrotron can dominate. We note that $\pi\gamma$ interactions actually refer to inelastic collisions characterized by the cross section obtained in Eq. (\ref{sigmapig}). For high energies, where the pion decay rate becomes subdominant, such $p\gamma$ interactions give the dominant cooling rate as compared to the corresponding pion-IC interactions. The latter take place at a rate similar to the muon IC cooling rate, and refer to the elastic case where there is only a pion and a photon in the final state.}

The case of the charged kaons is quite similar to that of the pions. They decay about twice as fast as pions, and the $K^{\pm}\gamma$ interactions are assumed to occur at the same rate as for pions. However, the synchrotron cooling rate for kaons is a factor $(m_\pi/m_K)^3 \simeq 2.3\times 10^{-2}$ lower than the one for pions.

 Muons are produced by the decay of pions, and we compute their injection $Q'_\mu(E'_\mu)$ as in \cite{reynoso2009}, making use of the formulas of \cite{lipari2007}, which depend on the distributions of pions $N'_\pi$ and account for the kinematics of the decay. The relevant cooling rates for muons are shown in Fig. \ref{fig9:mu-rates} for $L_0=10^{50}{\rm erg/s}$ and $L_0=10^{51}{\rm erg/s}$, {with $\Gamma=100$ in the top panels and $\Gamma=300$ in the bottom ones}. In the figure, it can be seen that decay takes place faster than the cooling for $E'_\mu\lesssim 10^{4}{\rm GeV}$, while for higher energies, the synchrotron and IC become significant. {It can also be seen that the IC cooling of muons at high energies is sensitive to the value of $a_{ep}$, except for the cases with $\Gamma=300$ and $L_0=10^{50}{\rm erg/s}$.}
 
 The pion, muon, and kaon distributions are obtained using Eq.(\ref{solution_Eq_transport}), making the replacement $T^{-1}_{\rm esc}\rightarrow T^{-1}_{\rm esc}+T^{-1}_{i,\rm d}$ where the decay timescales for each of these particle types are
 \be
  T_{\pi,\rm d}(E'_\pi)&\simeq & 2.6\times 10^{-8}{\rm s}\left(\frac{E'_\pi}{m_\pi c^2}\right) \\
  T_{\mu,\rm d}(E'_\mu)&\simeq &2.2\times 10^{-6}{\rm s}\left(\frac{E'_\mu}{m_\mu c^2}\right)\\
  T_{K,\rm d}(E'_K)&\simeq & 1.2\times 10^{-8}{\rm s}\left(\frac{E'_K}{m_K c^2}\right).  
 \ee  
 We show the obtained $N'_\pi(E'_\pi)$, $N'_\mu(E'_\mu)$, and $N'_K(E'_K)$ in the top, middle, and bottom panels of Fig. \ref{fig10:NpiNmuNK} for $\Gamma=100$ and of Fig. \ref{fig11:NpiNmuNK} for $\Gamma=300$. When comparing the resulting pion distributions, we observe that the different outcomes are sensitive to the proton distribution corresponding to each case as well as to the relevance of both $\pi\gamma$ and synchrotron emission at high energies. The same goes for the muon distributions, which directly depend on the corresponding pion distribution, and since they have a slower decay rate, the IC and synchrotron losses affect their distribution at lower energies than for pions. 
\

{We note that for $\Gamma=100$ and $L_0=10^{51}{\rm erg/s}$, the synchrotron cooling can be important at high energies for muons and even pions. Specifically, in the top-right panel of Fig. \ref{fig10:NpiNmuNK}, the obtained pion distributions for $\epsilon_B=1$ and $\epsilon_B=0.1$ extend up to similar energies despite the fact that in the former case, protons are accelerated to higher energies, but this is compensated by more efficient pion synchrotron losses than in the latter case. Similarly, in the middle-right panel of the same figure, the muon distributions extend to lower energies as $\epsilon_B$ increases, as a consequence of more severe synchrotron losses. These losses are also significant for $\Gamma=100$ and $\epsilon_B=1$, as can be seen in the middle-left panel of Fig. {\ref{fig10:NpiNmuNK}}.}

{In the other cases, since synchrotron cooling of pions or muons does not dominate, increasing $\epsilon_B$ just leads to a distribution that extends to higher energies due to a more efficient acceleration of parent protons. As for kaons, their distributions peak at higher energies than for pions, in agreement with \cite{hummer2010}, and  $K^\pm \gamma$ interactions are the most relevant cooling mechanism. Therefore, even for $\Gamma=100$, $L_0=10^{51}{\rm erg/s}$, and $\epsilon_B=1$, synchrotron cooling does not significantly affect their distribution, as can be seen for $a_{ep}=100$ in the bottom-right panel of Fig. \ref{fig10:NpiNmuNK}. If $a_{ep}=1$, instead, the cases with $\epsilon_B=0.1$ and $\epsilon_B=1$ become similar because the increase of the magnetic field is compensated by faster $K\gamma$ interactions with photons of the electron synchrotron emission. }
{We also note that different values of the proton-to-electron ratio affect the muon distributions more significantly if $\Gamma=100$, but only slightly for $\Gamma=300$ and $L_0=10^{51}{\rm erg/s}$.}

\begin{figure*}[]
    \centering
    \begin{subfigure}[t]{0.49\textwidth}
        \centering                          
        \includegraphics[width=0.5\linewidth,trim= 170 30 180 50]{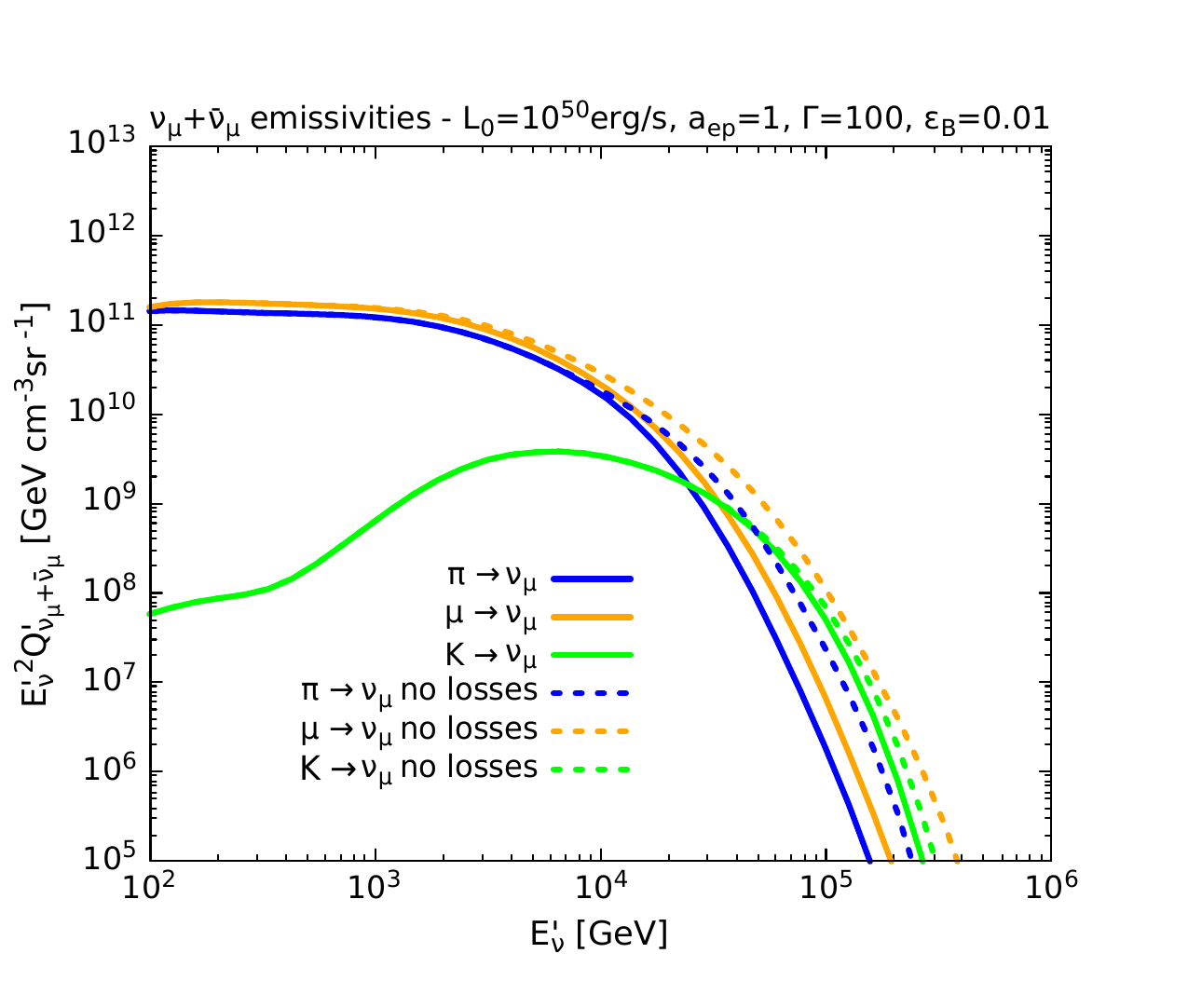} 
    \end{subfigure}
    \hfill
    \begin{subfigure}[t]{0.49\textwidth}
        \centering
        \includegraphics[width=0.5\linewidth,trim= 170 30 180 50]{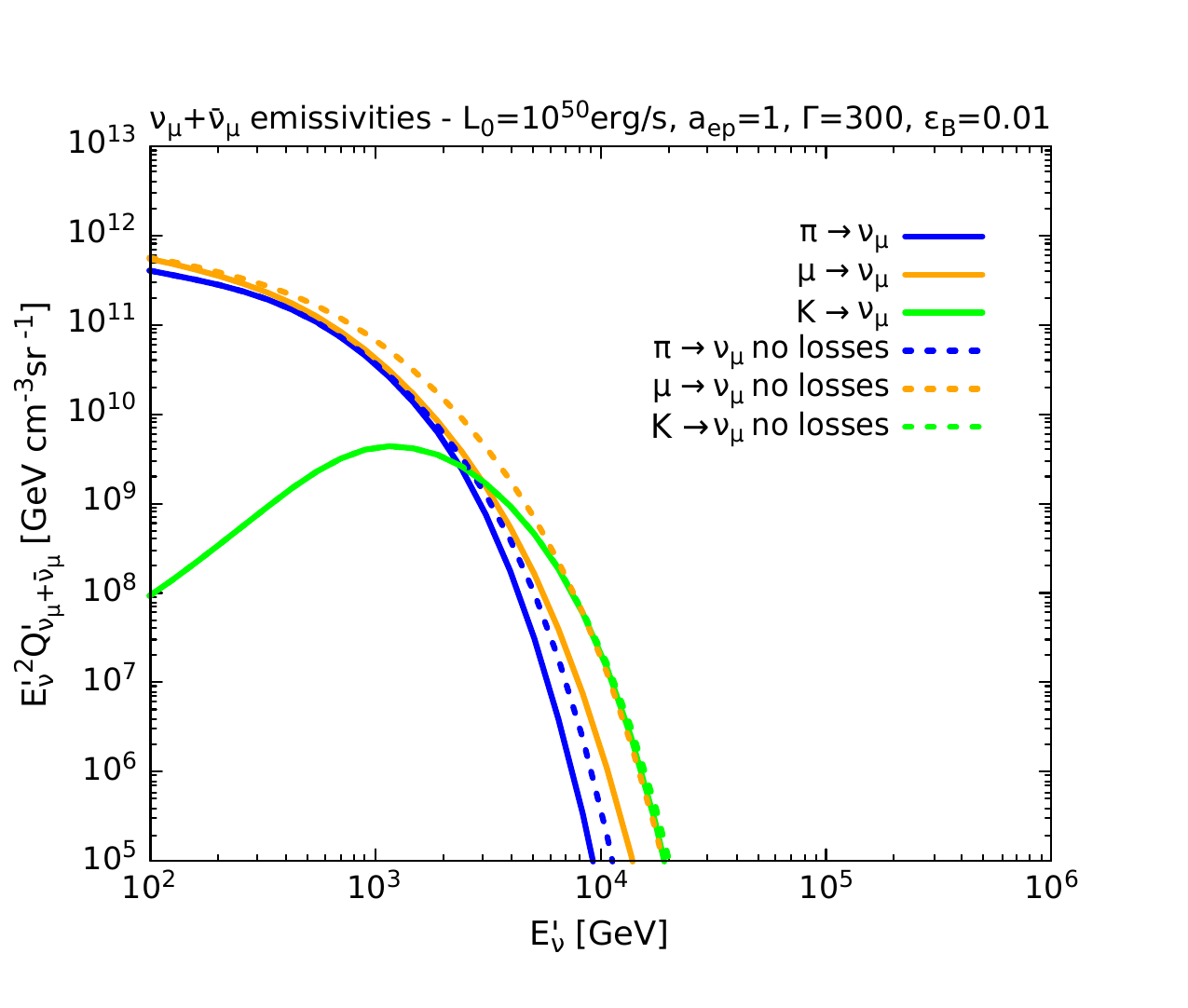} 
    \end{subfigure}
    
    \begin{subfigure}[t]{0.49\textwidth}
        \centering                          
        \includegraphics[width=0.5\linewidth,trim= 170 30 180 30]{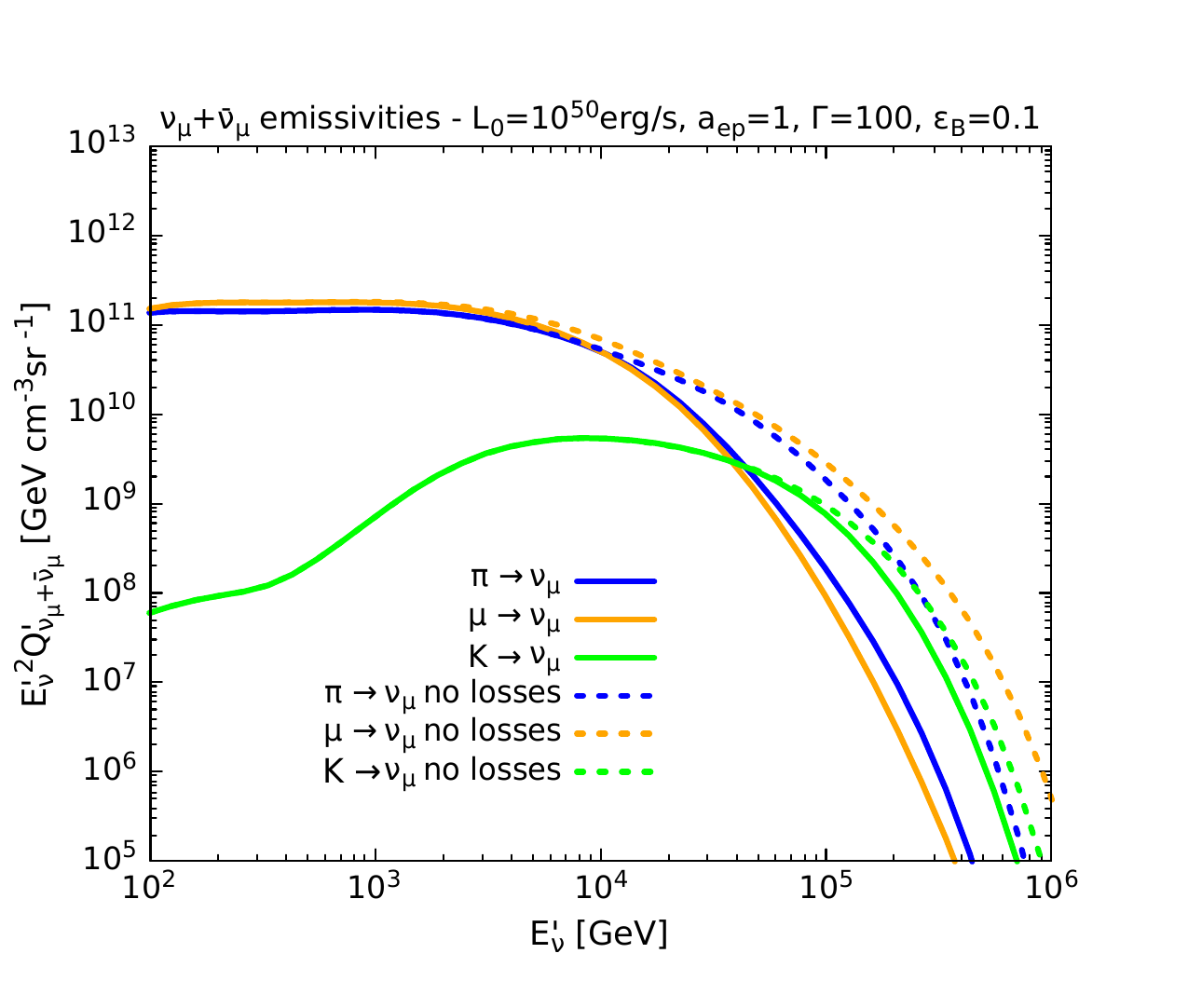} 
    \end{subfigure}
    \hfill
    \begin{subfigure}[t]{0.49\textwidth}
        \centering
        \includegraphics[width=0.5\linewidth,trim= 170 30 180 30]{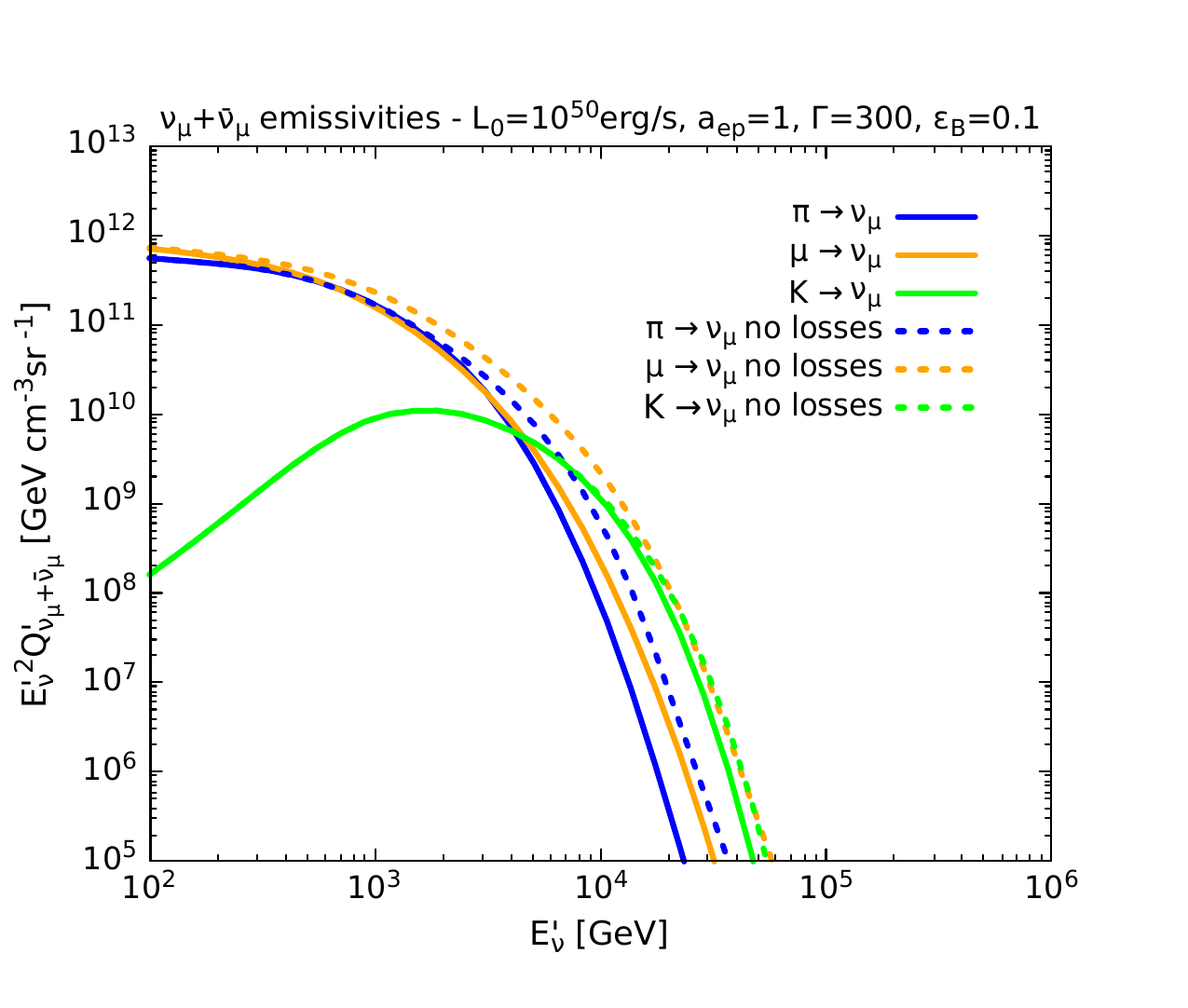} 
    \end{subfigure}     

    \begin{subfigure}[t]{0.49\textwidth}
        \centering                          
        \includegraphics[width=0.5\linewidth,trim= 170 30 180 30]{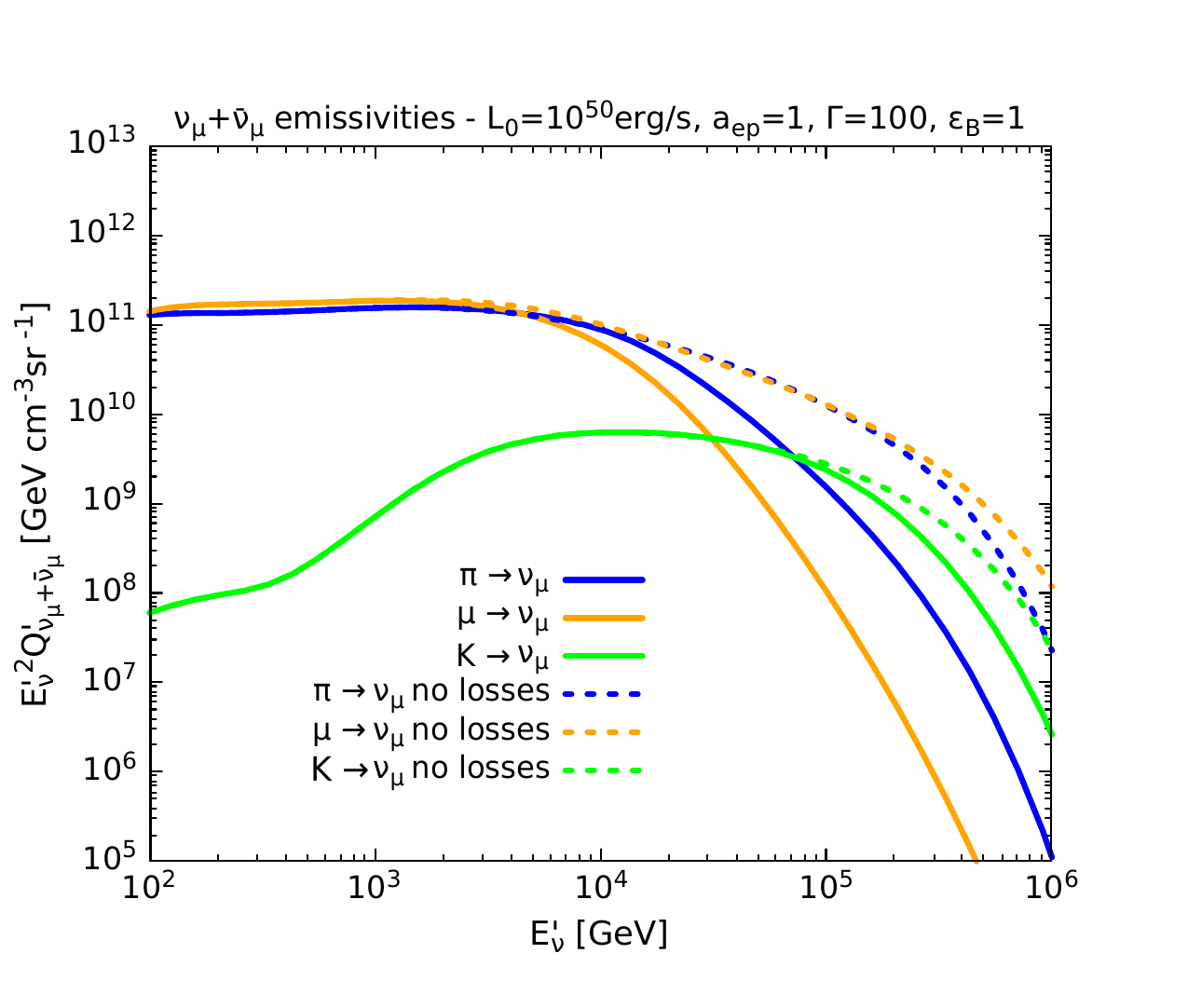} 
    \end{subfigure}
    \hfill
    \begin{subfigure}[t]{0.49\textwidth}
        \centering
        \includegraphics[width=0.5\linewidth,trim= 170 30 180 30]{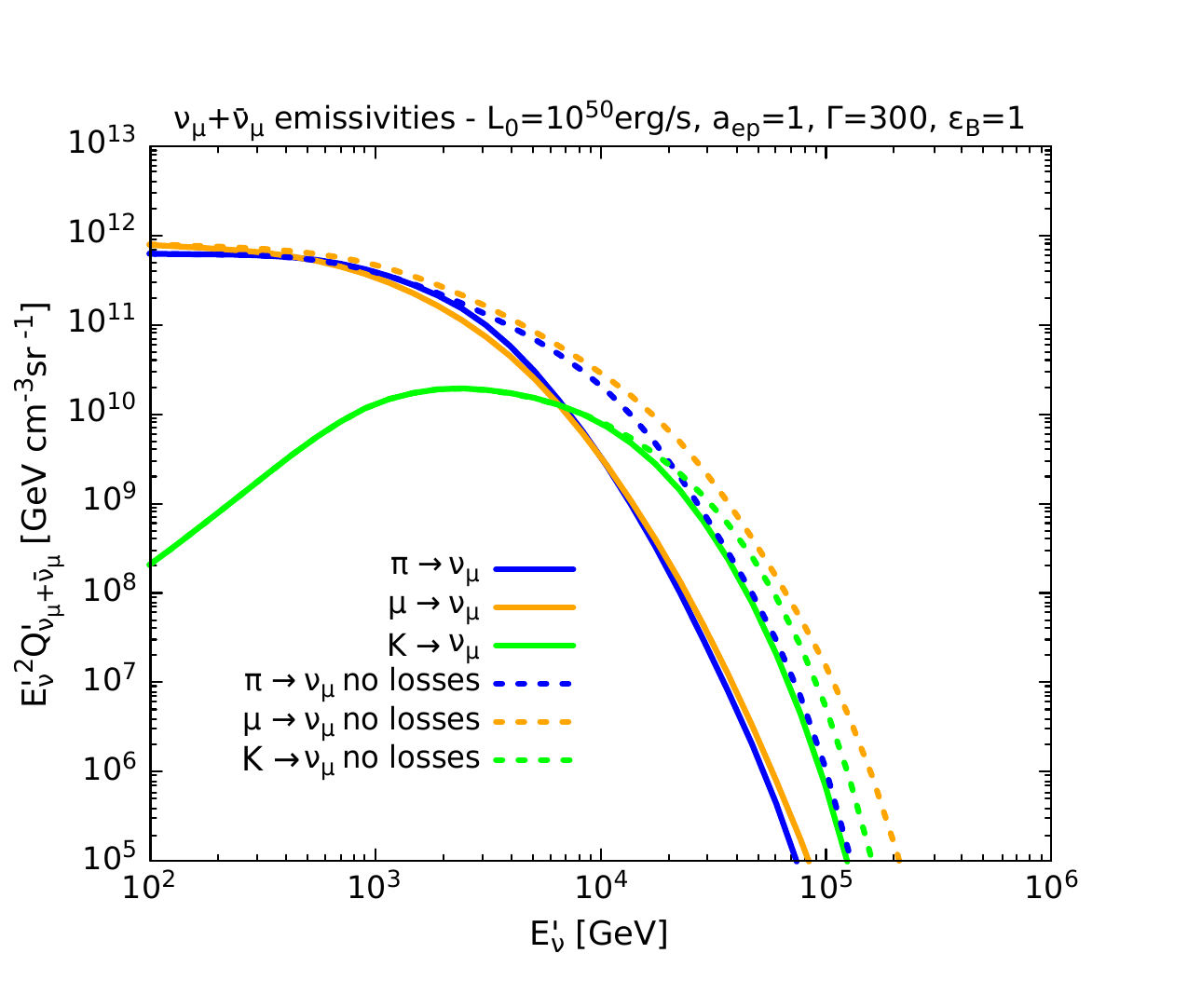} 
    \end{subfigure}     
    \caption{Emissivities of muon neutrinos and antineutrinos from the decay of pions (blue curves), muons (orange curves), and kaons (green curves) for $L_0=10^{50}{\rm erg/s}$. The dashed curves represent the obtained output if all  losses of the parent particles are neglected, while the solid curves mark the results if all the losses discussed are considered. The left plots correspond to $a_{ep}=1$ and the right plots to $a_{ep}=100$. }\label{fig12:Qnus}
\end{figure*} 
\begin{figure*}[]
    \centering
    \begin{subfigure}[t]{0.49\textwidth}
        \centering                          
        \includegraphics[width=0.5\linewidth,trim= 170 30 180 50]{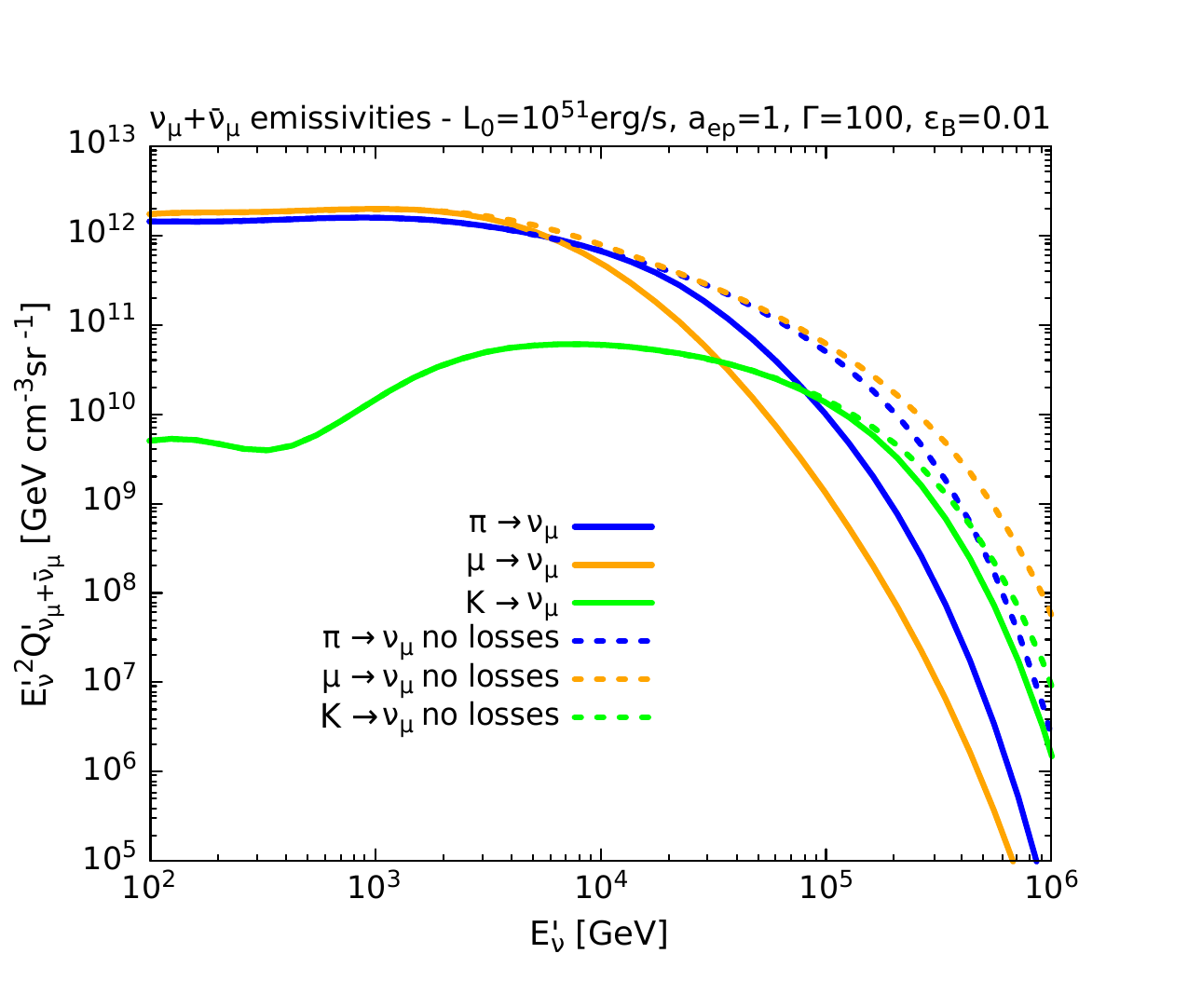} 
    \end{subfigure}
    \hfill
    \begin{subfigure}[t]{0.49\textwidth}
        \centering
        \includegraphics[width=0.5\linewidth,trim= 170 30 180 50]{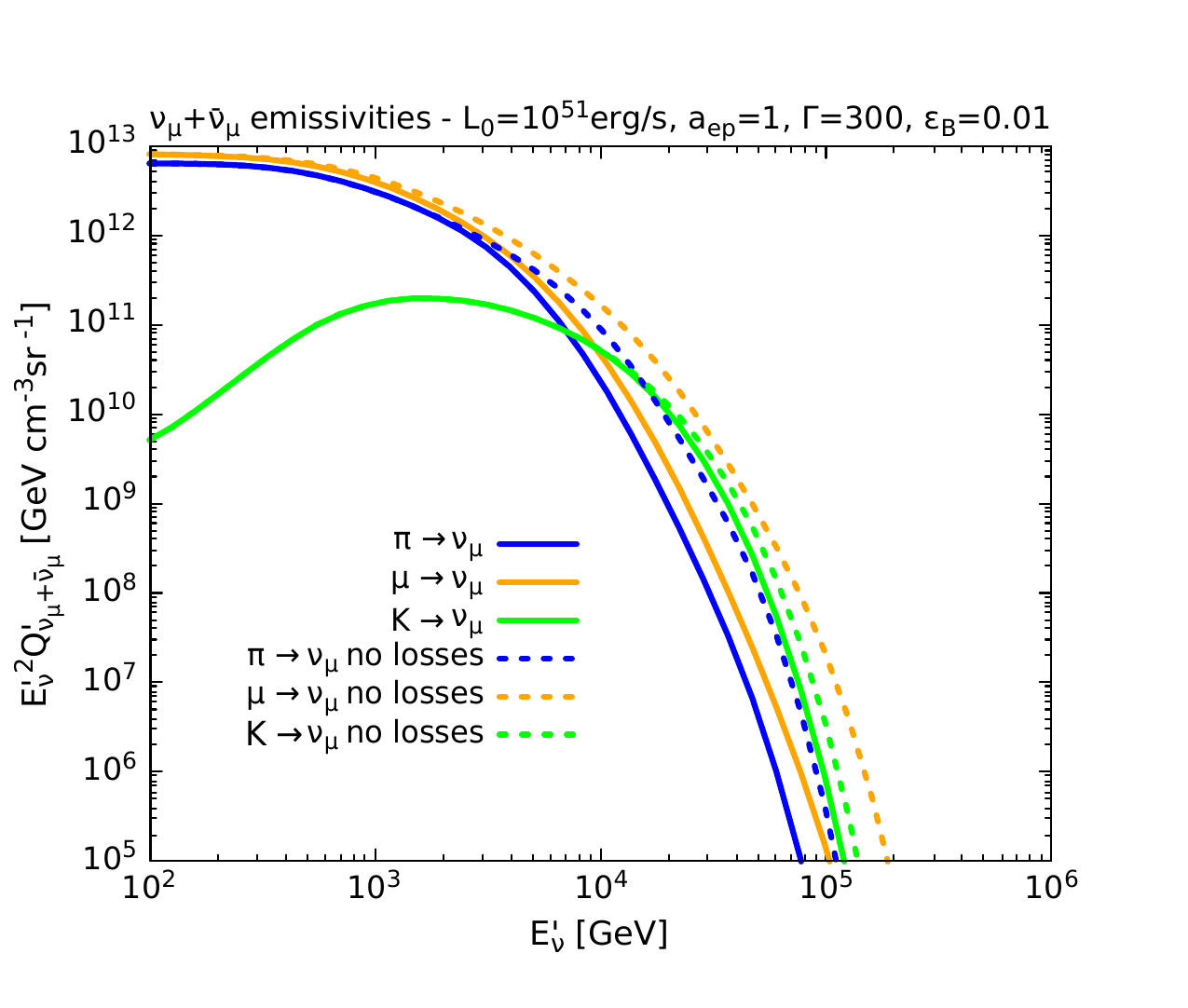} 
    \end{subfigure}
    
    \begin{subfigure}[t]{0.49\textwidth}
        \centering                          
        \includegraphics[width=0.5\linewidth,trim= 170 30 180 30]{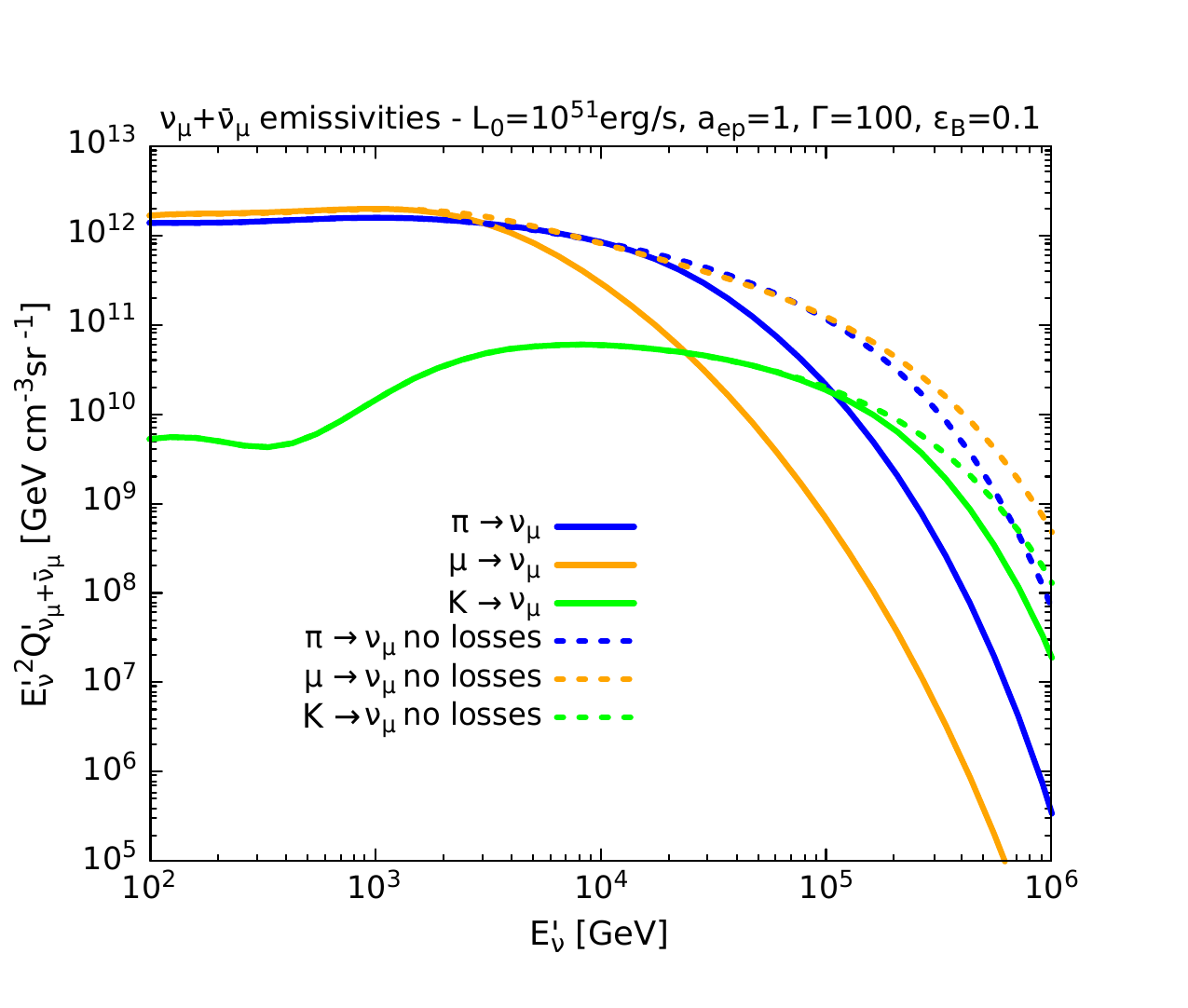} 
    \end{subfigure}
    \hfill
    \begin{subfigure}[t]{0.49\textwidth}
        \centering
        \includegraphics[width=0.5\linewidth,trim= 170 30 180 30]{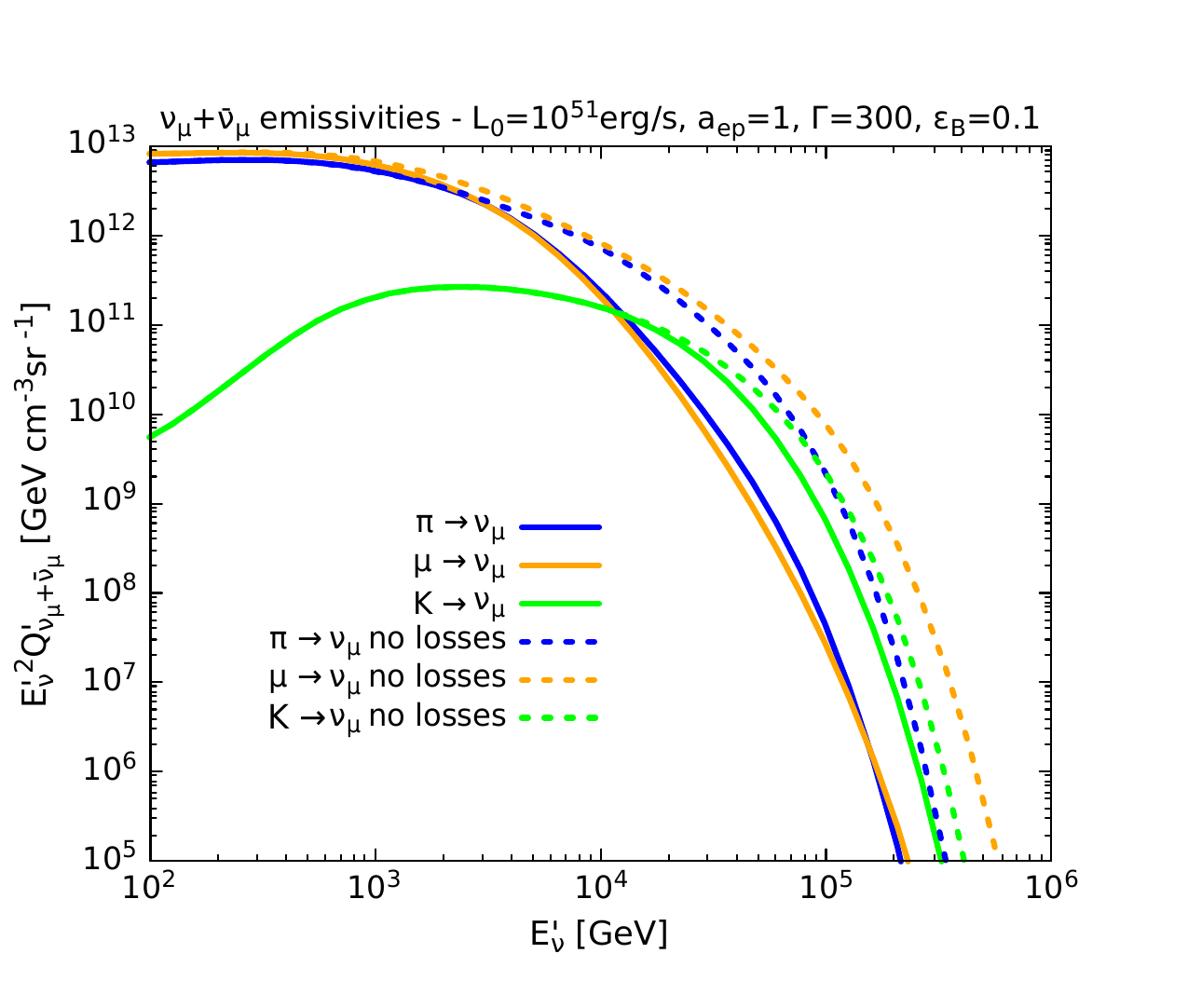} 
    \end{subfigure}     

    \begin{subfigure}[t]{0.49\textwidth}
        \centering                          
        \includegraphics[width=0.5\linewidth,trim= 170 30 180 30]{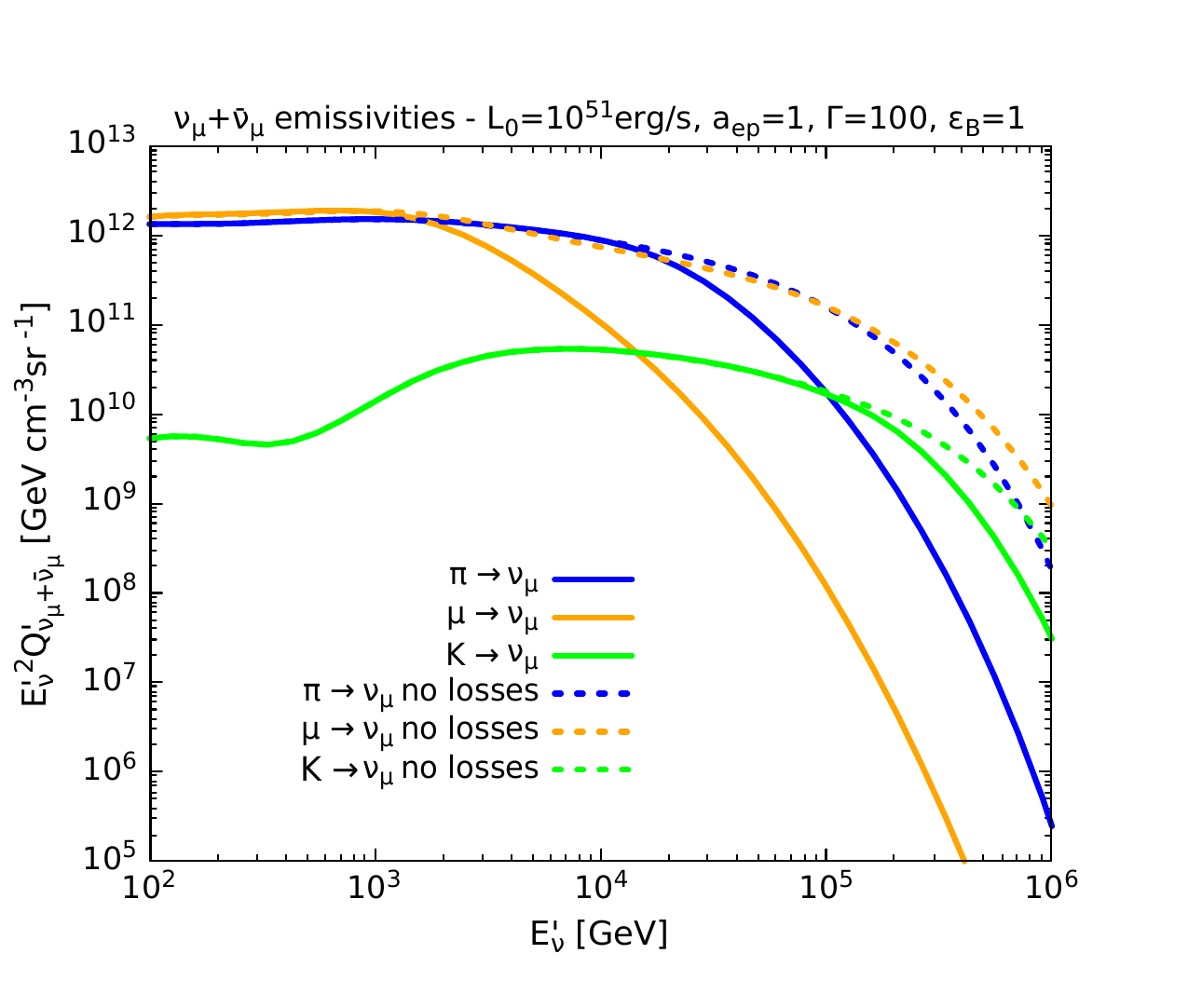} 
    \end{subfigure}
    \hfill
    \begin{subfigure}[t]{0.49\textwidth}
        \centering
        \includegraphics[width=0.5\linewidth,trim= 170 30 180 30]{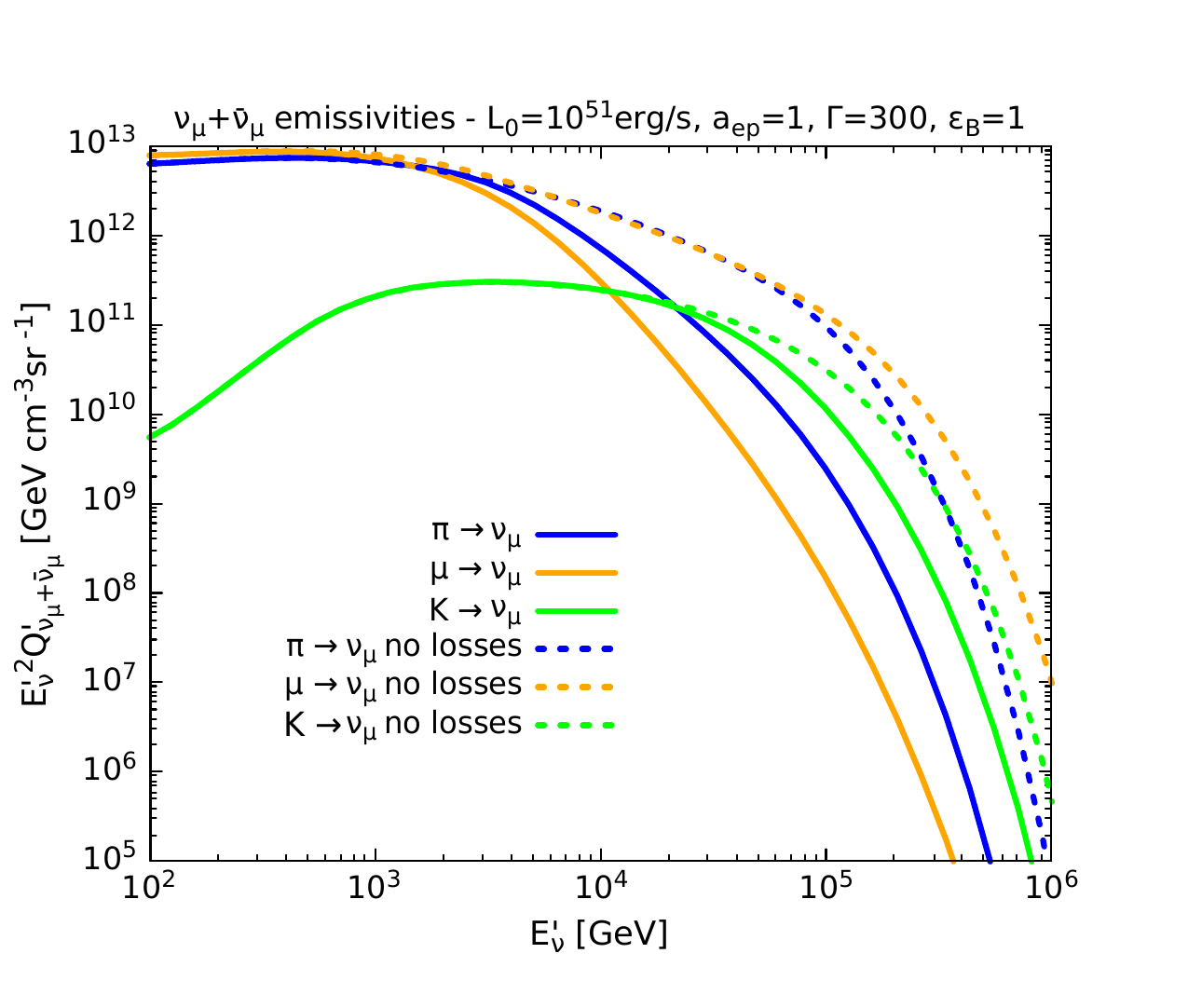} 
    \end{subfigure}     
    \caption{Emissivities of muon neutrinos and antineutrinos from the decay of pions (blue curves), muons (orange curves), and kaons (green curves) for $L_0=10^{51}{\rm erg/s}$. The dashed curves represent the obtained output if all  losses of the parent particles are neglected, while the solid curves mark the results if all the losses discussed are considered. The left plots correspond to $a_{ep}=1$ and the right plots to $a_{ep}=100$.}\label{fig13:Qnus}
\end{figure*} 

\subsection{Neutrinos}

Once the pion, muon, and kaon distributions are obtained, we can compute the injection or emissivity of neutrinos plus antineutrinos, as we are interested in studying the overall flavor composition. The contribution to the emissivity of muon neutrinos and antineutrinos resulting from the pion decays $\pi^+\rightarrow \mu^+\nu_\mu$ and $\pi^-\rightarrow \mu^-\bar{\nu}_\mu$ can be computed as \citep[e.g.,][]{lipari2007}:
\be 
Q'_{\pi\rightarrow\nu_\mu}(E'_\nu)=\int_{E_{\pi,\rm min}}^{\infty}dE_\pi \frac{N'_\pi(E_\pi)}{E_\pi(1-r_\pi)T_{\pi,\rm d}(E_\pi)},
\ee
where $E_{\pi,\rm min}=\frac{E'_\nu}{1-r_\pi}$, with $r_\pi= (m_\mu/m_\pi)^2$. A similar expression corresponds to the kaon decays  $K^+\rightarrow \mu^+\nu_\mu$ and $K^-\rightarrow \mu^-\bar{\nu}_\mu$ weighted by the branching ratio $Br_{K\rightarrow \mu {\nu}_\mu}\simeq 0.63$, making the replacements $N'_\pi\rightarrow N'_K$, $T_{\pi,\rm d}\rightarrow T_{K,\rm d}$, and $r_\pi\rightarrow r_K$.  

The contribution to the muon flavor at the production stage generated by the decays of muons $\mu^+\rightarrow e^+\nu_e\bar{\nu}_\mu$ and $\mu^-\rightarrow e^-\bar{\nu}_e{\nu}_\mu$ is accounted by the following emissivity \citep{lipari2007}:
\begin{multline}
Q'_{\mu\rightarrow\nu_\mu}(E'_\nu)= \sum_{i=1}^4\int_{E'_\nu}^{\infty}\frac{dE_\mu}{E_\mu}\frac{N'_{\mu_i}(E_\mu)}{T_{\mu,\rm d}(E_\mu)} \\ \times \left[\frac{5}{3}-3x^2+\frac{4}{3}x^3+
\left(3x^2-\frac{1}{3}-\frac{8x^3}{3}\right)h_i\right],  
\end{multline}
where $x=E'_\nu/E_\mu$, and the $i$ index refers to the different muons depending on their charge and helecity (i.e., $\mu_{1,2}=\mu_L^{-,+}$, $\mu_{3,4}=\mu_R^{-,+}$), with  $h_{1,2}=-h_{3,4}=-1$.
Similarly, the injection of electron neutrinos and antineutrinos can be computed as: 
\begin{multline}
Q'_{\mu\rightarrow\nu_e}(E'_\nu)= \sum_{i=1}^4\int_{E'_\nu}^{\infty}\frac{dE_\mu}{E_\mu}\frac{N'_{\mu_i}(E_\mu)}{T_{\mu,\rm d}(E_\mu)} \\ \times \left[2-6x^2+4x^3+
\left( 2-12x+18x^2-8x^3 \right)h_i\right].  
\end{multline}

In Figs. \ref{fig12:Qnus} and \ref{fig13:Qnus}, we show the different contributions to neutrino emissivities  
for $L_0=10^{50}{\rm erg/s}$ and $L_0=10^{51}{\rm erg/s}$, respectively, {both for $a_{ep}=1$. In these figures, the left panels correspond to $\Gamma=100$ and the ones on the right to $\Gamma=300$}. The cases with $\epsilon_B=\{0.01,0.1,1\}$ are shown in the top, middle, and bottom panels, respectively, and the dashed lines refer to the outcomes that would be obtained if losses of pions, muons, and kaons were neglected. It can be seen that even in the cases of low magnetic field {(i.e., Fig. \ref{fig12:Qnus} for $L_0=10^{50}{\rm erg/s}$, $\epsilon_B=0.01$, and $\Gamma=300$),} the neutrino emission is reduced at high energies due to interactions of the parent particles with the soft photon background, as mentioned above. In the cases of higher magnetic fields, synchrotron losses generate further depletion of the neutrino emissivity at high energies. In particular, the muon contribution is the most affected. Kaons are only moderately affected, and the neutrinos resulting from their decays give a small correction at the highest energies, similarly to what has been obtained in other previously studied cases \citep[e.g.,][]{hummer2010,hummer2012}.

\section{Diffuse neutrino flux and flavor ratios}\label{sec:fluxandflavor}

After considering in detail the cooling effect of the parent particles that generate the neutrinos, we are next interested in obtaining the diffuse neutrino flux produced by the cumulative CGRBs taking place at different redshifts in the universe. This would allow us to assess the contribution of these sources to the total neutrino flux that reaches the Earth, as well as their flavor composition. We assume that the generation rate of CGRBs $R_{\rm CGRB}(z)$ is proportional to the star formation rate $\rho_\star(z)$ \citep[e.g.,][]{he2018,fasano2021}: 
\be 
R_{\rm CGRB}(z)= A_{\rm CGRB}\rho_\star(z),
\ee  
where $A_{\rm CGRB}$ is a normalization constant. Following \cite{madau2014}, we adopt
\be
\rho_\star(z)= 0.015 M_\odot{\rm yr}^{-1}{\rm Mpc}^{-3}\frac{(1+z)^{2.7}}{1+[(1+z)/2.9]^{5.6}}.
\ee
The source evolution function $R_{\rm CGRB}(z)$ is used to weight the neutrino spectrum produced by a single typical CGBRB, 
\be
\frac{dN_{\nu,z}(E_{\nu,z})}{dE_{\nu,z}}= \Delta\Omega\, Q_{\nu,z}(E_{\nu,z})\Delta V\, T_{\rm CGRB} \left(\frac{\Delta\Omega}{4\pi}\right), 
\ee
where $E_{\nu,z}=E_\nu(1+z)$ is the local neutrino energy at redshift $z$ corresponding to an energy $E_\nu$ for an observer on Earth. We introduce the ratio $\Delta\Omega/(4\pi)$ to account for the fact that these sources are beamed, and we use an isotropic equivalent luminosity to obtain the neutrino emissivity $Q'_\nu(E'_\nu)$ at the comoving frame of source. The latter is then transformed to the local frame according to
\be
Q_{\nu,z}(E_{\nu,z})&\simeq & D\, Q'_\nu(E'_\nu)  \nonumber \\
      &\simeq &    2\Gamma Q'_\nu(E_{\nu,z}/(2\Gamma)). 
\ee
Given that the comoving neutrino density can be expressed as \citep[e.g.,][]{murase2007}
\be
dn_\nu(E_\nu)= R_{\rm CGRB}(z)(1+z)^3 \frac{dt}{dz}dz \, \frac{dN'_\nu(E'_\nu)}{dE'_\nu}dE'_\nu (1+z)^{-3},
\ee
and considering that the differential neutrino flux at $z=0$ is $\varphi(E_\nu)= \frac{c}{4\pi}\frac{dn_\nu}{dE_\nu}$, it follows that
\begin{multline}
 \varphi_{\nu_\alpha}(E_\nu)= \frac{c}{4\pi\,H_0}\int_{0}^{z_{\rm max}}\frac{dz \,R_{\rm CGRB}(z)}{\sqrt{\Omega_\Lambda+\Omega_{\rm m}(1+z)^3}}\\
  \left[\frac{dN_{\nu_e,z}(E_\nu(1+z)}{dE_{\nu,z}}P_{e\alpha}+ \frac{dN_{\nu_\mu,z}(E_\nu(1+z)}{dE_{\nu,z}}P_{\mu\alpha} \right].
\end{multline}
Here, we adopt $H_0=70\,{\rm km\ s^{-1}\,Mpc^{-1}}$, and we also include the effect of neutrino oscillation during their propagation by multiplying the emitted neutrino spectrum of the flavors $e$ and $\mu$ by  $P_{e\alpha}$ and $P_{\mu\alpha}$, which represent the probabilities of oscillation to the flavor of interest to be observed on Earth, $\alpha=\left\{e,\mu,\tau \right\}$. These probabilities depend on three mixing angles ($\theta_{12}$, $\theta_{23}$, $\theta_{13}$) and a phase ($\delta_{\rm cp}$) that characterize that the mixing matrix $U_{\alpha\beta}$. We adopt the following values obtained by \cite{esteban2020} through global fits to data from neutrino baseline experiments:
\be
\theta_{12}&=& 32.4^\circ\\
\theta_{23}&=& 49^\circ\\
\theta_{13}&=& 8.6^\circ\\
\delta_{\rm cp}&=& 197^\circ,
\ee
and they yield the following values for the oscillation probabilities: $P_{ee}\simeq 0.552$, $P_{e\mu}\simeq 0.171$, $P_{e\tau}\simeq 0.276$, and $P_{\mu\tau}\simeq 0.375$. 

\begin{figure*}
    \centering
    \begin{subfigure}[t]{0.49\textwidth}
        \centering                          
        \includegraphics[width=0.5\linewidth,trim= 160 30 180 40]{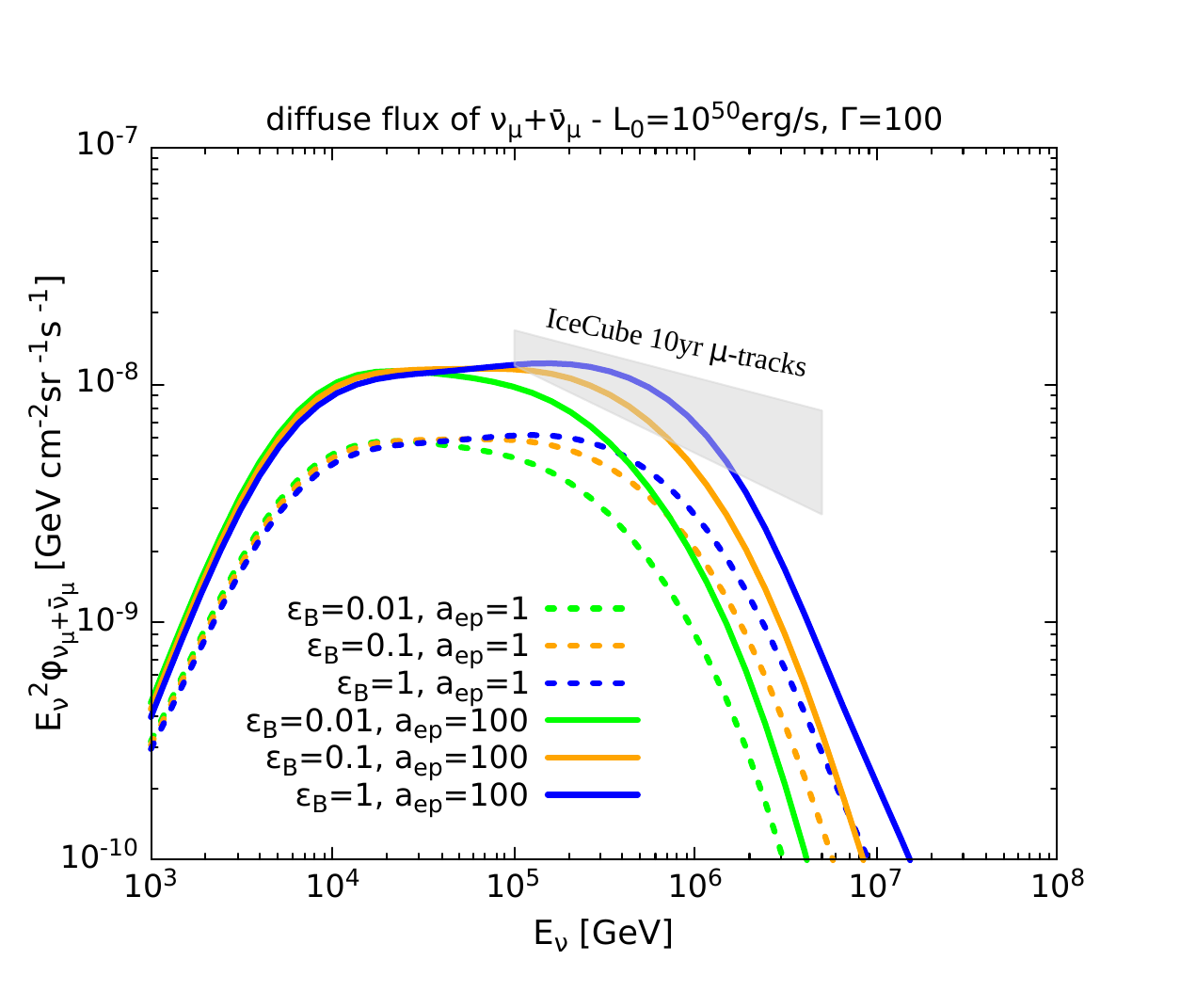} 
    \end{subfigure}
    \hfill
    \begin{subfigure}[t]{0.49\textwidth}
        \centering
        \includegraphics[width=0.5\linewidth,trim= 160 30 180 40]{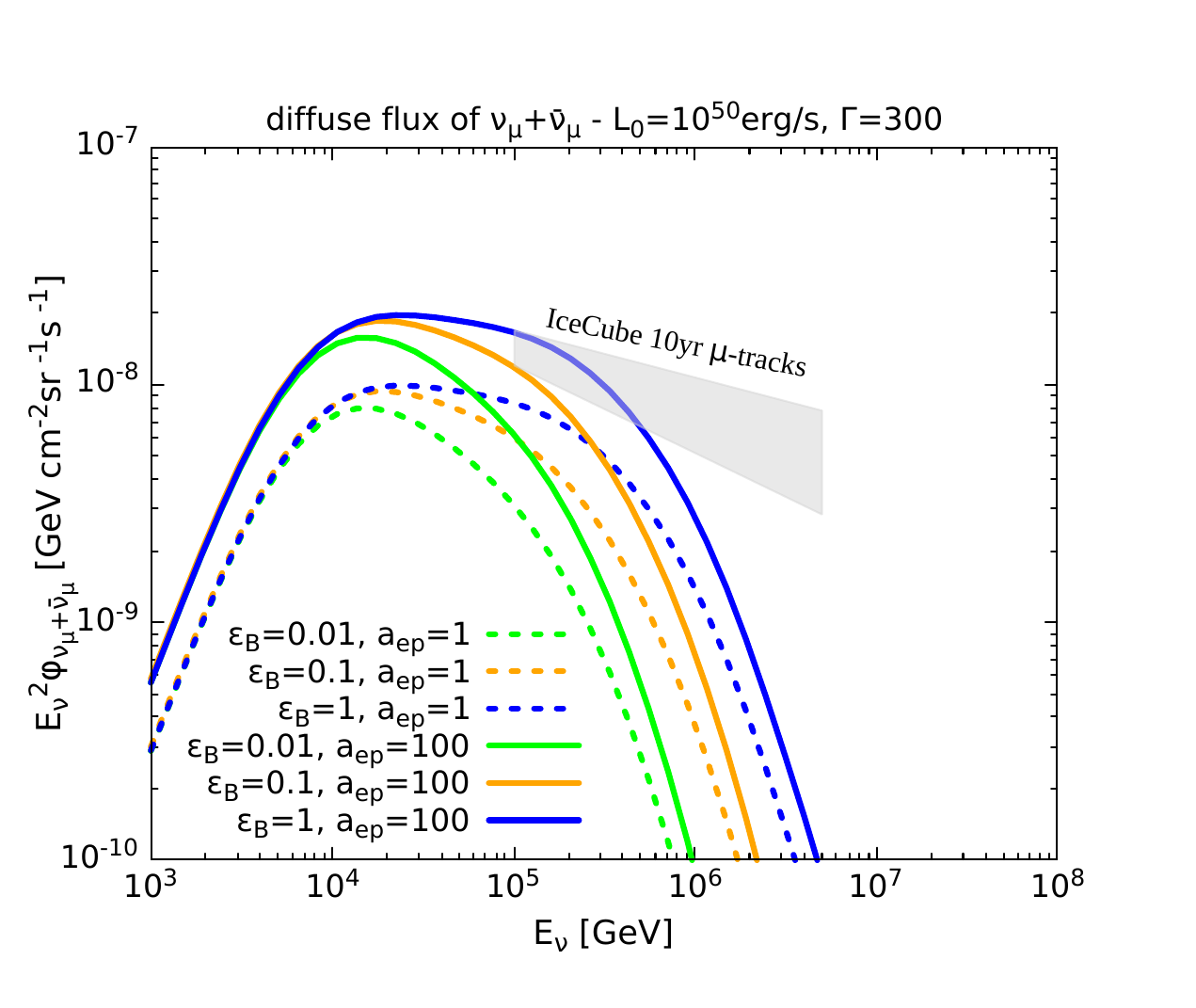} 
    \end{subfigure}

    \centering
    \begin{subfigure}[t]{0.49\textwidth}
        \centering                          
        \includegraphics[width=0.5\linewidth,trim= 160 30 180 40]{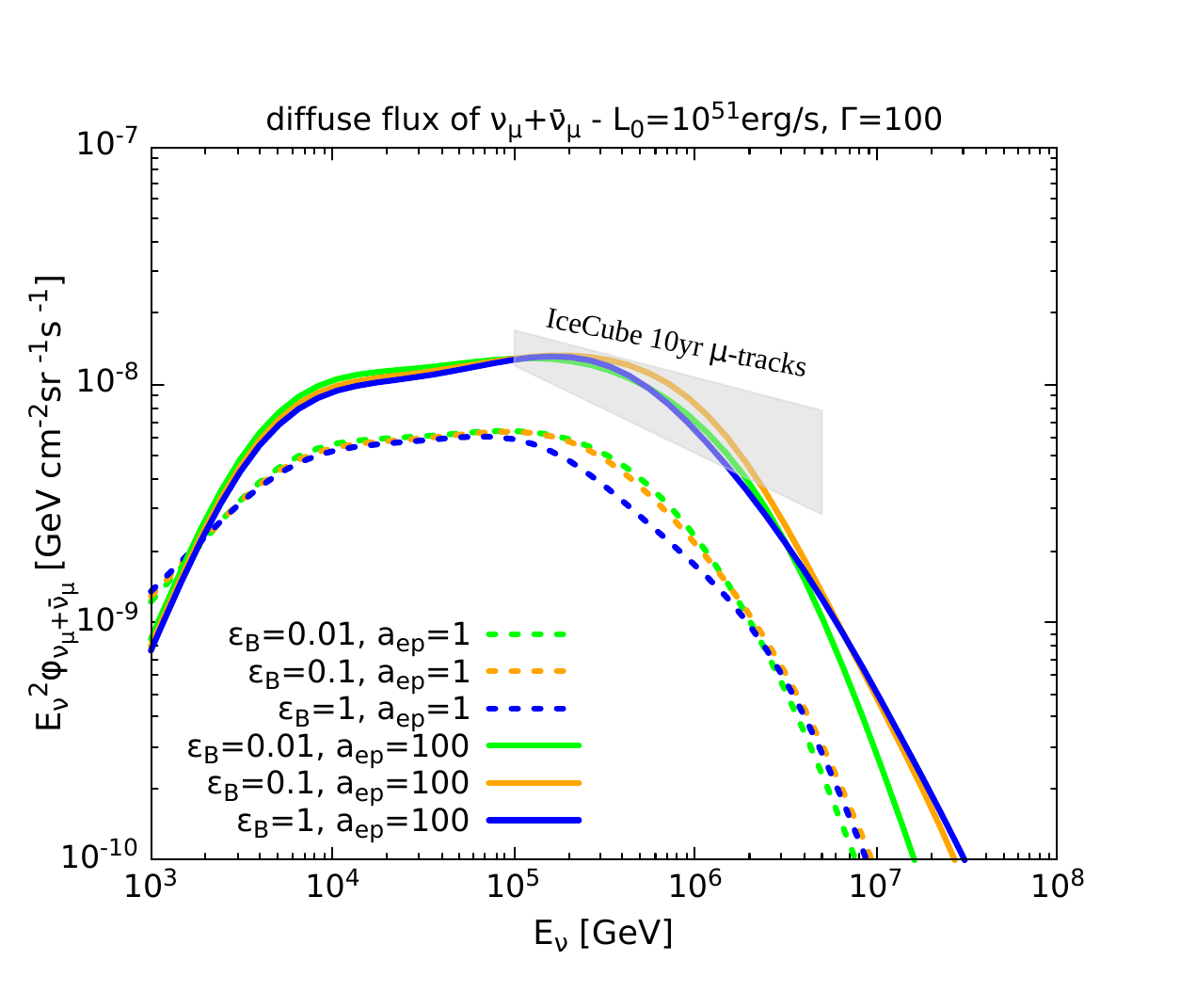} 
    \end{subfigure}
    \hfill
    \begin{subfigure}[t]{0.49\textwidth}
        \centering
        \includegraphics[width=0.5\linewidth,trim= 160 30 180 40]{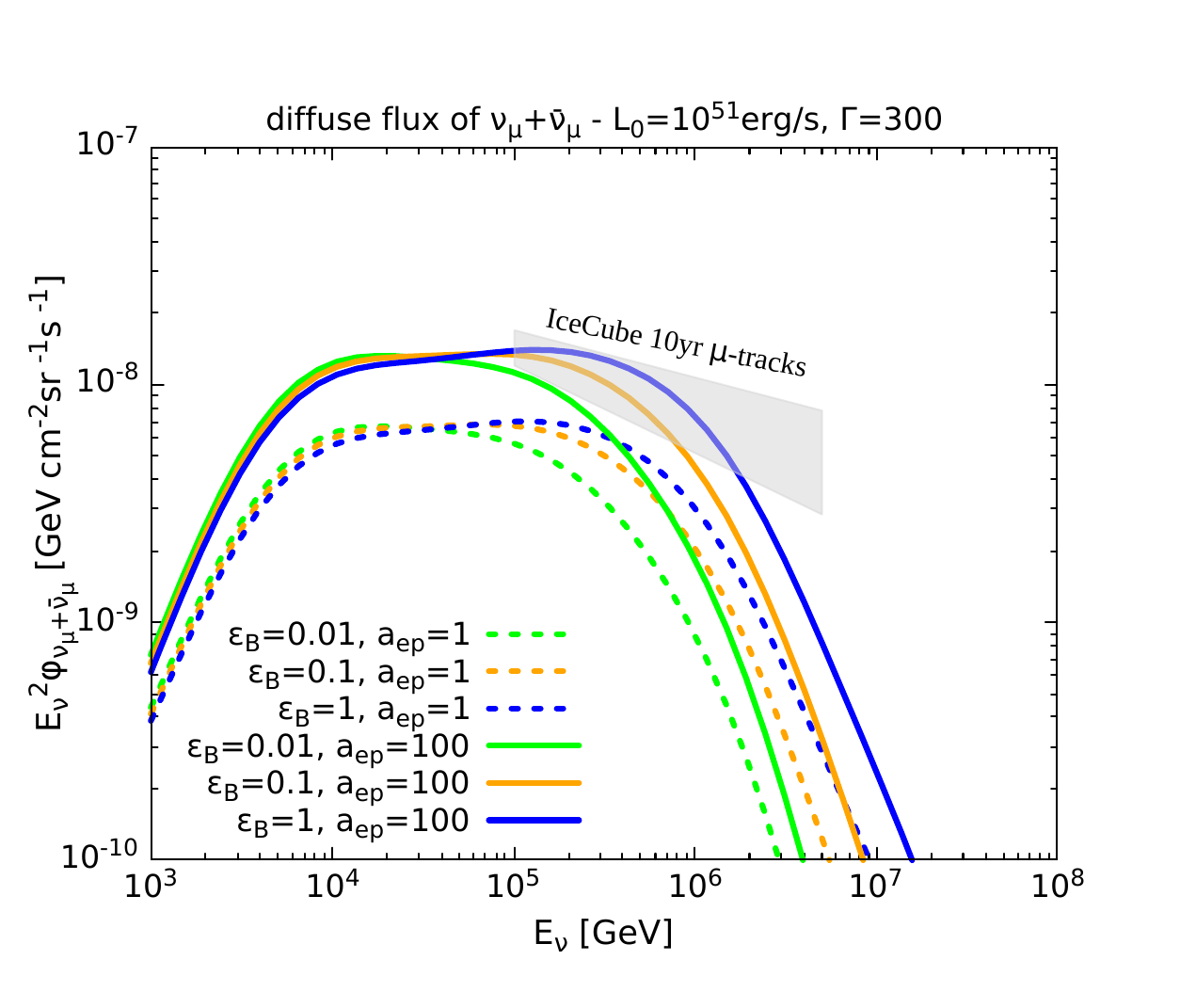} 
    \end{subfigure}
    \caption{Diffuse flux of muon neutrinos and antimuons for $L_0=10^{50}{\rm erg/s}$ (top panels) and $L_0=10^{51}{\rm erg/s}$ (bottom panels). The results obtained with $\Gamma=100$ and $\Gamma=300$ are shown in the left and right panels, respectively. {The green, orange, and blue curves refer to the cases of $\epsilon_B=0.01$, $\epsilon_B=0.1$, and $\epsilon_B=1$, respectively. The dashed curves correspond to $a_{ep}=1$ and the solid ones to $a_{ep}=100$.}}\label{fig14:numu}
                \end{figure*}

In Fig. \ref{fig14:numu}, we show the resulting muon neutrino and antineutrino fluxes for $L_0=10^{50}{\rm erg/s}$ and $L_0=10^{51}{\rm erg/s}$, {with $\Gamma=100$ in the left panels and $\Gamma=300$ in the right ones.} We include the cases of different values of the magnetic fields and of the proton-to-electron ratio $a_{ep}$ in the figure. The shape and relative importance of the resulting fluxes for the different cases studied can be understood by looking at the distributions of the pions, muons and kaons corresponding to each case. 

The normalizing constant mentioned above is fixed {at $A_{\rm CGRB}=8\times 10^{-8}M_\odot^{-1}$ for $L_0=10^{51}{\rm erg/s}$, $A_{\rm CGRB}=8\times 10^{-7}M_\odot^{-1}$ for $L_0=10^{50}{\rm erg/s}$ and $\Gamma=100$, and $A_{\rm CGRB}=4\times 10^{-7}M_\odot^{-1}$ for $L_0=10^{50}{\rm erg/s}$ and $\Gamma=300$. These values were obtained in order to have our resulting illustrative fluxes for $a_{ep}=100$ at the level of the fit obtained with IceCube 10-year data for muon tracks \citep{stettner2019}.}

%
\begin{figure*}
    \centering
    \begin{subfigure}[t]{0.49\textwidth}
        \centering                          
        \includegraphics[width=0.5\linewidth,trim= 140 30 150 35]{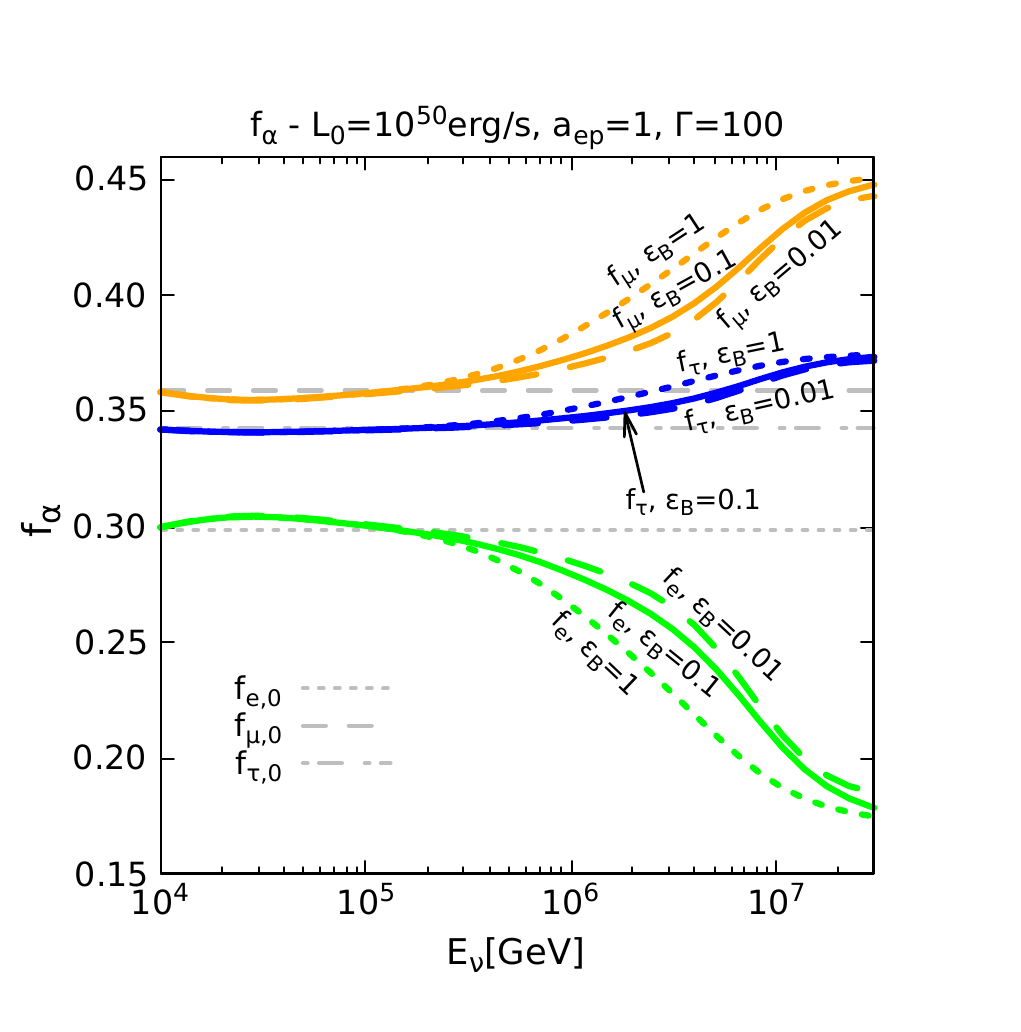} 
    \end{subfigure}
    \hfill
    \begin{subfigure}[t]{0.49\textwidth}
        \centering
        \includegraphics[width=0.5\linewidth,trim= 140 30 150 35]{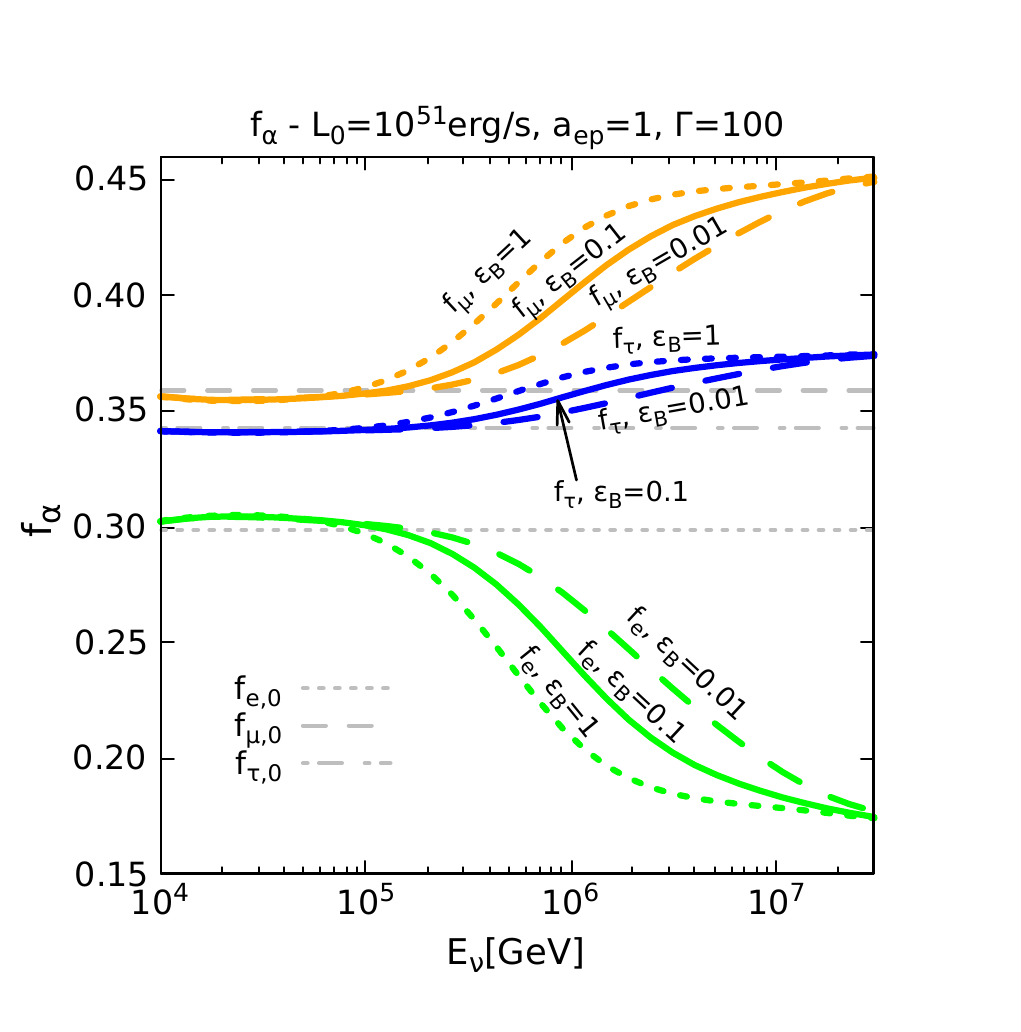} 
    \end{subfigure}
    
    \begin{subfigure}[t]{0.49\textwidth}
        \centering                          
        \includegraphics[width=0.5\linewidth,trim= 140 30 150 35]{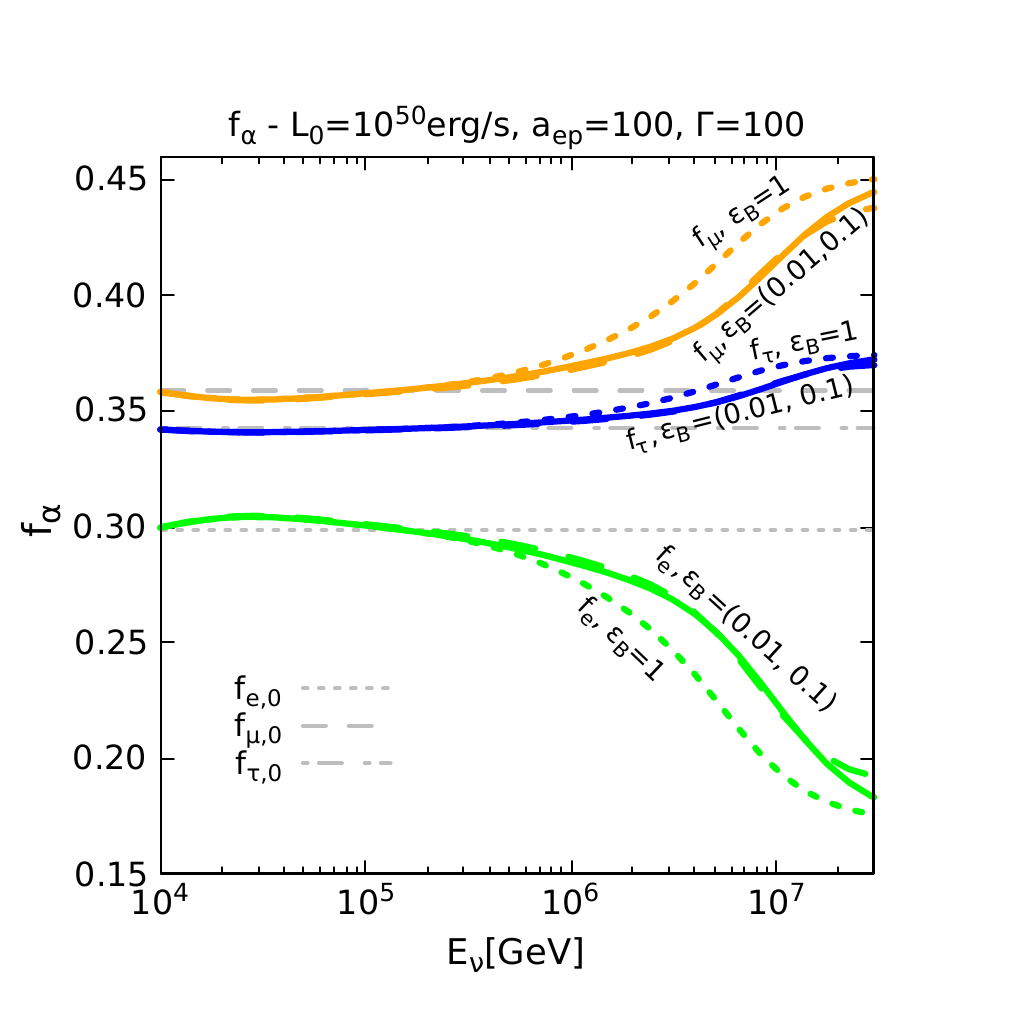} 
    \end{subfigure}
    \hfill
    \begin{subfigure}[t]{0.49\textwidth}
        \centering
        \includegraphics[width=0.5\linewidth,trim= 140 30 150 35]{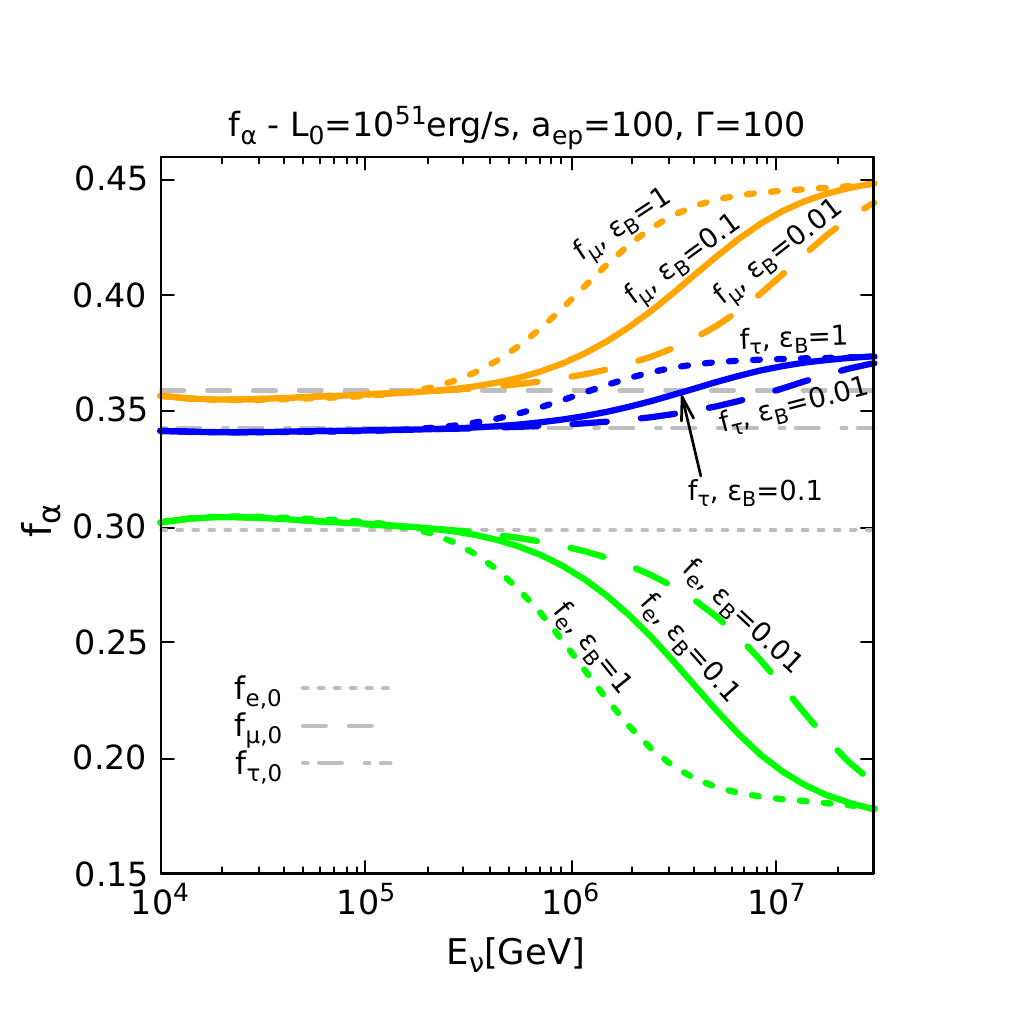} 
    \end{subfigure}     
    \caption{Energy dependent neutrino flavor ratios for $L_0=10^{50}{\rm erg/s}$ (left panels) and $L_0=10^{51}{\rm erg/s}$ (right panels) {for $\Gamma=100$. The top plots correspond to $a_{ep}=1$ and the bottom ones to $a_{ep}=100$. The green, orange, and blue curves correspond to the electron, muon, and tau flavor ratios, respectively. The long-dashed, solid, and short-dashed lines refer to the results for $\epsilon_B=0.01$, $\epsilon_B=0.1$, and $\epsilon_B=1$, respectively. The flavor ratios expected without considering losses are indicated with gray dotted lines for the electron flavor, gray long-dashed lines for the muon flavor, and gray long-short dashed lines for the tau flavor.}}
    \label{fig15:falphaG100}
\end{figure*}
\begin{figure*}
    \centering
    \begin{subfigure}[t]{0.49\textwidth}
        \centering                          
        \includegraphics[width=0.5\linewidth,trim= 140 30 150 35]{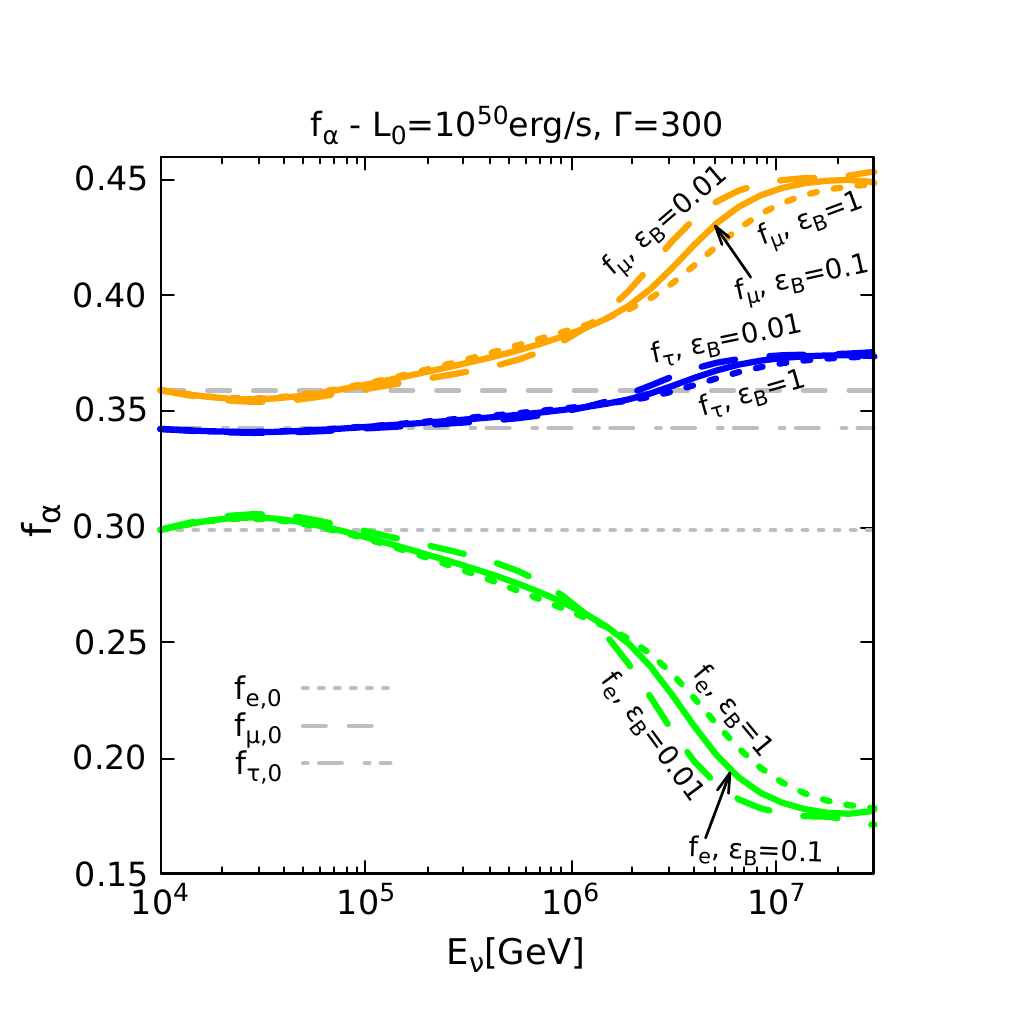} 
    \end{subfigure}
    \hfill
    \begin{subfigure}[t]{0.49\textwidth}
        \centering
        \includegraphics[width=0.5\linewidth,trim= 140 30 150 35]{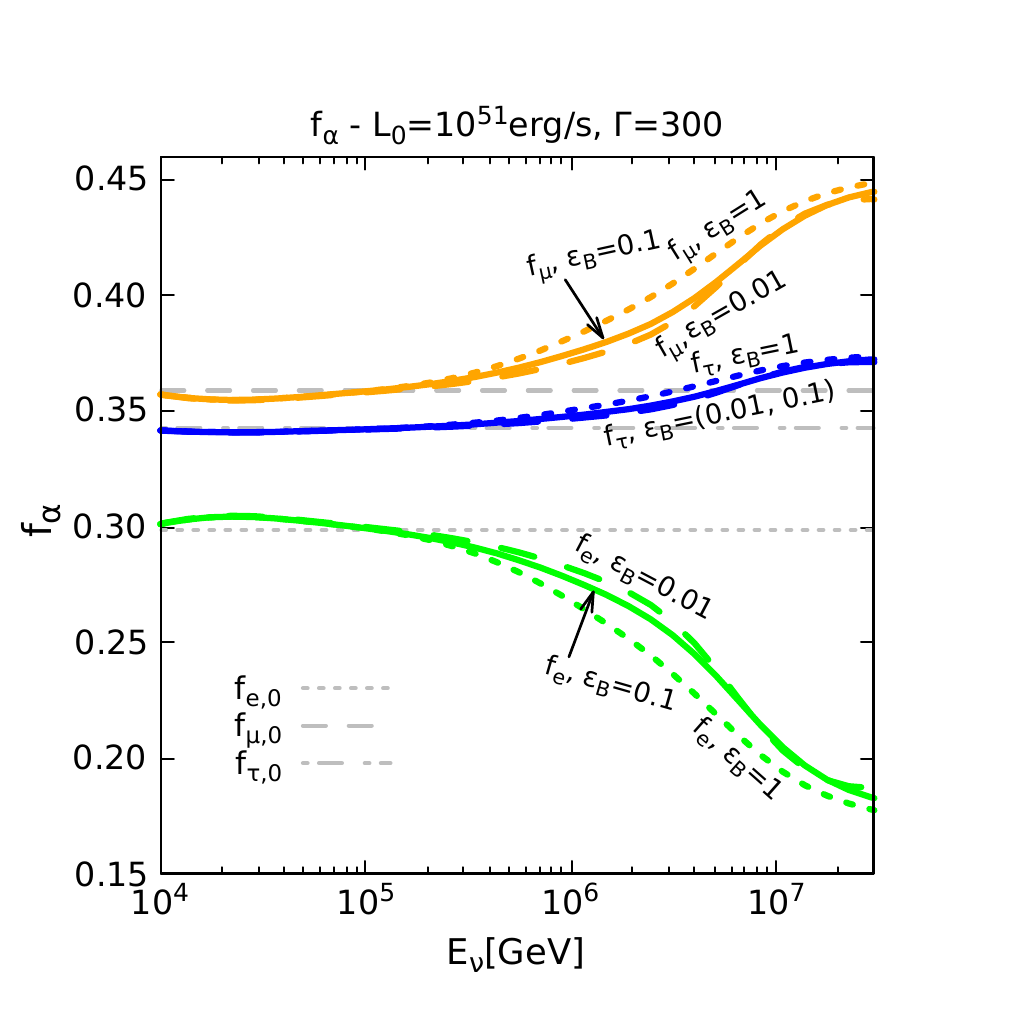} 
    \end{subfigure}
    \caption{Energy dependent neutrino flavor ratios for $L_0=10^{50}{\rm erg/s}$ (left panels) and $L_0=10^{51}{\rm erg/s}$ (right panels). {Both plots are for $\Gamma=300$. The green, orange, and blue curves correspond to the electron, muon, and tau flavor ratios, respectively. The long-dashed, solid, and short-dashed lines refer to the results for $\epsilon_B=0.01$, $\epsilon_B=0.1$, and $\epsilon_B=1$, respectively. The flavor ratios expected without considering losses are indicated with gray dotted lines for the electron flavor, gray long-dashed lines for the muon flavor, and gray long-short dashed lines for the tau flavor.}}    \label{fig16:falphaG300}
\end{figure*}
\begin{figure*}
    \centering
    \begin{subfigure}[t]{0.49\textwidth}
        \centering                          
        \includegraphics[width=0.5\linewidth,trim= 150 30 150 35]{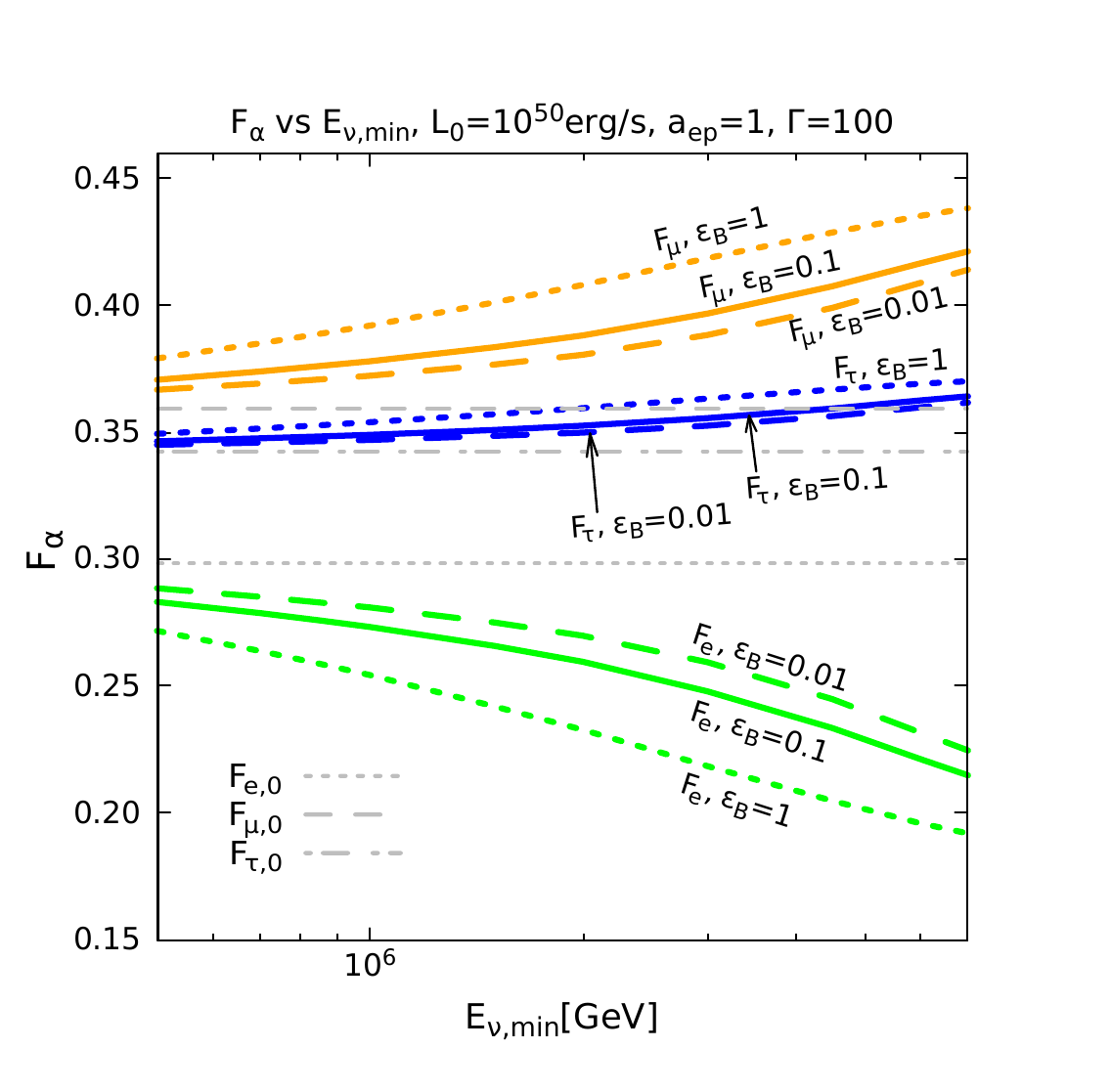} 
    \end{subfigure}
    \hfill
    \begin{subfigure}[t]{0.49\textwidth}
        \centering
        \includegraphics[width=0.5\linewidth,trim= 150 30 150 35]{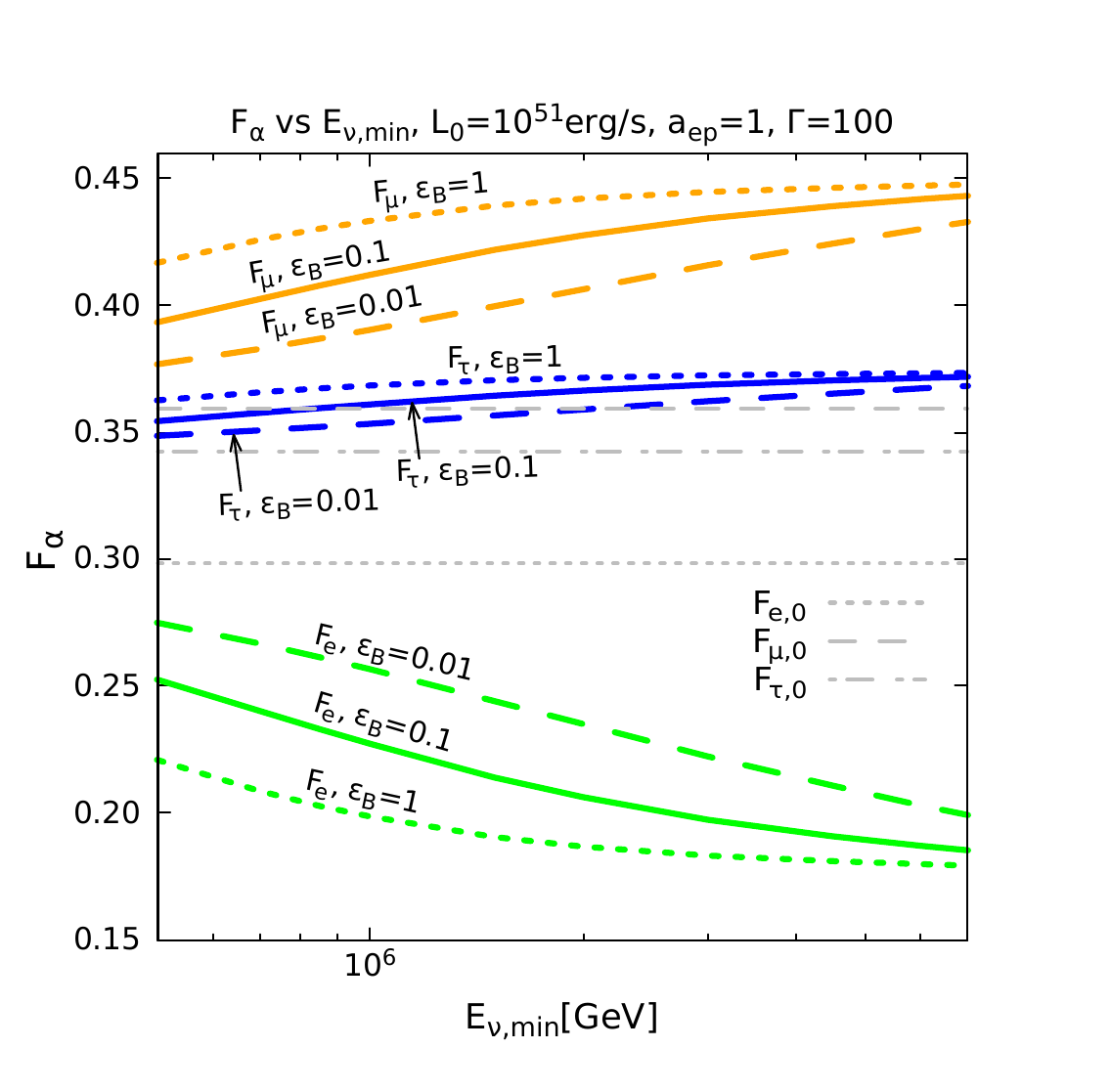} 
    \end{subfigure}
    
    \begin{subfigure}[t]{0.49\textwidth}
        \centering                          
        \includegraphics[width=0.5\linewidth,trim= 150 30 150 35]{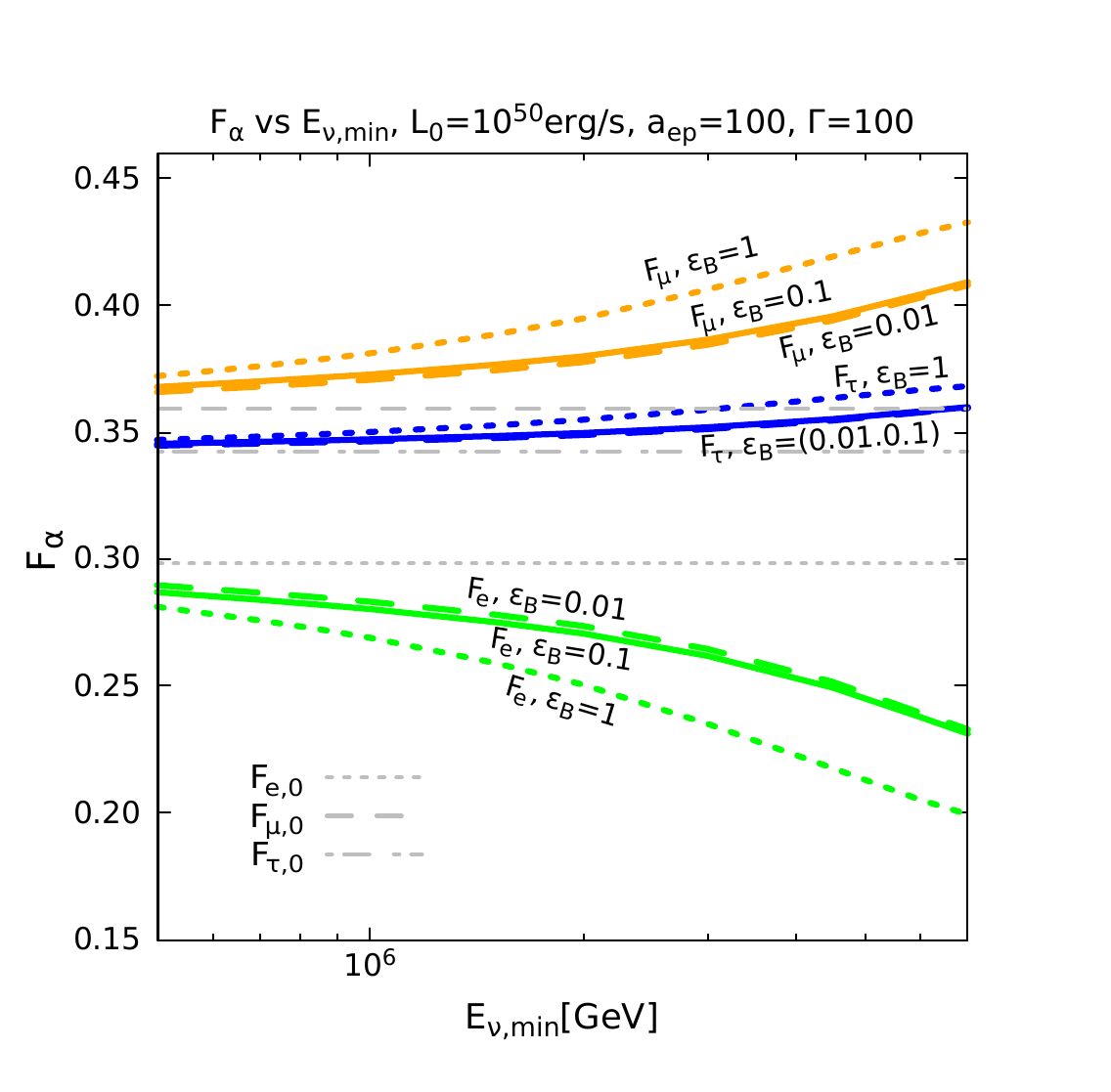} 
    \end{subfigure}
    \hfill
    \begin{subfigure}[t]{0.49\textwidth}
        \centering
        \includegraphics[width=0.5\linewidth,trim= 150 30 150 35]{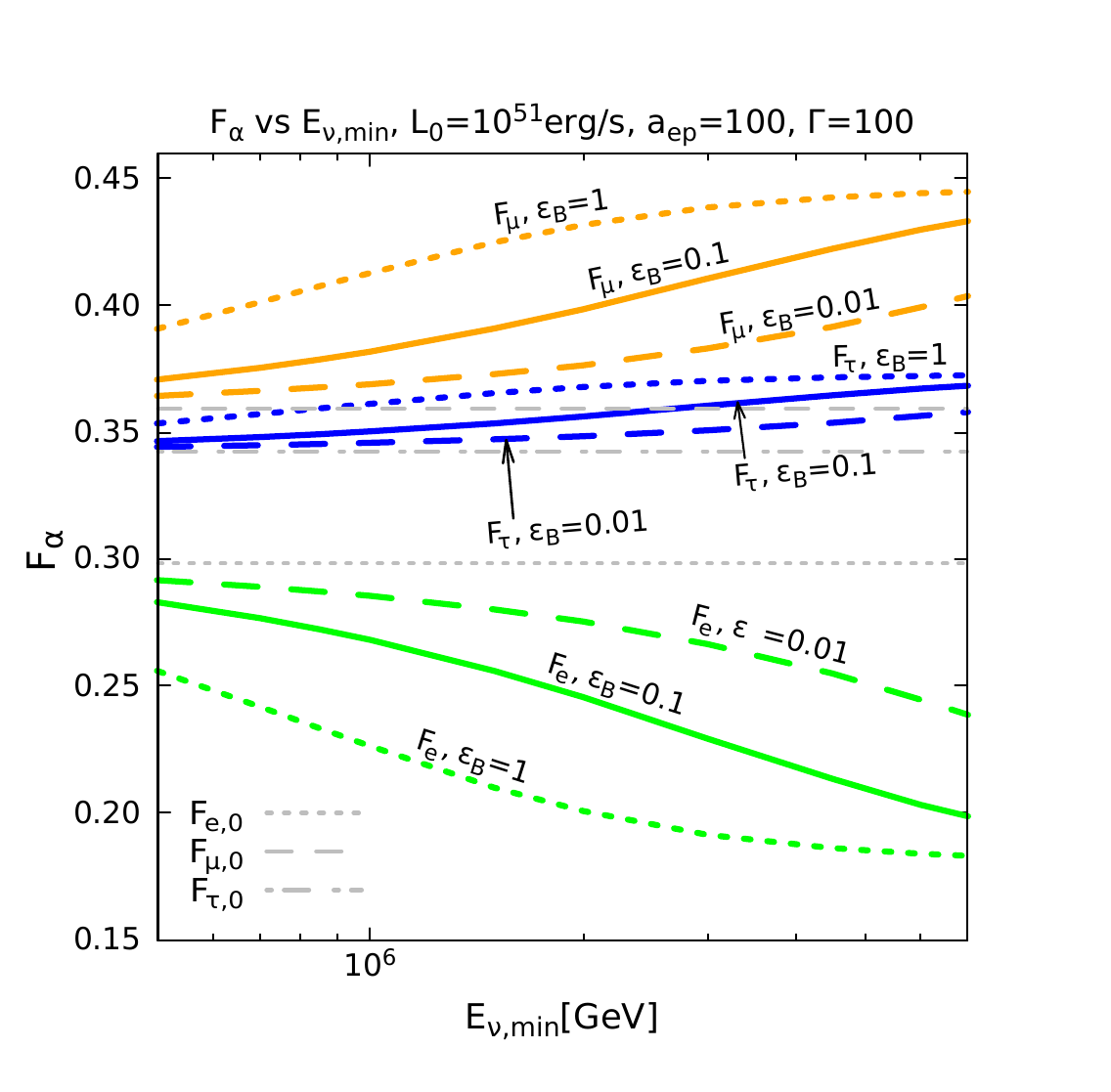} 
    \end{subfigure}     
    \caption{Flavor ratios of the integrated fluxes of neutrinos with energy above $E_{\nu,\rm min}$ for $L_0=10^{50}{\rm erg/s}$ (left panels) and $L_0=10^{51}{\rm erg/s}$ (right panels). All plots are for $\Gamma=100$. {The top plots correspond to $a_{ep}=1$ and the bottom ones to $a_{ep}=100$. The green, orange, and blue curves correspond to the electron, muon, and tau flavor ratios, respectively. The long-dashed, solid, and short-dashed lines refer to the results for $\epsilon_B=0.01$, $\epsilon_B=0.1$, and $\epsilon_B=1$, respectively. The flavor ratios expected without considering losses are indicated with gray dotted lines for the electron flavor, gray long-dashed lines for the muon flavor, and gray long-short dashed lines for the tau flavor.}}\label{fig17:FalphaG100}
\end{figure*} 
\begin{figure*}
    \centering
    \begin{subfigure}[t]{0.49\textwidth}
        \centering                          
        \includegraphics[width=0.5\linewidth,trim= 125 30 177 0]{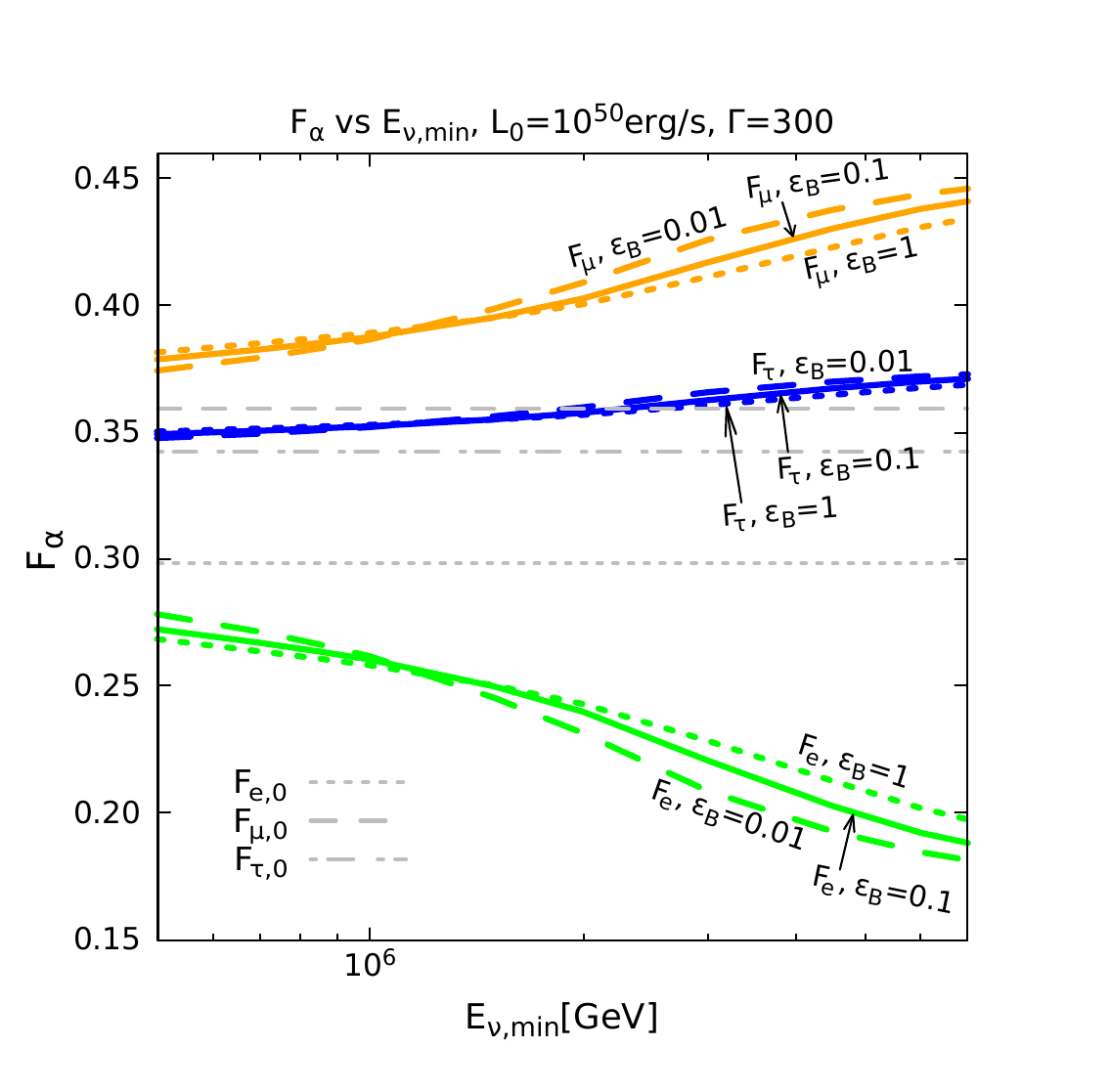} 
    \end{subfigure}
    \hfill
    \begin{subfigure}[t]{0.49\textwidth}
        \centering
        \includegraphics[width=0.5\linewidth,trim= 125 30 177 0]{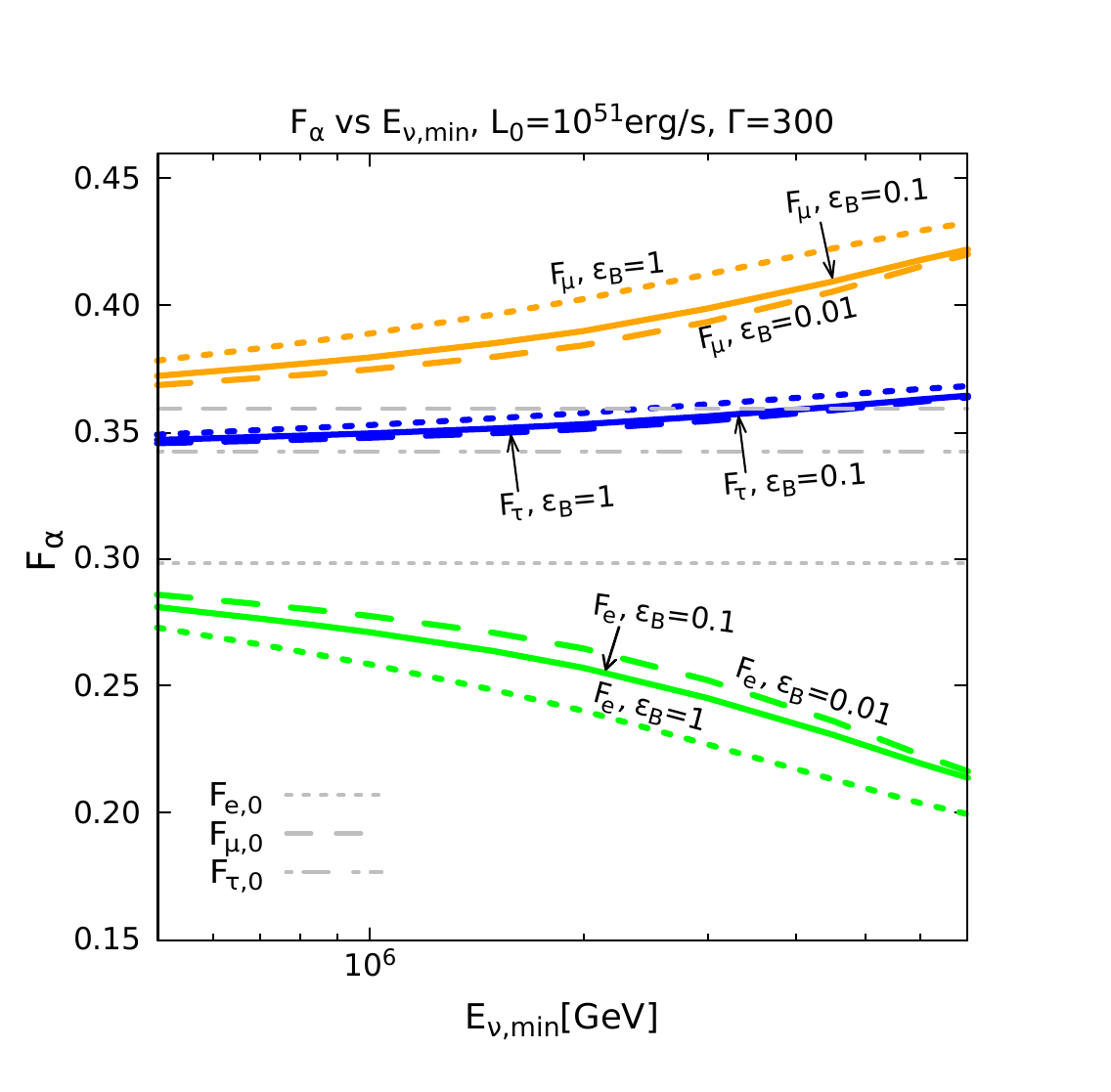} 
    \end{subfigure}
    \caption{Flavor ratios of the integrated fluxes of neutrinos with energy above $E_{\nu,\rm min}$ for $L_0=10^{50}{\rm erg/s  }$ (left panels) and $L_0=10^{51}{\rm erg/s}$ (right panels). {Both plots are for $\Gamma=100$. The green, orange, and blue curves correspond to the electron, muon, and tau flavor ratios, respectively. The long-dashed, solid, and short-dashed lines refer to the results for $\epsilon_B=0.01$, $\epsilon_B=0.1$, and $\epsilon_B=1$, respectively. The flavor ratios expected without considering losses are indicated with gray dotted lines for the electron flavor, gray long-dashed lines for the muon flavor, and gray long-short dashed lines for the tau flavor.}}\label{fig18:FalphaG300}
\end{figure*}

{We note that even for $A_{\rm CGRB}=8\times 10^{-7}M_\odot^{-1}$}, the corresponding local rate of choked jet events is $R_{\rm CGRB}(z=0)\approx 12\,{\rm yr^{-1}Gpc^{-3}}$, which is within the range of plausible values when taking into account that a certain fraction $f_{\rm jet}$ of the total core-collapse SNe harbor jets \citep[see, e.g.,][]{senno2016,denton2018,he2018,fasano2021}. {The precise fraction is actually unknown, but it plays a crucial role in establishing a link between CGRBs and such SNe and therefore in associating the location of neutrino events to a possible location of an SN. Recent studies in this area {\citep[e.g.,][]{guetta2020,chang2022}} suggest that the mentioned fraction should be low, $f_{\rm jet}\lesssim 0.1$, and although no correlation was confirmed for SNe and neutrino events, CGRBs can still be the dominant sources of the detected diffuse neutrino flux. }

To make a comparison of the neutrino fluxes of the different flavors, it is useful to obtain the energy dependent flavor ratios,
$$
f_{\alpha}(E_\nu)= \frac{\varphi_{\nu_{\alpha}}(E_\nu)}{\varphi_{\nu_{e}}(E_\nu)+ \varphi_{\nu_{\mu}}(E_\nu)+ \varphi_{\nu_{\tau}}(E_\nu)}.
$$ 
This includes all the effects of energy losses suffered by the parent particles. {We plot the corresponding ratios in Figs.  \ref{fig15:falphaG100} and \ref{fig16:falphaG300} for $\Gamma=100$ and $\Gamma=300$, respectively.}
We also consider, for reference, the standard estimate derived from the fact that for pion decays, a ratio of $2:1$ of muon neutrinos to electron neutrinos is produced at the sources, and neglecting all losses and without considering the decay spectra of the different particles, it follows that
\be
f_{e,0}&= &2 P_{e\mu}+ P_{ee}\simeq 0.30 \\
f_{\mu,0}&= &2 P_{\mu\mu}+ P_{e\mu}\simeq 0.36 \\
f_{\tau,0}&= &2 P_{\mu\tau}+ P_{e\tau}\simeq 0.34.
\ee
{This estimate gives roughly $(f_{e,0}:f_{\mu,0}:f_{\tau,0})\propto (1:1:1)$  \citep[e.g.,][]{bustamante2019,bustamente2020}.} 
Therefore, it can be seen in the mentioned figures that as the neutrino energy increases, the flavor ratios obtained as a result of our full calculation differ more from the expected outcome if losses are neglected. Specifically, the losses discussed (interactions with soft photons, and synchrotron emission) cause the electron flavor ratio to decrease to $\approx 0.17$ , the muon ratio to increase up to $\approx 0.45$, and the tau flavor to increase slightly to $0.36$. The smooth transition to these values takes place at different energies for the different cases considered.
 
{For $L_0=10^{51}{\rm erg/s}$ in particular, the flavor ratios are modified at energies lower than for $L_0=10^{50}{\rm erg/s}$, given that the interaction rates are higher in the former cases than in the latter ones. When comparing the top and lower panels of Fig. \ref{fig15:falphaG100} for $\Gamma=100$,  the flavor ratios are shown to be affected at slightly lower energies for $a_{ep}=1$ than for $a_{ep}=100$, and it is the interactions with more electron synchrotron radiation that makes the difference.}

{As discussed above, for $\Gamma=100$ and $L_0=10^{51}{\rm erg/s}$, synchrotron cooling of pions and muons become more relevant for higher values of $\epsilon_B$, and this affects the flavor ratios at lower energies. However, for $\Gamma=100$ and $L_0=10^{50}{\rm erg/s}$, synchrotron cooling of muons is only relevant if $\epsilon_B=1$, and therefore the cases with lower magnetic fields have similar flavor ratios (long dashed and solid lines in left panels of Fig. \ref{fig15:falphaG100}) and are affected mostly by pion and muon interactions with soft photons.} 

{As for the cases with $\Gamma=300$, the flavor ratios we present in Fig. \ref{fig16:falphaG300} refer to $a_{ep}=1$. A different situation arises in the cases with $L_0=10^{50}{\rm erg/s}$ shown in the left panel as compared those with $L_0=10^{51}{\rm erg/s}$ shown in the right panel. In the former cases, unlike the rest of the cases studied, the corresponding muon distributions extend only to energies below $\sim 10^5{\rm GeV}$ (see Fig. \ref{fig11:NpiNmuNK}, middle-left panel), and cooling is mostly dominated by the muon IC process, even for the case of $\epsilon_B=1$. In fact, at high energies, the IC cooling rate is in the Klein-Nishina regime (i.e., decreasing with the energy). This implies that increasing $\epsilon_B$ only increases the acceleration rate and therefore the maximum energy of the parent protons and of the produced pions and muons. However, since muons cool more slowly at higher energies in these particular cases (for $\Gamma=300$ and $L_0=10^{50}{\rm erg/s}$), the resulting curves for the flavor ratios for low values of $\epsilon_B$ appear slightly more affected by interactions at high energies as compared those for higher magnetic fields. This situation is different for $\Gamma=300$ and $L_0=10^{51}{\rm erg/s}$ as wells as in the rest of the cases studied, as the IC cooling of muons does not always dominate if the magnetic field is increased because synchrotron cooling can become important at high energies, and this results in similar flavor ratios to the obtained, for instance, using $\Gamma=100$ and $L=10^{51}{\rm erg/s}$. }

{Given the relatively low number of detected neutrino events of astrophysical origin, a more useful observable to be extracted from experiments may be the flavor ratios of the integrated fluxes. Depending on the energy range of integration, the effects discussed for the energy dependent ratios may also be appreciated in these flavor ratios.}
If we integrate the fluxes on the neutrino energy above a minimum energy $E_{\nu,\rm min}$
$$
\Phi_{\nu_\alpha}(E_{\nu,\rm min})=\int_{E_{\nu,\rm min}}^\infty dE_\nu \varphi_{\nu_\alpha}(E_\nu),
$$
the corresponding flavor ratios can be obtained as
\be
F_{\alpha}(E_{\nu,\rm min})= \frac{\Phi_{\nu_\alpha}(E_{\nu,\rm min})}{\Phi_{\nu_e}(E_{\nu,\rm min})+\Phi_{\nu_\mu}(E_{\nu,\rm min})+\Phi_{\nu_\tau}(E_{\nu,\rm min})}. 
\ee
 We plot the results in Figs. \ref{fig17:FalphaG100} and \ref{fig18:FalphaG300}. The plots show that the effects of cooling can be better observed by concentrating on the neutrino fluxes of the different flavors at energies higher than $\sim 5\times 10^{5}{\rm GeV}$. Otherwise, the more abundant events of lower energies tend to mask the departure from the standard values of the flavor ratios corresponding to no cooling of the secondary particles.

\section{Discussion}\label{sec:discussion}

We have studied in detail the processes leading to the production of high energy neutrinos in CGRBs by making use of steady-state kinetic equations to account for the cooling mechanisms of all the intervening particle populations, namely, electrons, protons, pions, kaons, and muons. Typically, the target photons considered for protons accelerated at the internal shocks are those that escape from the shocked jet head. In the present work, we also considered as targets the synchrotron photons emitted by electrons accelerated at the internal shocks as well as the protons. This implies a rise in the $p\gamma$ cooling rate {only in the cases with a powerful jet $L_0=10^{51}{\rm erg/s}$ and $\Gamma=100$}, and therefore means a decrease in the maximum proton energy. 

We have accounted for IC cooling to obtain the particle distributions of electrons as well as of muons. Pions and kaons produced in $p\gamma$ interactions are also affected by interactions with the mentioned target of soft photons. We found that these interactions can indeed cause important losses of the mentioned particles and therefore a decrease in the neutrino flux at high energies, even in the cases where the magnetic field is relatively low and no significant losses by synchrotron emission are expected. As a consequence of these losses, the flavor ratios become modified: The electron flavor ratio decreases, and the muon flavor ratio increases. This effect gradually manifests in the energy dependent flavor ratios at high energies, as can be seen in Figs. \ref{fig15:falphaG100}, \ref{fig16:falphaG300}, \ref{fig17:FalphaG100}, and \ref{fig18:FalphaG300}, where the different curves represent possible signatures of key physical conditions in the jets of CGRBs, such as the magnetic field, proton-to-electron ratio, Lorentz factor, and the jet power. {For these parameters, we adopted typically plausible values, according to several works describing the same general context. In particular, by requiring the flow to be optically thin at the internal shock region and that this region be placed inside the collimation shock, we made sure that the sets of parameters we used would imply shocks that are not mediated by radiation, since that allows for efficient particle acceleration. }

{A treatment like the present one could be applied to other compact sources, such as  tidal disruption events, where the magnetic field can be high and therefore significant cooling of pions and/or muons can take place \citep[e.g.,][]{senno2017}. The cores of AGN could also be studied by adapting the present model, such as in the case of NGC 1068. As described by \cite{murase2022}, the possible neutrino emission site could be relatively close to the central BH (e.g., in a corona). Another possibility is that it is located further away where protons would be accelerated through shocks generated by a wind from the accretion disk \citep[][]{lamastra2016}. In this latter case, a sub-gauss magnetic field would yield the expected no-loss flavor ratios, $\approx(1:1:1)$, while in the case of a corona, the magnetic field would be expected to be higher (possibly $B \gtrsim 1000\,{\rm G}$), leading to flavor ratios that would depart from the no-loss result for high neutrino energies. }

Still, in order to be able to measure the energy dependent flavor ratios, as well as the flavor ratios of the integrated fluxes for different minimum energies, more observation time and larger neutrino telescopes are necessary. {In general, such resources would allow important information on the physical conditions at the neutrino 
sources to be obtained. In the specific case of CGRBs,} these data would help probe the consequences of the generation of such explosive, gamma-ray hidden events, with the physical conditions normally expected for them, providing new clues as to whether these types of sources can indeed contribute significantly to the diffuse neutrino flux.

\vspace{6pt} 
\begin{acknowledgements}
  We thank ANPCyT and Universidad Nacional de Mar del Plata for their financial support through grants PICT 2021-GRF-T1-00725 and EXA1102/22, respectively.
\end{acknowledgements}

\end{document}